\documentclass[review]{elsarticle}

\usepackage[colorlinks,citecolor=blue,linktoc=all,linkcolor=cyan]{hyperref}
\usepackage{graphicx}

\usepackage[T1]{fontenc}
\usepackage{mathrsfs}             
\usepackage{slashed}              
\usepackage{amsmath}
\usepackage{amssymb}
\usepackage{amsbsy}  
\usepackage{amsfonts} 

\usepackage{wasysym} 
\usepackage{booktabs} 

\numberwithin{equation}{section}
\numberwithin{table}{section}
\numberwithin{figure}{section}

\journal{Progress in Particle and Nuclear Physics}

\usepackage{titlesec}
\usepackage{sectsty}
\titleformat{\section}{\normalfont\Large\bfseries}{\thesection}{1em}{}
\titleformat{\subsection}{\normalfont\large\bfseries}{\thesubsection}{1em}{}
\titleformat{\subsubsection}{\normalfont\normalsize\bfseries}{\thesubsubsection}{1em}{}

\begin{document}

\begin{frontmatter}

\title{Solar neutrino physics}

%
%

\author[1]{Xun-Jie Xu\corref{mycorrespondingauthor}}
\cortext[mycorrespondingauthor]{Corresponding author}
\ead{xuxj@ihep.ac.cn}
\address[1]{Institute of High Energy Physics, Chinese Academy of Sciences, Beijing 100049, China}
\author[2,3]{Zhe Wang}
\ead{wangzhe-hep@tsinghua.edu.cn}
\author[2,3]{Shaomin Chen}
\ead{chenshaomin@tsinghua.edu.cn}
\address[2]{Center for high energy physics,Tsinghua University, Beijing 100084, China}
\address[3]{Department of Engineering Physics, Tsinghua University, Beijing 10084, China}
 
\begin{abstract}
As a free, intensive, weakly interacting, and well directional messenger, solar neutrinos have been driving both solar physics 
and neutrino physics developments for more than half a century. Since more extensive and advanced neutrino experiments 
are under construction, being planned or proposed, we are striving toward an era of precise and comprehensive 
measurement of solar neutrinos in the next decades. 
In this article, we review recent theoretical and experimental progress achieved in solar neutrino 
physics. 
We present not only an introduction to neutrinos from the standard solar model and the standard flavor evolution, but also a compilation of a variety of new physics that could affect and hence be probed by solar neutrinos.
After reviewing the latest 
techniques and issues involved in the measurement of solar neutrino spectra and background reduction, we provide our 
anticipation on the physics gains from the new generation of neutrino experiments.
\end{abstract}


\end{frontmatter}

\newpage

\thispagestyle{empty}
\tableofcontents

\newpage
\section{Introduction}

\textit{``How does the Sun shine? Does the neutrino have a mass?
	Can solar neutrinos be used to test the theory of stellar evolution?
	To explore the unification of strong, weak, and electromagnetic forces?''}
John N. Bahcall raised these questions at the very beginning of his famous book~\cite{Bahcall1989}. Starting
from Eddington's speculation in 1920~\cite{Eddington1920}\footnote{It is worth mentioning that while the contraction hypothesis (i.e.,~the solar energy was from gravitational contraction) was prevailing by then,  Eddington discussed abundantly contradictory consequences of the the contraction hypothesis in his paper~\cite{Eddington1920} and conceived that the ``sub-atomic'' energy might actually power the Sun.}, 
followed by the establishment of the theory of stellar nucleosynthesis (in the 1930s) and  decades of experimental observations of solar neutrinos (since the 1960s)
and theoretical efforts,  our understanding of the Sun over a century  has evolved and eventually
led to a surprising and profound discovery---neutrino masses---which are of crucial importance to the most fundamental physics.

The theory of stellar nucleosynthesis anticipates that the Sun produces an enormous amount of neutrinos from nuclear fusion. They can be used as a unique probe to the solar energy production mechanism, inspiring R. Davis to carry out his pioneering experiment at Homestake in 1968~\cite{Davis:1968cp} via the Pontecorvo-Alvarez~\cite{Pontecorvo,Alvarez}
inverse $\beta$ decay: $\nu_{e}~+{}^{37}\text{Cl}\rightarrow e^{-}~+{}^{37}\text{Ar}$. The first result~\cite{Davis:1968cp} came out as an upper limit of 3 Solar Neutrino Units (SNU, $1$~SNU$\equiv10^{-36}$~events/atom/sec)
which is lower than the theoretical prediction published at the same time~\cite{Bahcall:1968hc}\footnote{In Ref.~\cite{Bahcall:1968hc}, the theoretical values were  predicted to be 21,
11, 7.7, 4.4, and 11 SNU for five models considered by Bahcall {\it et al}.
The differences among these models include the proton-proton reaction
rate, the hydrogen/helium fraction, the metallicity, the central temperature
and density, which were not well determined by then. Despite the model dependence,
it was clear that the theoretical values were all higher than the
experimental upper limit, 3 SNU.}.
From 1970 to 1994, the Homestake experiment continued data taking
with improved techniques to discriminate the signal from backgrounds,
and eventually obtained a precise measurement: $2.56\pm0.16\pm0.16$~SNU~\cite{Cleveland:1998nv},
which is only one-third of the prediction~\cite{Bahcall:1987jc}.
Similar deficits were also confirmed by gallium experiments (GALLEX/GNO~\cite{Gallex1992}
and SAGE~\cite{sage2002}) and water Cherenkov detectors (Kamiokande~\cite{kamiokande1989},
Super-Kamiokande~\cite{sk1998}, and SNO~\cite{sno2002}). The discrepancy
between the observation of solar neutrinos and the prediction became the so-called ``solar neutrino problem.''

Like Arthur B. McDonald stated in his Nobel lecture~\cite{McDonald:2016ixn},
\textit{``possible reasons for the discrepancy could have been that
	the experiment or the theory was incorrect.''} 
The discrepancy motivated many theoretical efforts to address the solar neutrino problem.
From the 1970s to the early 1990s, a variety of theoretical interpretations
to the solar neutrino problem were  proposed and investigated, including
neutrino oscillation\footnote{The concept of neutrino oscillation was first proposed  by Pontecorvo
	in 1957~\cite{Pontecorvo:1957cp,Pontecorvo:1957qd}, a decade earlier
	than Davis's Homestake experiment. The original consideration was
	$\nu\leftrightarrow\overline{\nu}$ oscillation, while oscillation
	due to flavor mixing was later considered by Pontecorvo, Maki, Nakagawa
	and Sakata in the 1960s~\cite{Maki:1962mu,Pontecorvo:1967fh}.
}
 with the Mikheyev-Smirnov-Wolfenstein (MSW) effect~\cite{Wolfenstein:1977ue,Mikheev:1986gs,Mikheev:1986wj}, 
oscillation in vacuum~\cite{Glashow:1987jj},
spin or spin-flavor precession due to neutrino magnetic moments~\cite{Cisneros:1970nq,Okun:1986na,Lim:1987tk,Akhmedov:1988uk},
flavor conversion due to non-standard interactions (NSI) of massless neutrinos~\cite{Wolfenstein:1977ue,Wolfenstein:1979ni,Roulet:1991sm,Guzzo:1991hi},
neutrino decays~\cite{Bahcall:1972my,Berezhiani:1987gf}, etc. Eventually,
neutrino oscillation with the MSW effect and a large mixing angle
(LMA)  became the standard solution (MSW-LMA) to the solar neutrino problem.

Neutrino oscillation implies that solar neutrinos, initially being produced as $\nu_e$, may change their flavors to $\nu_{\mu}$ or $\nu_{\tau}$ along their path to the Earth. As a consequence, only a fraction of the neutrinos appear as $\nu_{e}$ in the detector.
The survival probability of
$\nu_{e}$ in the MSW-LMA solution is approximately $\sin^{2}\theta_{12}\approx 0.3$ at high energies ($\sim10$ MeV, e.g.,~for $^{8}$B neutrinos)
and increases to $\cos^{4}\theta_{12}+\sin^{4}\theta_{12}\approx 0.55$ at low energies ($\lesssim1$
MeV, e.g.,~for pp neutrinos). Both values have so far been consistent with the
observations.   The flavor conversion to $\nu_{\mu}$ and $\nu_{\tau}$
has been indirectly probed by neutral current (NC) and electron scattering
events in SNO~\cite{sno2002}  and Super-Kamiokande~\cite{sk2001}. 
Moreover, the values of $\theta_{12}$
and $\Delta m_{21}^{2}$ in the LMA regime have been confirmed by the
KamLAND experiment, which measured $\theta_{12}$ and $\Delta m_{21}^{2}$
in long-baseline reactor neutrino oscillation, independent of solar neutrino observations.
Further measurements
from the Borexino experiment, which is dedicated to solar neutrino
observations and has identified pp, $^7$Be, pep, and CNO components~\cite{Borexino:2007kvk,BOREXINO:2014pcl,Borexino:2017rsf,BOREXINO:2018ohr,BOREXINO:2020aww}, agree well with the MSW-LMA solution. 

Next-generation large underground detectors such as Hyper-Kamiokande~\cite{Hyper-Kamiokande:2018ofw}, JUNO~\cite{JUNO:2015zny},
DUNE~\cite{DUNE2016}, JNE~\cite{Jinping:2016iiq}, THEIA~\cite{Theia:2019non},
as well as dark matter (DM) detectors\footnote{Solar neutrinos can be detected by DM detectors mainly via the coherent elastic neutrino-nucleus scattering (CE$\nu$NS) process~\cite{Freedman:1973yd}. For DM detection, this is regarded as a background known as the solar neutrino floor. Currently, multi-ton scale liquid xenon detectors such as XENONnT, PandaX, and LUX are approaching the solar neutrino floor---see Sec.~\ref{sub:DM} for further details.}, will usher in an era of precision measurement of solar neutrinos.
Given the experimental prospect and the verified theory, there is a crucial
question for future experiments:  what can be explored in the
precision measurement of solar neutrinos? 
The answer varies from different
perspectives: 

From the astrophysical viewpoint, a full spectrum of solar neutrinos
of all components (e.g., pep, hep, $^{13}{\rm N}$, $^{15}{\rm O}$,
etc.) is valuable to the study of solar and stellar physics. Some of the spectral components are not measured precisely while some have not been detected yet. 
In particular, the observation
of CNO neutrinos has just started\,---\,very recently the first observation is achieved by Borexino~\cite{BOREXINO:2020aww, Borexino:2022pvu}. The measurement
of CNO neutrino fluxes will be of great importance to the poorly known
metallicity of the Sun and also to the study of heavier
($\gtrsim1.3M_{\astrosun}$) stars in which the CNO cycle is believed
to dominate the energy production. As the nearest star, the Sun provides
the unique opportunity to measure neutrinos precisely  from stellar
nucleosynthesis.

From the perspective of particle physics, neutrino masses point toward
new physics, while a variety of new physics might affect  solar
neutrino observations. The precision measurement of solar neutrino
spectra allows us to search for new physics signals.   Currently, the
low- and high-energy regimes of the MSW-LMA solution have been measured~\cite{Borexino:2011ufb,Super-Kamiokande:2008ecj,SNO:2011hxd},
while the transition between the two regimes (known as the {\it up-turn})
is not seen yet. The {\it up-turn} is sensitive to new physics such
as Non-Standard Interactions (NSI) or sterile neutrinos. In addition,
the Sun, due to its large mass and close distance, provides an exceptional 
environment for the study of dark matter, which might cause observable effects
on solar neutrinos. As will be comprehensively summarized in this review, all relevant new physics scenarios call for extensive investigations of solar neutrinos.

This paper aims at a timely review of theoretical and experimental progress in solar neutrino physics.  We will introduce the
standard solar neutrino physics, compile a variety of relevant
new physics studied in the literature, review the latest techniques
and issues involved in the measurement of solar neutrino spectra and
background reduction, and discuss the physics gains from future experiments. Past reviews have
their respective focuses on, e.g.,~solar models~\cite{Gann:2021ndb,Antonelli:2012qu},
the detection~\cite{Wurm:2017cmm}, the experimental progress~\cite{Ianni:2017iyw,Antonelli:2012qu,Wurm:2017cmm},
or neutrino phenomenology and new physics~\cite{Maltoni:2015kca}.
Here, we shall highlight the feature of this review, which is to present a comprehensive summary of new physics, together with the prospects of full-spectrum precision measurements. We hope such a combination might be helpful for both
theorists interested in probing their theories using solar neutrinos
and experimentalists looking for new physics goals for their experiments.

\section{Solar neutrino physics\label{sec:th}}

\subsection{Neutrino fluxes in the standard solar model}

\begin{figure}[h]
	\centering
	
	\includegraphics[width=0.98\textwidth]{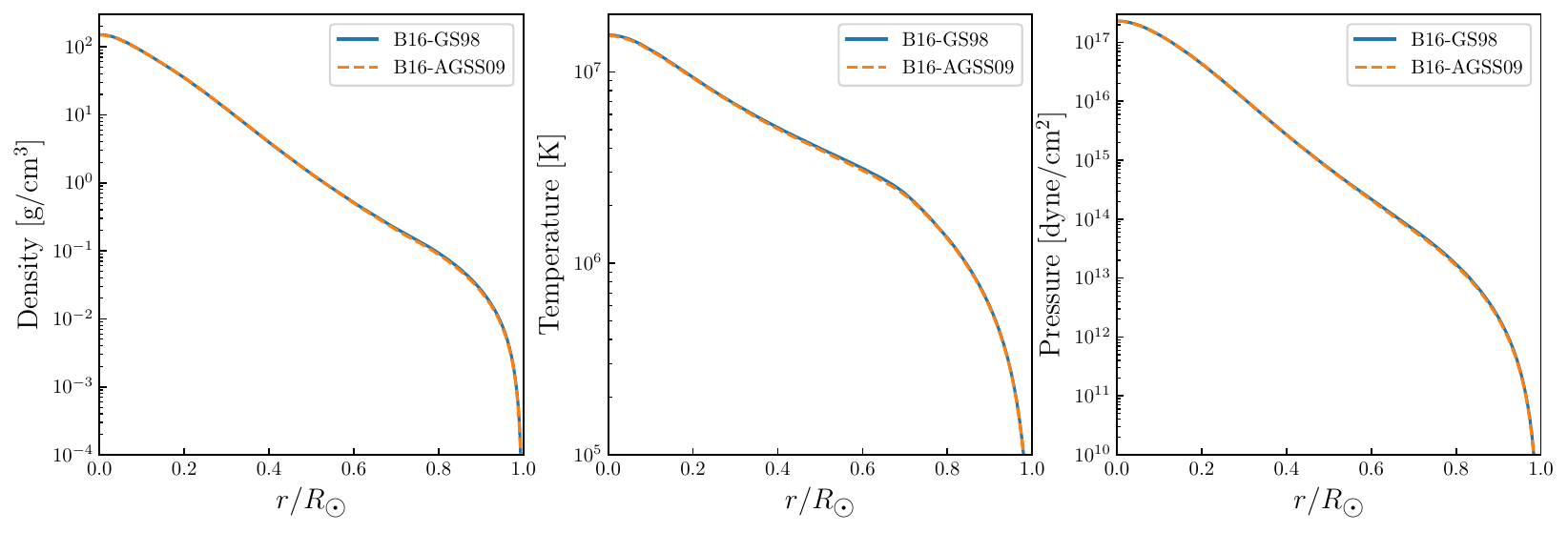} 
	\caption{Solar density (left), temperature (middle), and pressure (right) profiles
		in two standard solar models B16-GS98 and B16-AGSS09~\cite{Vinyoles:2016djt}.
		Here $r$ denotes the distance to the solar center and $R_{\astrosun}$
		denotes the solar radius. The main differences between the two models are to be explained in Sec.~\ref{sub:open}. \label{fig:solar-model}
	}
\end{figure}

The standard solar model (SSM) is constructed upon equations of hydrostatic equilibrium,
 energy transport, and energy production rates via nuclear reactions---see Ref.~\cite{Bahcall1989} for a pedagogical introduction and Ref.~\cite{Christensen-Dalsgaard:2020imv} for a recent review.
 With boundary conditions imposed from solar radius and luminosity, and additional inputs, including the Sun's age, elemental abundances, and radiative opacity, 
 these equations can be solved to predict accurate profiles of the solar density, temperature, pressure, and neutrino fluxes.
 Figure~\ref{fig:solar-model} depicts the distributions of these quantities obtained in recent
calculations~\cite{Vinyoles:2016djt}. 

\subsubsection{The pp chain and the CNO cycle}

\begin{figure}
	\centering 
	
	\includegraphics[scale=0.7]{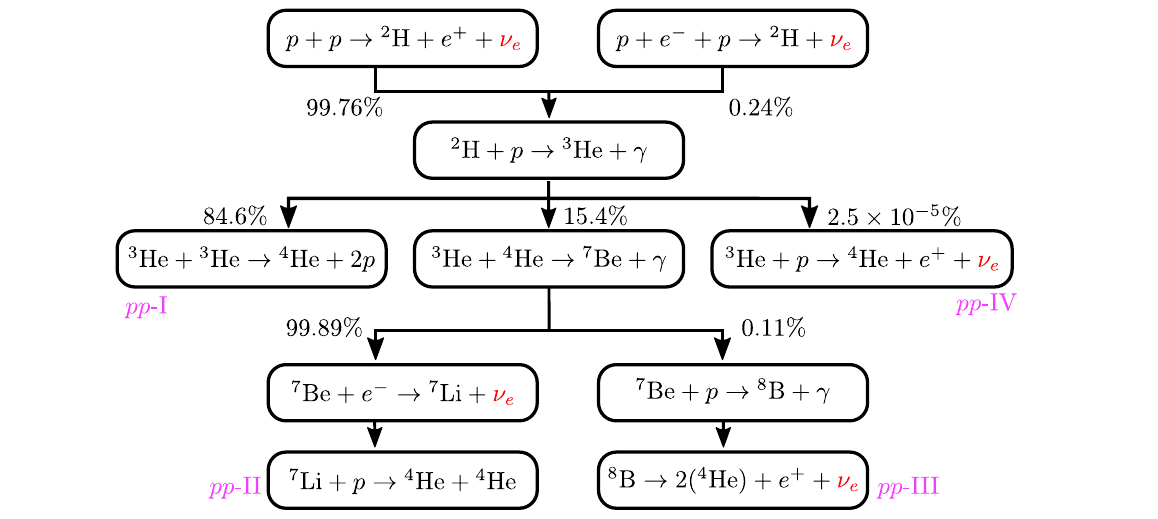} \caption{Reactions in the solar pp chain. Neutrinos ($\nu_{e}$) produced in
		the five reactions in the top-down order are referred to as pp, pep,
		hep,  $^{7}{\rm Be}$, $^{8}{\rm B}$ neutrinos, respectively. The
		theoretical branching percentages are taken from Ref.~\cite{Haxton:2012wfz}.
		\label{fig:ppchain}}
	
	\includegraphics[scale=0.7]{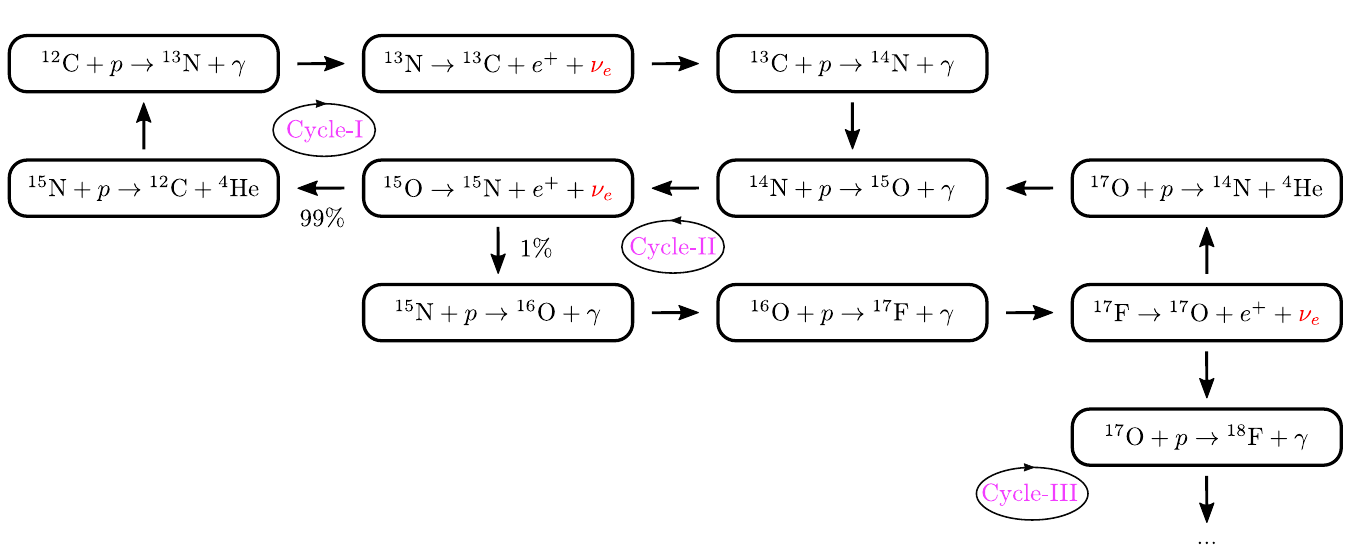} \caption{Reactions in the CNO cycle. Neutrinos ($\nu_{e}$) are produced from
		decays of $^{13}{\rm N}$, $^{15}{\rm O}$, and $^{17}{\rm F}$ in
		the first two cycles, Cycle-I and Cycle-II, with the latter suppressed
		relatively by $\sim1\%$. Subsequent cycles such as  Cycle-III produce
		heavier nuclear elements but their contributions are almost negligible
		for the Sun~\cite{Haxton:2005rw}.      \label{fig:cnocycle}}
\end{figure}

There are two sets of nuclear reactions responsible for neutrino and
energy productions in the Sun, the pp chain and the CNO cycle, as
illustrated in Figs.~\ref{fig:ppchain} and \ref{fig:cnocycle}.
The pp chain powers about 99\% of the total solar energy,  whereas the
CNO cycle accounts for the remaining $\sim 1\%$. 
For stars with masses greater
than $1.3M_{\astrosun}$, the CNO cycle dominates the energy production~\cite{Salaris}. 
The total neutrino flux from
the Sun should be consistent with the solar luminosity in photons, if all fusion processes are known.

As depicted in Fig.~\ref{fig:ppchain}, five reactions
in the pp chain produce neutrinos. They are referred
to, according to
the initial particles in the reactions,   as pp, pep, hep, $^{7}$Be, and $^{8}$B neutrinos. The pp chain consists of four
sub-chains, marked as pp-I to pp-IV in the figure. Note that all the sub-chains end up with
$^{4}$He. Therefore, despite some heavier elements appearing at intermediate stages, the pp chain burns hydrogen
only to helium. The first three sub-chains (pp-I, pp-II, and pp-III)
generate most of the energy (and hence most of the neutrinos) produced in the pp-chain.  The last sub-chain (pp-IV) contributes
a very insignificant amount ($10^{-5}$) to  the energy production
but produces the most energetic solar neutrinos (hep neutrinos), with
energy up to 18.77 MeV~\cite{SOLAR2011}.

In the CNO cycle\footnote{
Bethe first studied the CNO cycle for stellar energy production in 1939~\cite{Bethe1939}. It should be noted, however, that neutrinos were absent in the nuclear reactions Bethe used since the existence of neutrinos was still in question at the time.}, carbon and nitrogen serve as catalysts, meaning their abundances
are almost unchanged after a complete cycle of reactions. As shown in
Fig.~\ref{fig:cnocycle}, $^{12}{\rm C}$, after capturing a proton,
is converted to $^{13}{\rm N}$, which decays and produces $^{13}{\rm C}$,
followed by similar reactions converting $^{13}{\rm C}\to{}^{14}{\rm N}\to{}^{15}{\rm O}\to{}^{15}{\rm N}$.
Then the final element, $^{15}{\rm N}$, is dominantly converted back
to $^{12}{\rm C}$.  In this cycle, which we refer to as Cycle-I,
neutrinos are produced via $\beta^{+}$ decays of $^{13}{\rm N}$
and $^{15}{\rm O}$. The net effect, hence,  is that hydrogen is converted
to helium with energy and neutrino emission. Besides the dominant Cycle-I,
$^{15}{\rm N}$ can be converted to $^{16}{\rm O}$ with a small branching
ratio, entering a subdominant cycle, Cycle-II, in which $^{17}{\rm F}$
is produced and provides an additional source of neutrino emission.
Due to the small branching ratio,  Cycle-II is suppressed  roughly by two orders of magnitude. Consequently, the $^{17}{\rm F}$ neutrino
flux is lower than $^{13}{\rm N}$ and $^{15}{\rm O}$ neutrino fluxes
by two orders of magnitude---see Tab.~\ref{tab:flux}.

The radioactive elements $^{13}{\rm N}$, $^{15}{\rm O}$, and $^{17}{\rm F}$
can also produce monochromatic neutrino lines via electron capture
(e.g.,~$^{13}{\rm N}+e^{-}\to{}^{13}{\rm C}+\nu_{e}$), which has
not been extensively investigated so far~\cite{Bahcall:1989ygn,Stonehill:2003zf,Villante:2014txa}.
The corresponding fluxes are suppressed by $\sim10^{-4}$
compared to their $\beta^{+}$ decay neutrino fluxes~\cite{Villante:2014txa}:
\begin{equation}
\Phi_{e^{13}{\rm N}}=7.9\times10^{-4}\Phi_{^{13}{\rm N}}\thinspace,\ \ \Phi_{e^{15}{\rm O}}=3.9\times10^{-4}\Phi_{^{15}{\rm O}}\thinspace,\ \ \Phi_{e^{17}{\rm F}}=5.8\times10^{-4}\Phi_{^{17}{\rm F}}\thinspace.\label{eq:eX}
\end{equation}

	
\begin{table}
	\centering
	\caption{\label{tab:flux} Solar neutrino fluxes from two calculations, Bahcall-Serenelli-Basu~(BSB)~\cite{Bahcall:2005va}
		and Barcelona 2016~(B16)~\cite{Vinyoles:2016djt}, based on solar
		chemical composition data from GS98~\cite{Grevesse:1998bj}, AGS05~\cite{Asplund:2004eu},
		and AGSS09~\cite{Asplund:2009fu}.}
	\begin{tabular}{ccccc}
		\toprule 
		$\nu_{e}$ flux {[}${\rm cm}^{-2}{\rm s}^{-1}${]} & BSB05-GS98\cite{Bahcall:2005va} & BSB05-AGS05\cite{Bahcall:2005va} & B16-GS98~\cite{Vinyoles:2016djt}  & B16-AGSS09~\cite{Vinyoles:2016djt}\tabularnewline
		\midrule 
		$\Phi_{{\rm pp}}/10^{10}$ & $5.99(1\pm0.009)$ & $6.06(1\pm0.007)$ & $5.98(1\pm0.006)$ & $6.03(1\pm0.005)$\tabularnewline
		$\Phi_{{\rm pep}}/10^{8}$ & $1.42(1\pm0.015)$ & $1.45(1\pm0.011)$ & $1.44(1\pm0.01)$ & $1.46(1\pm0.009)$\tabularnewline
		$\Phi_{{\rm hep}}/10^{3}$ & $7.93(1\pm0.155)$ & $8.25(1\pm0.155)$ & $7.98(1\pm0.30)$ & $8.25(1\pm0.30)$\tabularnewline
		$\Phi_{^{7}{\rm Be}}/10^{9}$ & $4.84(1\pm0.105)$ & $4.34(1\pm0.093)$ & $4.93(1\pm0.06)$ & $4.50(1\pm0.06)$\tabularnewline
		$\Phi_{^{8}{\rm B}}/10^{6}$ & $5.69(1_{-0.147}^{+0.173})$ & $4.51(1_{-0.113}^{+0.127})$ & $5.46(1\pm0.12)$ & $4.50(1\pm0.12)$\tabularnewline
		$\Phi_{^{13}{\rm N}}/10^{8}$ & $3.05(1_{-0.268}^{+0.366})$ & $2.00(1_{-0.127}^{+0.145})$ & $2.78(1\pm0.15)$ & $2.04(1\pm0.14)$\tabularnewline
		$\Phi_{^{15}{\rm O}}/10^{8}$ & $2.31(1_{-0.272}^{+0.374})$ & $1.44(1_{-0.142}^{+0.165})$ & $2.05(1\pm0.17)$ & $1.44(1\pm0.16)$\tabularnewline
		$\Phi_{^{17}{\rm F}}/10^{6}$ & $5.83(1_{-0.420}^{+0.724})$ & $3.25(1_{-0.142}^{+0.166})$ & $5.29(1\pm0.20)$ & $3.26(1\pm0.18)$\tabularnewline
		\bottomrule
	\end{tabular}

\end{table}

Table~\ref{tab:flux} summarizes the neutrino fluxes calculated by  Bahcall-Serenelli-Basu~(BSB) \cite{Bahcall:2005va} and the Barcelona group in 2016~(B16)~\cite{Vinyoles:2016djt} for different solar models. The differences of these models and related open issues will be discussed in Sec.~\ref{sub:open}.

\subsubsection{Energy spectral shapes and production rate distributions}

\begin{figure}
	\centering \includegraphics[width=0.8\textwidth]{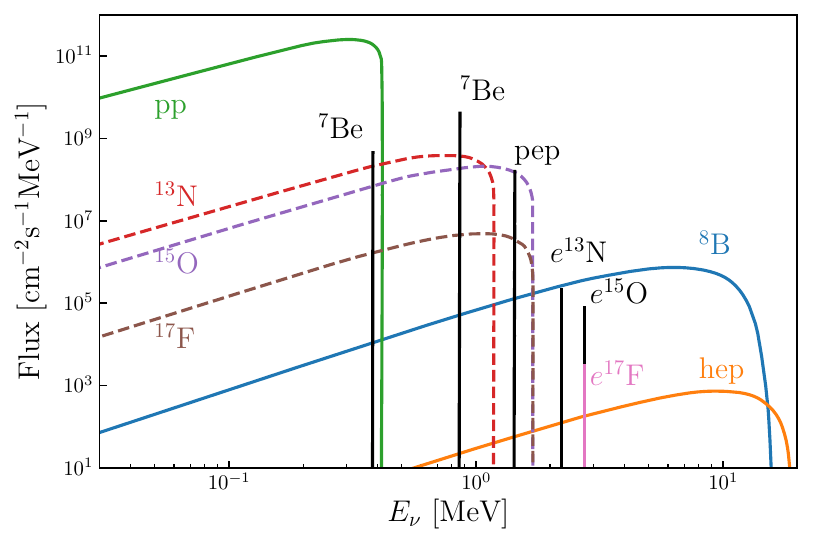}
	\caption{The solar neutrino spectra predicted by the SSM. Monochromatic spectra
		are given in units of ${\rm cm}^{-2}{\rm s}^{-1}$. Dashed lines
		denote neutrinos from the CNO cycle ($^{13}$N, $^{15}$O, $^{17}$F).
		The figure is updated using the data from Ref.~\cite{Vinyoles:2016djt}
		(B16-GS98, see also Tab.~\ref{tab:flux}), Bahcall's energy spectrum~\cite{Bahcall1989},
		and the electron capture rates (for $e\thinspace^{13}{\rm N}$, $e\thinspace^{15}{\rm O}$,
		$e\thinspace^{17}{\rm F}$) in Eq.~(\ref{eq:eX}).}
	\label{fig:solarnuflux}
\end{figure}

\begin{figure}
	\centering \includegraphics[width=0.95\textwidth]{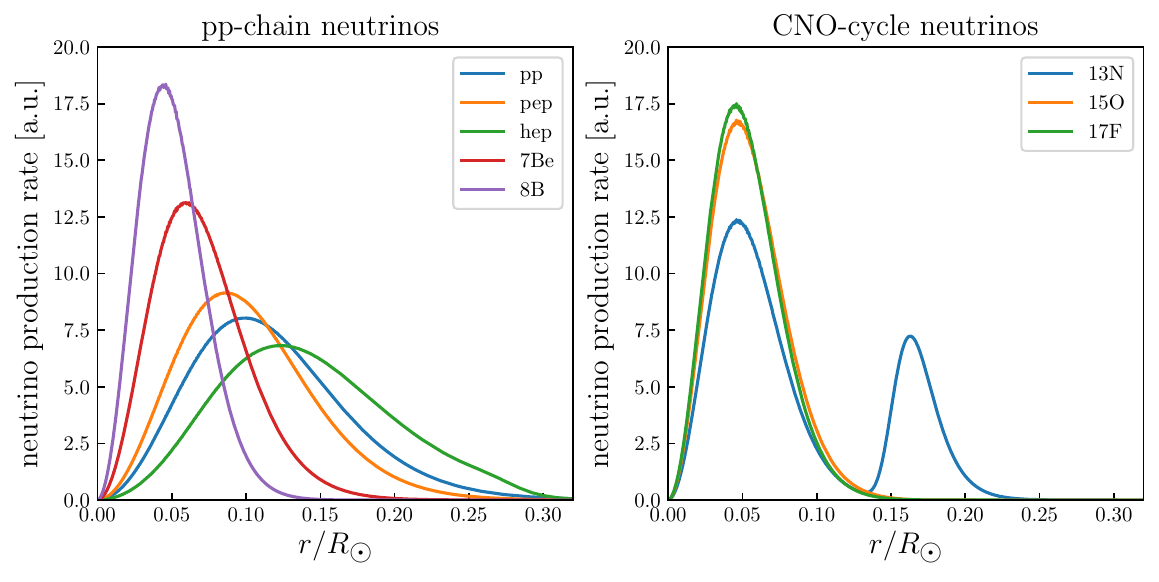} \caption{Neutrino production rates (in arbitrary unit {[}a.u.{]}) varying with
		the distance to the center,  $r$.  Data taken from Ref.~\cite{Vinyoles:2016djt}
		(B16-GS98).}
	\label{fig:nuradius}
\end{figure}

The shapes of solar neutrino energy distributions are not affected
by model-dependent uncertainties such as those caused by the solar
metallicity problem. They are determined mainly by the kinematics
of the corresponding nuclear reactions, with additional corrections
due to the Coulomb potentials of nuclei. Temperatures and densities hardly
affect the energy distributions because the energy released by nuclear reactions is around the MeV scale and is much higher than the core temperature
($\sim$keV) and the chemical potential. In Fig.~\ref{fig:solarnuflux}, 
we present the energy distributions using Bahcall's results for the spectral shapes~\cite{Bahcall1989} and normalizing them according to B16-GS98 in Tab.~\ref{tab:flux}, as well as the ratios in Eq.~\eqref{eq:eX}  for $e\thinspace^{13}{\rm N}$, $e\thinspace^{15}{\rm O}$, and 
$e\thinspace^{17}{\rm F}$ neutrinos.

For reactions with two particles in the final states (e.g.,~all the aforementioned electron-capture reactions and  $p+e^{-}+p\to\thinspace^{2}{\rm H}+\nu_{e}$), the energy spectra
of $\nu_{e}$ are monochromatic.  The pep neutrinos have an energy
of 1.442 MeV, while the $^{7}$Be neutrino spectrum consists of two
lines: 0.861 MeV (90\%) when $^{7}$Li is in the ground state and
0.383 MeV (10\%) when $^{7}$Li is excited.   The monochromatic
energy spectra of $e\thinspace{}^{13}{\rm N}$, $e\thinspace{}^{15}{\rm O}$, 
and $e\thinspace{}^{17}{\rm F}$ have neutrino energies at  2.220,
2.754, and 2.761 MeV~\cite{Stonehill:2003zf}. The widths of these monochromatic
lines are around the keV scale, caused by the thermal motion of initial
state particles. 

For a reaction with three or more particles in the final states (e.g.,~$p+p\to\thinspace^{2}{\rm H}+e^{+}+\nu_{e}$),
the energy spectrum is continuous, with the endpoint ($E_{\nu}^{\max}$) determined by the  difference between the initial and final total
masses. The shape of this continuous
spectrum is approximately given by~\cite{Bahcall:1986qc}
\begin{equation}
\frac{d\Phi}{dE_{\nu}}\propto E_{e}p_{e}E_{\nu}^{2}\times F_{{\rm Fermi}}(Z,p_{e})\thinspace,\label{eq:spectra}
\end{equation}
where  $E_{e}$ and $E_{\nu}$ are the energies of $e^{+}$ and $\nu_{e}$
in the final states ($E_{e}=E_{\nu}^{\max}+m_{e}-E_{\nu}$); $p_{e}=\sqrt{E_{e}^{2}-m_{e}^{2}}$
is the momentum of $e^{+}$; $Z$ denotes the nucleus charge; and $F_{{\rm Fermi}}(Z,p_{e})$  is the
Fermi function which takes into account the influence of the Coulomb
potential on the outgoing $e^{+}$. When the neutrino energy is not
close to the endpoint (so that the positron keeps energetic, $E_{e}\gg m_{e}$),
one can ignore the Fermi function and take $F_{{\rm Fermi}}(Z,p_{e})\approx1$.
Therefore for a continuous spectrum with $E_{\nu}^{\max}\gg m_{e}$
(such as $^{8}{\rm B}$ neutrinos), the spectral shape is approximately
given by $E_{e}^{2}E_{\nu}^{2}\approx(E_{\nu}^{\max}-E_{\nu})^{2}E_{\nu}^{2}$.
For more accurate results, we refer to Tabs.~6.2-6.4 in Ref.~\cite{Bahcall1989}.

	The spectrum with the highest endpoint is the hep neutrino spectrum, ending at 18.77 MeV. Next to it, $^{8}$B neutrinos have a less energetic spectrum, for which we refer to Ref.~\cite{Winter2006} for an up-to-date study of the spectrum and uncertainties.
It should be noted that the usual endpoint, 15.04 MeV, is for $^{8}$B decaying to an excited state of $^{8}{\rm Be}$ ($2^{+}$)
which is an allowed transition. Due to the very short lifetime of
this state, the forbidden decay of $^{8}$B to the ground state of
$^{8}{\rm Be}$ can happen, though at a suppressed branching ratio~\cite{Winter2006,Bhattacharya2006}.
It would extend the $^{8}$B spectrum to higher energies (16.95
MeV) and might be a background for future measurements of hep neutrinos~\cite{Bacrania2007}.

Although 18.77 MeV is the highest energy of solar neutrinos generated via nuclear fusion, the Sun can emit neutrinos of much higher energies produced via	cosmic rays scattering off the solar atmosphere, 
known as {\it solar atmospheric neutrinos}~\cite{Moskalenko:1991hm,Seckel:1991ffa,Moskalenko:1993ke,Ingelman:1996mj,Arguelles:2017eao,Ng:2017aur,Edsjo:2017kjk}.
The energy spectrum of solar atmospheric neutrinos roughly follows the power law $d\Phi/dE_{\nu} \propto E_{\nu}^{-3}$ within $10<E_{\nu}/{\rm GeV}<10^5$~\cite{Arguelles:2017eao}. At $E_{\nu}=300$ GeV, the magnitude of the solar atmospheric neutrino flux is around $d\Phi/dE_{\nu}\sim 3\times 10^{-13}{\rm cm}^{-2}\,{\rm s}^{-1}{\rm GeV}^{-1}$~\cite{Ng:2017aur}. Despite the low flux, solar atmospheric neutrinos would be an irreducible background for future searches for DM annihilation in the Sun.

Solar neutrinos are produced mainly within the solar core, with a radius of $0.2\sim0.25R_{\astrosun}$. The production rates are
sensitive to the temperature and density, which decrease rapidly with
the distance to the center, $r$. In addition, the chemical composition, which also varies with $r$, has a significant influence. 
Figure~\ref{fig:nuradius} shows the distributions of the neutrino
production rates, taking the B16-GS98 model from Ref.~\cite{Vinyoles:2016djt}.
	The double-peak structure of the $^{13}{\rm N}$ curve is because the reaction $^{14}\text{N}+p\to{}^{15}\text{O}+\gamma$
	is not effective in the region $0.13<r/R_{\astrosun}<0.25$. It 
	breaks the equilibrium of the CN cycle in the region, causing an out-of-equilibrium
	production rate of $^{13}\text{N}$ neutrinos---see, e.g.~\cite{Villante:2020adi}.

The production rate distributions in Fig.~\ref{fig:nuradius} are used in the state-of-the-art calculation of solar neutrino oscillation, in which one needs to integrate the oscillation probability weighted by the varying production rates.

\subsubsection{Open issues in solar models\label{sub:open}}


Heavy elements, or metals\footnote{In astronomy, elements heavier than hydrogen and helium are viewed as ``metals'', and metallicity refers to the abundance of these elements.}, in the Sun can affect the radiative opacity and/or participate in nuclear reactions (e.g., C, N, O). Consequently, the temperature profile and neutrino fluxes are affected by them. The effects of heavy elements depend on not only the total abundance of metals (metallicity) but also the detailed chemical compositions and distributions. 

Currently, there are two competing classes of solar models with prominent differences in metallicity.
Models based on the chemical compositions obtained by Grevesse \& Sauval in 1998 (GS98)~\cite{Grevesse:1998bj} have comparatively high metallicity (often referred to as high-$Z$ models). 
Models based on the chemical compositions obtained by \cite{Asplund:2004eu} (AGS05) or \cite{Asplund:2009fu} (AGSS09) are referred to as low-$Z$ models, which incorporate new developments in simulations (e.g.,~changing from 1D to 3D, using non-local thermodynamic equilibrium, etc.) but are unfortunately in tension with helioseismological measurements. The sound speed at $0.3 \lesssim r/R_{\astrosun}\lesssim 0.7$ computed in low-$Z$ models is significantly higher than the helioseismic constraints---see, e.g.~\cite{Vinyoles:2016djt}. This is known as the ``solar metallicity problem.''

High-$Z$ models predict higher $^{8}{\rm B}$, $^{13}{\rm N}$, $^{15}{\rm O}$, and $^{17}{\rm F}$ neutrino fluxes  than low-$Z$ models,  as shown in Tab.~\ref{tab:flux}. For $^{13}{\rm N}$, $^{15}{\rm O}$, and $^{17}{\rm F}$ neutrinos, they can be understood from that higher abundances of elements in the CNO cycles should lead to higher production rates of CNO neutrinos. 
Besides, higher metallicity causes higher radiative opacity, increasing the temperature gradient and the interior temperature. This effect also increases the nuclear reaction rates and the neutrino fluxes. For $^{8}{\rm B}$ neutrinos, the higher flux of high-$Z$ models is mainly affected by the higher opacity.

Precision measurements of solar neutrino fluxes will play an important role in solving the solar metallicity problem. However, the problem is much more complex than just metallicity, as it involves complicated correlations among many different ingredients in constructing solar models. 
Recently the authors of Ref.~\cite{Magg:2022rxb} presented a new analysis of the solar chemical composition and obtained higher metallicity, which could reduce the tension with helioseismic constraints. But this work was later questioned by Ref.~\cite{Buldgen:2022nso}, in which the authors showed that additional tensions would arise in such   models. 

\subsection{Solar neutrino propagation in matter and vacuum}

Being produced at the core of the Sun, solar neutrinos first propagate through the solar medium to the surface and then fly in vacuum to the Earth. 
The matter effect in the Sun is crucial to high-energy (above a few MeV) neutrinos.
If arriving at night, solar neutrinos also pass through 
the Earth, 
causing a modulation signal (often known as  the day-night asymmetry) due to the matter effect in the Earth. 


\subsubsection{The MSW-LMA solution\label{sub:standard-sol}}
The evolution of neutrino flavors during the propagation in matter
is governed by the following Schrödinger equation:
\begin{equation}
i\frac{d}{dL}\nu=H\nu\thinspace,\label{eq:schro}
\end{equation}
with
\begin{eqnarray}
&  & H=\frac{1}{2E_{\nu}}U_{{\rm PMNS}}\left(\begin{array}{ccc}
m_{1}^{2}\\
& m_{2}^{2}\\
&  & m_{3}^{2}
\end{array}\right)U_{{\rm PMNS}}^{\dagger}+\left(\begin{array}{ccc}
V_{e}\\
& 0\\
&  & 0
\end{array}\right),\label{eq:Hamilt}\\[2mm]
&  & \nu=\left(\begin{array}{c}
\nu_{e}\\
\nu_{\mu}\\
\nu_{\tau}
\end{array}\right)=U_{\rm PMNS}\left(\begin{array}{c}
\nu_{1}\\
\nu_{2}\\
\nu_{3}
\end{array}\right).\label{eq:mix}
\end{eqnarray}
Here  $L$ denotes the propagation distance, $U_{{\rm PMNS}}$
is the PMNS mixing matrix, $V_{e}\equiv\sqrt{2}G_{F}n_{e}$ is the
MSW effective potential induced by coherent forward scattering of
neutrinos with electrons in matter, and $n_{e}$ is the electron number
density, which is $L$ dependent. 
	Throughout, we adopt the standard parametrization of the PMNS matrix~\cite{ParticleDataGroup:2022pth}, 
	\begin{equation}
	U_{{\rm PMNS}}=\left(\begin{array}{ccc}
	c_{12}c_{13} & s_{12}c_{13} & s_{13}e^{-i\delta_{{\rm CP}}}\\
	-s_{12}c_{23}-c_{12}s_{13}s_{23}e^{i\delta_{{\rm CP}}} & c_{12}c_{23}-s_{12}s_{13}s_{23}e^{i\delta_{{\rm CP}}} & c_{13}s_{23}\\
	s_{12}s_{23}-c_{12}s_{13}c_{23}e^{i\delta_{{\rm CP}}} & -c_{12}s_{23}-s_{12}s_{13}c_{23}e^{i\delta_{{\rm CP}}} & c_{13}c_{23}
	\end{array}\right),\label{eq:UPMNS}
	\end{equation}
which is parametrized by three mixing angles ($\theta_{12}$, $\theta_{13}$,  $\theta_{23}$) with the abbreviation
$(c_{ij},\ s_{ij})\equiv(\cos\theta_{ij},\ \sin\theta_{ij})$ and one Dirac CP phase ($\delta_{{\rm CP}}$). Among the four parameters, $\delta_{{\rm CP}}$ and $\theta_{23}$ are irrelevant to
	the standard solar neutrino oscillation as long as $\nu_{\mu}$ and
	$\nu_{\tau}$ are indistinguishable at detection.
Neutrino oscillations also involve two mass squared differences  defined as $\Delta m_{21}^{2}\equiv m_{2}^{2}-m_{1}^{2}$
and $\Delta m_{31}^{2}\equiv m_{3}^{2}-m_{1}^{2}$. The values of  relevant oscillation parameters ($\theta_{12}$, $\theta_{13}$, $\Delta m_{21}^{2}$, $\Delta m_{31}^{2}$) and the status of their measurements will be reviewed in Sec.~\ref{sub:solar-measurement}.

It is sometimes useful to define the effective mixing matrix $U^m$ in matter by the following re-diagonalization of $H$:
 \begin{equation}
 H=\frac{1}{2E_{\nu}} U^m{\rm diag}\left(\tilde{m}_{1}^{2},\ \tilde{m}_{2}^{2},\ \tilde{m}_{3}^{2} \right)U^{m\dagger}\thinspace,
 \label{eq:Um}
 \end{equation}
 where $\tilde{m}_{1,2,3}$ are effective neutrino masses in matter.

The survival probability of solar electron neutrinos can be obtained
by solving the Schrödinger equation in~\eqref{eq:schro}, either numerically\footnote{Numerical solutions obtained by straightforwardly solving the differential
	equation are highly oscillatory due to the long propagation distance ($L\Delta m_{31}^{2}/E_{\nu}, L\Delta m_{21}^{2}/E_{\nu} \gg 1$). The oscillatory part can be averaged
	out by integrating over $E_{\nu}$ within the finite energy resolution
	of a detector, or over $L$ since solar neutrinos are not produced
	at a point-like source---see Fig.~\ref{fig:nuradius}.
} 
 or analytically.
The latter employs the adiabatic approximation~(to be explained in Sec.~\ref{sub:adiabatic}) and leads to the following
result~\cite{Maltoni:2015kca}:
\begin{equation}
P_{ee} =\left(c_{13}c_{13}^{m}\right)^{2}\left(\frac{1}{2}+\frac{1}{2}\cos2\theta_{12}^{m}\cos2\theta_{12}\right)+\left(s_{13}s_{13}^{m}\right)^{2},\label{eq:p-ee}
\end{equation}
with  
\begin{align}
\cos2\theta_{12}^{m} & \approx\frac{\cos2\theta_{12}-\beta_{12}}{\sqrt{(\cos2\theta_{12}-\beta_{12})^{2}+\sin^{2}2\theta_{12}}}\thinspace,\label{eq:-1}\\
\left(s_{13}^{m}\right)^{2} & \approx s_{13}^{2}\left(1+2\beta_{13}\right),\label{eq:-2}\\
\beta_{12} & \equiv\frac{2c_{13}^{2}V_{e}^{0}E_{\nu}}{\Delta m_{21}^{2}}\thinspace,\label{eq:-3}\\
\beta_{13} & \equiv\frac{2V_{e}^{0}E_{\nu}}{\Delta m_{31}^{2}}\thinspace.\label{eq:-4}
\end{align}
Here $V_{e}^{0}$ denotes the value of $V_{e}$ at the core where $\nu_{e}$ is
produced and the superscript ``$m$'' denotes quantities
modified by the matter effect. 

 The above survival probability is often referred to as the MSW-LMA
(where LMA stands for Large-Mixing-Angle) solution in the literature. There are two noteworthy
limits which we would like to discuss briefly. 
\begin{itemize}
\item Low-$E_{\nu}$ limit (vacuum limit): \\
When $E_{\nu}$ is sufficiently small, the matter effect is negligible
($\beta_{12},\ \beta_{13}\approx0$), and Eq.~\eqref{eq:p-ee} simply
reduces to
\begin{equation}
P_{ee}\approx 1-\frac{1}{2} \sin ^2(2 \theta_{12}) = c_{12}^{4}+s_{12}^{4}\thinspace,\label{eq:-6}
\end{equation}
where we have neglected the effect of $\theta_{13}$.
The result is easy
to understand: when $\nu_{e}$ is produced, it consists of $c_{12}\nu_{1}+s_{12}\nu_{2}$
(assuming $\theta_{13}=0$). Each mass eigenstate propagates to the
Earth independently. Due to the long distance, they lose coherence.
At production, the probability of $\nu_{e}$ being $\nu_{1}$ ($\nu_{2}$)
is $c_{12}^{2}$ ($s_{12}^{2}$); at detection, the probability of
$\nu_{1}$ ($\nu_{2}$) being detected as $\nu_{e}$ is also $c_{12}^{2}$
($s_{12}^{2}$). Hence the survival probability of $\nu_{e}$ at detection
is given by $(c_{12}^{2})^{2}+(s_{12}^{2})^{2}$. 
\item High-$E_{\nu}$ limit:\\
When $E_{\nu}$ is large so that $\beta_{12}\gg1$ while $\beta_{13}$ remains small ($\beta_{13}\ll1$),
Eq.~\eqref{eq:p-ee} in the limit of $\theta_{13}=0$ reduces
\begin{equation}
P_{ee}\approx s_{12}^{2}\thinspace.\label{eq:-5}
\end{equation}
Nonzero $\theta_{13}$ can lead to a correction of $\sim 5\%$ to the result. 
Eq.~\eqref{eq:-5} can be seen from the adiabatic approximation. When $\nu_{e}$
is produced at the center with a high electron number density, it
is almost pure $\nu_{2}^{m}$ due to the strong matter effect ($\theta_{12}^{m}\approx90^{\circ}$).
As the density slowly decreases to zero, the evolution of all mass
eigenstates is adiabatic, which means $\nu_{2}^{m}$ will eventually
come out of the Sun as $\nu_{2}$. Since the probability of $\nu_{2}$
being detected as $\nu_{e}$ is $s_{12}^{2}$, the survival probability
in the  high-$E_{\nu}$ limit is simply $s_{12}^{2}$. 
\end{itemize}

\begin{figure}
\centering

\includegraphics[width=0.6\textwidth]{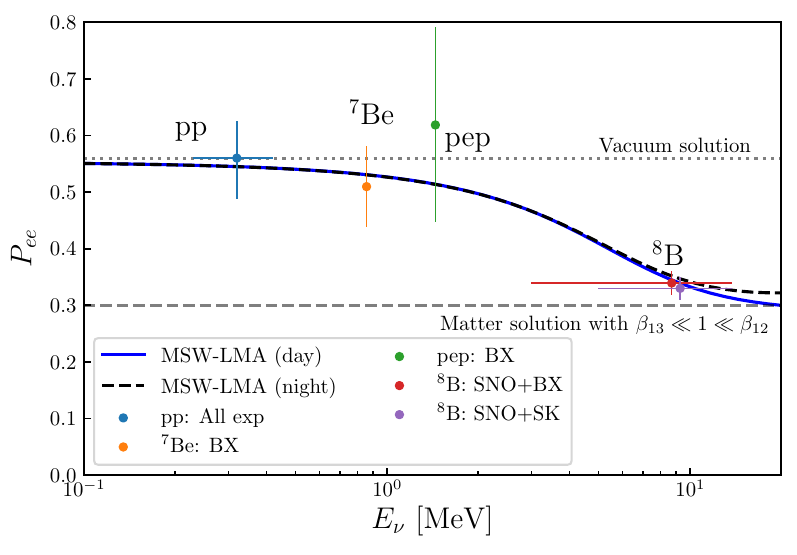}

\caption{The survival probability of solar electron neutrinos $P_{ee}$.	 
		The MSW-LMA (day) curves is obtained using Eq.~\eqref{eq:p-ee} with $\theta_{12}=33.4^{\circ},\ \theta_{13}=8.5^{\circ},\ \Delta m^2_{12}=7.5\times 10^{-5}{\rm eV}^2,\ \Delta m^2_{13}=2.4\times 10^{-3}{\rm eV}^2$. The difference between the day and night curves is computed using Eq.~\eqref{eq:day-night} with the average Earth matter density, 5.5 g/cm$^3$.  
		 The data points are taken from Refs.~\cite{Borexino:2011ufb,SNO:2011hxd,Super-Kamiokande:2008ecj}.
Note that the curve and the data points are only for illustration, neglecting some subleading order effects which are discussed in the text.
	 \label{fig:survival}}

\end{figure}

	Figure~\ref{fig:survival} shows how the survival probability
varies as a function of $E_{\nu}$ from the low-$E_{\nu}$ limit (which corresponds
to $P_{ee}\approx c_{12}^{4}+s_{12}^{4} \approx 0.55$) to the high-$E_{\nu}$ one ($P_{ee}\approx s_{12}^{4} \approx 0.3$).
The probability becomes significantly energy-dependent in the range $2\ {\rm MeV}\lesssim E_{\nu}\lesssim 6\ {\rm MeV}$.
This part, often called the {\it up-turn}, has not been well measured by current data but it is of crucial importance to the solar determination of $\Delta m_{21}^2$ as well as to searches of new physics effects.

Note that the curves in Fig.~\ref{fig:survival} are only for illustration without including a number of subleading-order corrections, which should be taken into account in real analyses. For instance, the radical spread of neutrino production rates in Fig.~\ref{fig:nuradius} is not included (i.e.~we simply assume all neutrinos are produced at $r=0$). When this is included, one should have different $P_{ee}$ curves for different flux components (pp, $^8$B, CNO, etc.). When solar neutrinos propagate through the Earth, the Earth matter effect can slightly enhance $P_{ee}$ at high $E_{\nu}$, causing a day-night difference in the observed event rates. This will be discussed in further detail in Sec.~\ref{sub:day-night}. The $P_{ee}$ curves are computed assuming that the evolution is purely adiabatic while non-adiabatic corrections do exist, though they are negligibly small in the standard framework of oscillations---see Sec.~\ref{sub:adiabatic}.
Finally, uncertainties of oscillation parameters also lead to variations of the $P_{ee}$ curve. Therefore it is  often shown in the literature as bands with finite widths.


\subsubsection{Measurements of solar neutrino parameters\label{sub:solar-measurement}}
\begin{figure}
	\centering 
	
	\includegraphics[width=0.8\textwidth]{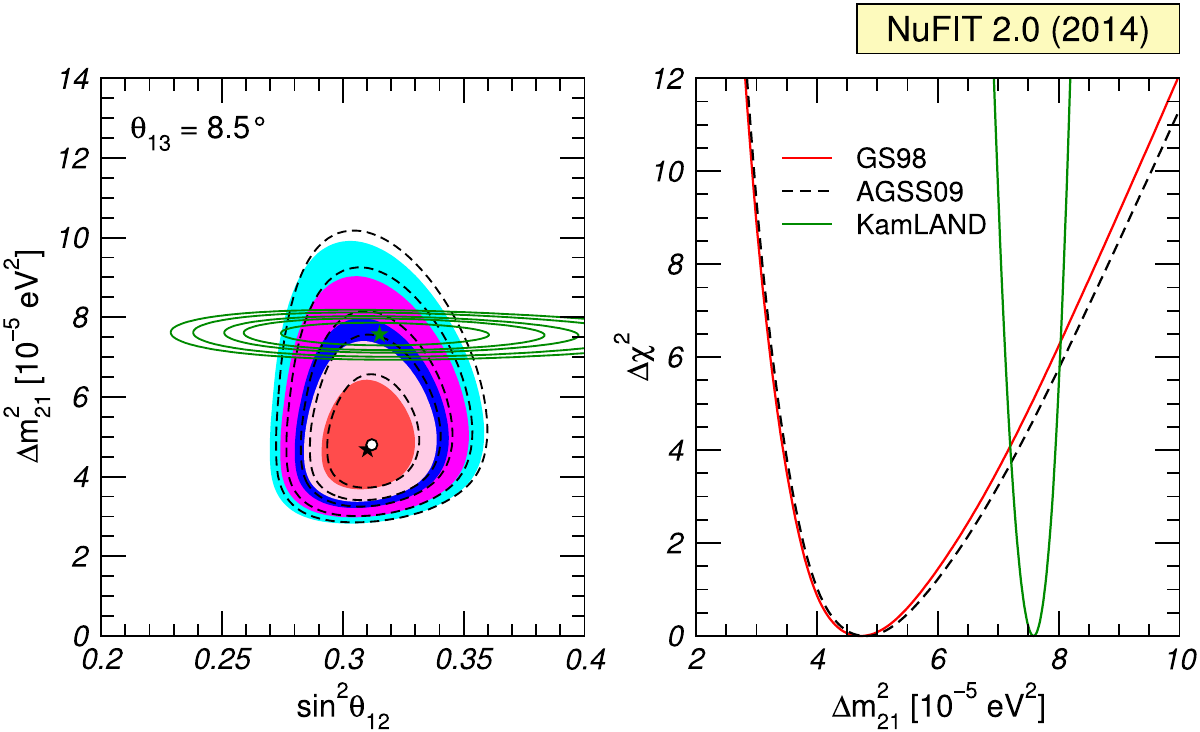} 
	
	\includegraphics[width=0.8\textwidth]{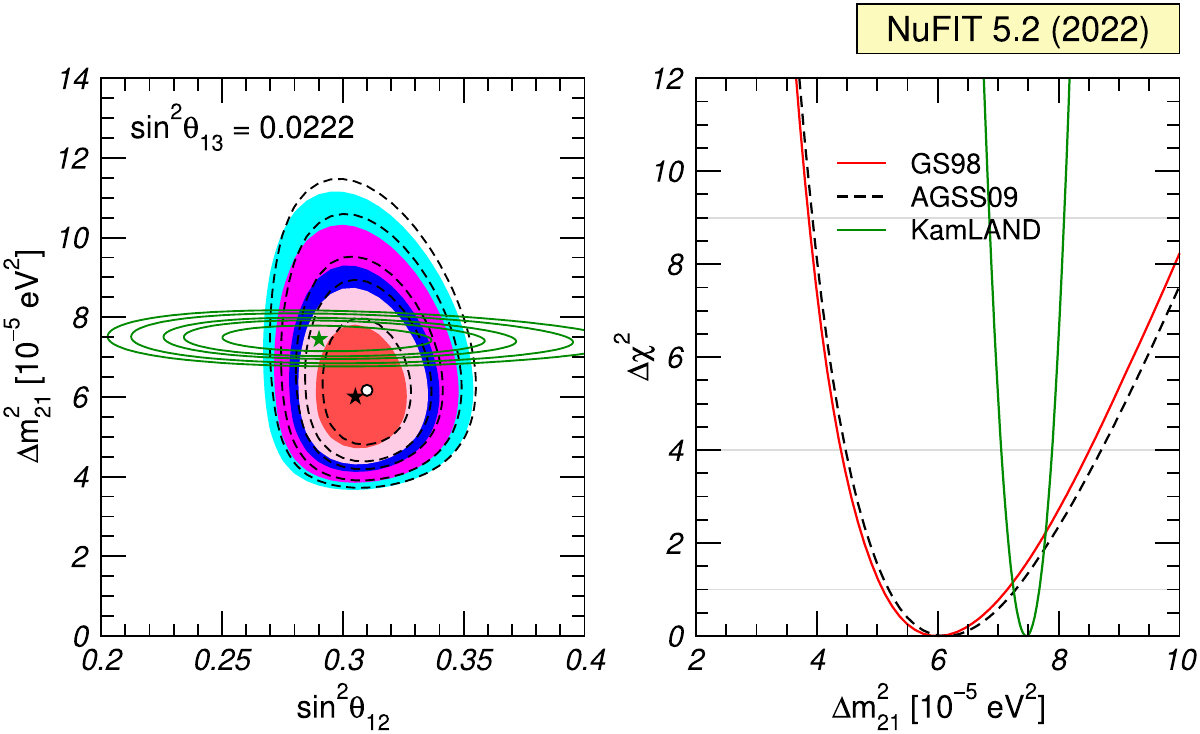} \caption{Past and present measurements of $\Delta m_{21}^{2}$ and $\theta_{12}$.
		The upper panels show a $\sim2\sigma$ tension between the KamLAND
		measurement and solar neutrino data in 2014. The lower panels show
		that this tension has reduced to $\sim1\sigma$ as of 2022. Figure
		taken from Refs.~\cite{nu-fit,Gonzalez-Garcia:2014bfa,Esteban:2020cvm}.
		\label{fig:tension}}
\end{figure}

According to Eq.~\eqref{eq:p-ee}, the survival probability of $\nu_{e}$
is mainly sensitive to $\theta_{12}$ and $\Delta m_{21}^{2}$, and
also weakly depends on $\theta_{13}$ and $\Delta m_{31}^{2}$. 
The former can be measured not only by solar neutrino observations but
also by long-baseline reactor neutrino experiments (KamLAND, JUNO). 
The latter, thanks to the high-statistics measurements of reactor
neutrino experiments (Daya-Bay, RENO, Double-Chooz, etc.) in recent
years, have been determined with excellent precision. The latest 3-$\nu$
oscillation fit to the global data as of November 2022 ({\tt NuFit~5.2}~\cite{nu-fit,Esteban:2020cvm})
reported the following results:
\begin{align}
\sin^{2}\theta_{12} & =0.303_{-0.012}^{+0.012}\thinspace,\ \ \ \ \ \ \ \ \ \ \ \ \ \ \ \ \ \ \ \ \ \ \ \ \ \ \ \Delta m_{21}^{2}/10^{-5}\text{eV}^{2}=7.41_{-0.20}^{+0.21}\thinspace,\label{eq:nu-fit1}\\
\sin^{2}\theta_{13} & /10^{-2}=2.225_{-0.059}^{+0.056}\left(2.223_{-0.058}^{+0.058}\right),\ \ \Delta m_{31}^{2}/10^{-3}\text{eV}^{2}=+2.507_{-0.027}^{+0.026}\left(-2.486_{-0.028}^{+0.025}\right),\label{eq:nu-fit2}
\end{align}
where the values in parentheses are for the inverted mass ordering if they depend on the mass ordering. 

Here we would like to comment on the complementarity between solar and reactor
(both long- and short-baseline) neutrino data in determining the above
parameters. As a long-baseline reactor neutrino experiment, KamLAND
can measure $\Delta m_{21}^{2}$ with good precision because the disappearance
probability $(1-P_{ee})$  is approximately proportional to $\sin^{2}(\Delta m_{21}^{2}L/4E_{\nu})$.
At KamLAND, the first two peaks of the probability as a function of
$L/E_{\nu}$ can be clearly seen~\cite{KamLAND:2013rgu}. By contrast,
solar neutrino observations are less constraining on $\Delta m_{21}^{2}$
because $\Delta m_{21}^{2}$ mainly governs the {\it up-turn}, which
as a transition from the high- to low-$E_{\nu}$ limits is not well
measured. The well-measured values of $P_{ee}$ in the two limits,
however, are almost independent of $\Delta m_{21}^{2}$ but can be
used to constrain $\theta_{12}$ very effectively. At high energies ($10 \sim 20$ MeV), the day-night asymmetry (see Sec.~\ref{sub:day-night}) could be used to determine  $\Delta m_{21}^{2}$, but this requires high-precision measurements of the $^8$B spectrum. Compared to long-baseline reactor experiments, it is more challenging to measure $\Delta m_{21}^{2}$ via solar neutrinos. 
Consequently,
the solar and KamLAND measurements are complementary to each other:
one measures $\Delta m_{21}^{2}$ better and the other is more constraining
on $\theta_{12}$. The 1-3 mixing parameters, $\theta_{13}$ and $\Delta m_{31}^{2}$,
could have potentially important influence on the determination of
$\theta_{12}$ and $\Delta m_{21}^{2}$, as implied by Eq.~\eqref{eq:p-ee}.
Since they have been measured with high precision by short-baseline
reactor neutrino experiments, their uncertainties are unlikely to
affect future measurements of solar neutrino parameters. 

It is worth mentioning that there was a long-standing tension between
solar and KamLAND measurements of $\Delta m_{21}^{2}$. More specifically,
the latest result of the KamLAND experiment reported in 2013 is $\Delta m_{21}^{2}=7.53_{-0.18}^{+0.18}\times10^{-5}\ \text{eV}^{2}$
\cite{KamLAND:2013rgu} while the Super-Kamiokande measurement, combining
SK-I, II, III, and IV data until February 2014, gave $\Delta m_{21}^{2}=4.8_{-0.8}^{+1.5}\times10^{-5}\ \text{eV}^{2}$
\cite{Super-Kamiokande:2016yck}, which is $\sim2\sigma$ lower than
the KamLAND value. Figure~\ref{fig:tension} shows the past and present
status of this tension. From the upper right panel, one can see that
the solid red curve (obtained using the GS98 solar model) crosses
the green curve at $\Delta\chi^{2}\approx4$ (corresponding to $2\sigma$).
Changing to a different solar model (AGSS09, dashed curves) cannot
effectively alleviate the tension. However, with the very recent update
of the Super-Kamiokande solar neutrino results reported in 2022~\cite{Koshio},
the solar best-fit value of $\Delta m_{21}^{2}$ shifted to $6\times10^{-5}\ \text{eV}^{2}$
and reduced the tension to $\sim1\sigma$, as shown in the lower panels
of Fig.~\ref{fig:tension}. Hence the tension 
has eased significantly.


\subsubsection{The adiabatic approximation\label{sub:adiabatic}}
The adiabatic approximation assumes that the matter density varies sufficiently slow so that if a neutrino is in an effective mass eigenstate (say, $\nu^{m}_i$, which is $n_e$ dependent) in the matter, then in the course of propagation it will remain $\nu_{i}^{m}$ and will not go into another matter eigenstate.
The only change is limited within the flavor composition of $\nu_{i}^{m}$, caused by the variation of $n_e$.
For example, 
a neutrino produced in the above high-$E_{\nu}$ limit is almost $\nu_2^m$, which after propagating through the solar medium to vacuum, becomes $\nu_2$.
Under the adiabatic approximation, the survival probability can be
computed by
\begin{equation}
P_{ee}=\sum_{i}|U^m_{ei}|^{2}|U_{ei}|^{2}\thinspace,\label{eq:n-11}
\end{equation}
where $U^m$ is introduced in Eq.~\eqref{eq:Um}. 

The validity of the adiabatic approximation and potential corrections have been addressed in Refs.~\cite{Haxton:1986dm,Parke:1986jy,Petcov:1987zj,deHolanda:2004fd}.
The condition of the adiabatic approximation can be formulated as~\cite{deHolanda:2004fd} 
\begin{equation}
\gamma\equiv\frac{4\Delta m_{21}^{2}E_{\nu}^{2}\sin2\theta_{12}}{\left(\Delta m_{21}^{4}\sin^{2}2\theta_{12}+\left(\Delta m_{21}^{2}\cos2\theta_{12}-2E_{\nu}V_{e}\right)^{2}\right)^{3/2}}\frac{dV_{e}}{dr}\ll 1\,.\label{eq:adia_valid}
\end{equation}
For the standard MSW-LMA solution of solar neutrinos, Eq.~\eqref{eq:adia_valid} is very well satisfied. One can take the solar density profile in Fig.~\ref{fig:solar-model} and evaluate the $\gamma$ parameter in Eq.~\eqref{eq:adia_valid}. The result depends on both $r$  and $E_{\nu}$. For $E_{\nu}=10$ MeV, we find that $\gamma$ peaks at $r\simeq 0.05R_{\astrosun}$ and the maximal value is $\gamma\simeq 8\times10^{-4}$.
The non-adiabatic correction to $P_{ee}$ is roughly given by
\begin{equation}
\delta P\sim\frac{\gamma^{2}}{4}\lesssim10^{-7}\,.\label{eq:adiabatic_P}
\end{equation}

Note that for large $E_{\nu}$, the solar resonance (corresponding to $\Delta m_{21}^{2}\cos2\theta_{12}=2E_{\nu}V_{e}$) can always be reached at certain $r$. At the solar resonance, $\gamma$ in Eq.~\eqref{eq:adia_valid}  can be simplified to 
\begin{equation}
\gamma_{{\rm resonance}}=\frac{4E_{\nu}^{2}}{\Delta m_{21}^{4}\sin^{2}2\theta_{12}}\frac{dV_{e}}{dr}\,,\label{eq:adia-res}
\end{equation}
which implies that
 the neutrino energy $E_{\nu}$ would have to be very high to cause significant non-adiabatic effects. For the standard solar profile and oscillation parameters, $\gamma \sim 0.1$ would require $E_{\nu}\gtrsim 1$ GeV. Alternatively, if hypothetical neutrino states with much smaller mass squared differences or mixing angles exist (e.g.~sterile neutrinos), then Eq.~\eqref{eq:adia-res} could also reach ${\cal O}(1)$ and therefore break the validity of the adiabatic approximation.
If the adiabatic approximation fails, one can numerically solve Eq.~\eqref{eq:schro}  to obtain the solution. 



\subsubsection{The Earth matter effect\label{sub:day-night}}

Due to the matter effect of the Earth, the survival probability of solar neutrinos arriving at nighttime is slightly different from that at daytime, causing the so-called day-night asymmetry. According to the calculations in Refs.~\cite{Blennow:2003xw,Ioannisian:2004jk}, the difference of $P_{ee}$ after averaging out oscillating parts can be approximately estimated as follows:
\begin{equation}
	\Delta P\equiv P_{ee}^{({\rm day})}-P_{ee}^{({\rm night})}\approx\frac{1}{2}c_{13}^{6}\frac{\cos 2\theta_{12}^{m} \sin^{2} 2\theta_{12} KV_{\oplus}}{K^{2}-2c_{13}^{2}\cos 2\theta_{12} V_{\oplus}K+V_{\oplus}^{2}}\,,
	\label{eq:day-night}
\end{equation}
where 
$\cos2\theta_{12}^{m}$ is given by Eq.~\eqref{eq:-1}, 
$K=\Delta m_{21}^{2}/(2E_{\nu})$, 
and $V_{\oplus}$
denotes the average value of $V_{e}$ in the Earth. 
In Fig.~\ref{fig:survival}, we plot the MSW-LMA night curve using Eq.~\eqref{eq:day-night} and Eq.~\eqref{eq:p-ee}. 
	Note that in the low-$E_{\nu}$ limit, $\Delta P$ in Eq.~\eqref{eq:day-night} is positive because $\cos2\theta_{12}^{m}\approx \cos2\theta_{12}$ is positive and the denominator is always positive. It turns negative when $E_{\nu}$ passes the solar resonance (around $2\sim 3$ MeV) above which $\cos 2\theta_{12}^{m} $ becomes negative [see Eq.~\eqref{eq:-1}]. Above the solar resonance, $|\Delta P|$ increases as $E_{\nu}$ increases, causing the $P_{ee}$ curve to split into two different (day and night) curves as shown in Fig.~\ref{fig:survival}.


In practical observation, the day-night asymmetry is usually measured by $A_{ DN}\equiv 2(R_D-R_N)/(R_D+R_N)$, where $R_{D}$/$R_{N}$ denotes the average day/night event rate. For the best-fit values in Eq.~\eqref{eq:nu-fit1}, the expected value of $A_{ DN}$ is $-1.7\%$ at Super Kamiokande~\cite{Super-Kamiokande:2016yck} (the day-night asymmetry also depends on the latitude of the observation site). 
In 2013,  Super-Kamiokande first reported an indication of day-night asymmetry at $2.7\ \sigma$ significance. 
The asymmetry parameter $A_{ DN}$ is measured to be  $A_{ DN}=(-3.2\pm1.1_{\rm stat.}\pm0.5_{\rm syst.})\%$~\cite{Super-Kamiokande:2013mie}. 
As of 2022, the significance of day-night asymmetry at  Super-Kamiokande  reached 3.1 $\sigma$ with $A_{ DN}=(-2.8\pm 0.9)\%$~\cite{Koshio}.
So far, acquiring sufficient statistics is still the main challenge in measuring the day-night asymmetry.



%



If measured in the future with high statistics, the day-night asymmetry would be a direct probe of the earth matter effect.  In addition, we note that there are already some discussions on the oscillation tomography of the Earth with solar neutrinos and future experiments~\cite{Akhmedov:2005yt,Ioannisian:2017dkx, Bakhti:2020tcj}. Solar neutrino detectors near the equator would be more suitable in this aspect as solar neutrinos could pass the innermost part of the Earth before arriving at the detectors.

%
%
%

\subsection{Search for new physics with solar neutrinos\label{sub:new-ph}}

Before the MSW-LMA became the standard solution to the solar neutrino problem, various new physics interpretations of solar neutrino data were proposed. To date, even though the standard solution has been well tested, many new physics scenarios remain indistinguishable. 
Since neutrinos are regarded as the portal to new physics beyond the SM, and a few experimental anomalies are still inconsistent with our current understanding of neutrinos, the search for new physics is of great importance in the era of precision measurement of solar neutrinos. Below we review a few popular new physics scenarios often considered in the literature. 


\subsubsection{Non-Standard Interactions (NSI)}

A variety of neutrino mass models predict new interactions of neutrinos.
As an effective field theory (EFT) approach, the so-called Non-Standard
interactions (NSI), which Wolfenstein first considered in his seminal paper on the matter effect~\cite{Wolfenstein:1977ue},  have attracted
increasing interest in recent years---see \cite{Bhupal2019,Davidson:2003ha,Ohlsson:2012kf,Farzan:2017xzy,Esteban:2018ppq}
for NSI reviews. 

There are two types of NSI often considered in the literature, namely
the NC-like and the  CC-like NSI, formulated as 
\begin{eqnarray}
{\cal L}_{\text{NC}} & = & -2\sqrt{2}G_{F}\left[\bar{\nu}_{\alpha}\gamma^{\mu}P_{L}\nu_{\beta}\right]\left[\bar{f}\gamma_{\mu}\left(\varepsilon_{\alpha,\beta}^{f,L}P_{L}+\varepsilon_{\alpha,\beta}^{f,R}P_{R}\right)f\right],\label{eq:lnclcc}\\
{\cal L}_{\text{CC}} & = & -2\sqrt{2}G_{F}\left[\bar{\nu}_{\alpha}\gamma^{\mu}P_{L}\ell_{\beta}\right]\left[\bar{f}\gamma_{\mu}\left(\varepsilon_{\alpha,\beta}^{ff',L}P_{L}+\varepsilon_{\alpha,\beta}^{ff',R}P_{R}\right)f^{\prime}\right],
\end{eqnarray}
where $G_{F}$ is Fermi's constant, $\alpha$ and $\beta$  denote
lepton flavors, and the $\varepsilon$'s quantify the strengths of
neutrino interactions with matter fermions $f$ and $f'$.  For NC-like NSI, $f$ can vary in $\{e,u,d\}$ while for CC-like NSI, $f$ and $f'$ vary in $\{u,d\}$ with $f\neq f'$. 
The matter effect of neutrino oscillation is only affected by the NC-like NSI. 
CC-like NSI could modify neutrino production and detection but they have been stringently constrained by meson decay data.

NSI can be generated in many extensions of the SM. One of the most
classic example is the type-II seesaw model~\cite{Konetschny:1977bn,Cheng:1980qt,Schechter:1980gr,Mohapatra:1980yp}
which introduces a Higgs triplet  interacting with charged leptons and
neutrinos. After integrating out the  Higgs triplet and performing
the Fierz transformation, it naturally gives rise to NSI in the lepton
sector~\cite{Malinsky:2008qn}, though the strengths are found to
lie below current detectability~\cite{Mandal:2022zmy}. Alternatively,
NSI could also be generated in $Z'$ models~\cite{Heeck:2018nzc},
radiative neutrino mass models~\cite{Babu:2019mfe}, or from the loop
effects~\cite{Bischer:2018zbd,Chauhan:2020mgv}.

There are two effects of NSI on solar neutrinos: they could modify
(i) propagation of neutrinos in the solar medium~\cite{Friedland:2004pp,Miranda:2004nb,Bolanos:2008km,0907.2630,Palazzo:2009rb,1101.3875,Gonzalez-Garcia:2013usa,1608.05897,1704.04711,1704.06222,2203.11772}
and (ii) neutrino scattering at detection~\cite{1705.00661,Khan:2019jvr,2003.12984,2111.03031,Coloma:2022umy,Schwemberger:2022fjl}.
The two aspects are explained below.  

\vspace{8pt}
\textbf{$\blacksquare$ Effect on propagation}

In the presence of NSI, when neutrinos propagate in matter, coherent
forward scattering of neutrinos with matter particles would be modified,
causing an effect on the flavor evolution. As first noticed by Wolfenstein~\cite{Wolfenstein:1977ue},
neutrino oscillation could occur in matter even for massless neutrinos,
provided that the neutral current had flavor off-diagonal interactions.
The effect of NSI on neutrino flavor evolution can be accounted for
by replacing the Hamiltonian in Eq.~(\ref{eq:Hamilt}) with 
\begin{equation}
H=\frac{1}{2E_{\nu}}U_{{\rm PMNS}}\left(\begin{array}{ccc}
m_{1}^{2}\\
& m_{2}^{2}\\
&  & m_{3}^{2}
\end{array}\right)U_{{\rm PMNS}}^{\dagger}+V_{e}\begin{pmatrix}1+\varepsilon_{ee} & \varepsilon_{e\mu} & \varepsilon_{e\tau}\\
\varepsilon_{e\mu}^{*} & \varepsilon_{\mu\mu} & \varepsilon_{\mu\tau}\\
\varepsilon_{e\tau}^{*} & \varepsilon_{\mu\tau}^{*} & \varepsilon_{\tau\tau}
\end{pmatrix},\label{eq:H-NSI}
\end{equation}
where 
\begin{equation}
\varepsilon_{\alpha\beta}\equiv\sum_{f}\frac{n_{f}}{n_{e}}\left(\varepsilon_{\alpha,\beta}^{f,L}+\varepsilon_{\alpha,\beta}^{f,R}\right)\thinspace\label{eq:n}
\end{equation}
includes contributions of all fermions in matter. Hence the summation
is weighted by $n_{f}$ which is the number density of fermion $f$. 

Again, like the standard case, one can compute the survival probability
in this case by numerically solving Eq.~\eqref{eq:schro} with the
Hamiltonian in Eq.~\eqref{eq:H-NSI}. Analytically, one can obtain
approximate solutions using the adiabatic assumption and neglecting
the small correction caused by nonzero $\theta_{13}$. The $\nu_{e}$
survival probability obtained in this way reads~\cite{Friedland:2004pp}:

\begin{equation}
P_{ee}\approx\frac{1}{2}+\frac{1}{2}\cos2\theta_{\epsilon}\cos2\theta_{12}\thinspace,\label{eq:p-ee-NSI}
\end{equation}
where 
\begin{align}
\cos2\theta_{\epsilon} & =\frac{\cos2\theta_{12}-x_{\epsilon}\cos2\alpha}{\sqrt{1+x_{\epsilon}^{2}-2x_{\epsilon}\left(\cos2\alpha\cos2\theta_{12}-\sin2\alpha\sin2\theta_{12}\cos2\phi\right)}}\thinspace,\label{eq:n-1}\\
\alpha & =\frac{1}{2}\arctan\frac{|\epsilon_{2}|}{1+\epsilon_{11}}\thinspace,\ \phi=\frac{1}{2}\arg(\epsilon_{2})\thinspace,\label{eq:n-2}\\
x_{\epsilon} & \equiv2V_{e}^{0}E_{\nu}\frac{\sqrt{(1+\epsilon_{1})^{2}+|\epsilon_{2}|^{2}}}{\Delta m_{21}^{2}}\thinspace.\label{eq:n-3}
\end{align}
The effective NSI parameters $\epsilon_{1}$ and $\epsilon_{2}$
in the absence of $\mu$-flavored NSI are defined as\footnote{The effective parameters $\epsilon_{1}$ and $\epsilon_{2}$ are introduced
	in many studies on solar neutrinos with NSI (see e.g., \cite{Friedland:2004pp,Miranda:2004nb,Bolanos:2008km,Palazzo:2009rb,Gonzalez-Garcia:2013usa},
	though their specific forms may vary) due to the commonly used reduction
	of the $3\times3$ matrix form of the Hamiltonian to a $2\times2$
	form, by performing a rotation between the second and third rows and
	columns of $H$. }
\begin{equation}
\epsilon_{1}\equiv\varepsilon_{ee}-\varepsilon_{\tau\tau}\sin^{2}\theta_{23}\thinspace,\ \epsilon_{2}\equiv-2\varepsilon_{e\tau}\sin\theta_{23}\thinspace.\label{eq:n-4}
\end{equation}
In the presence of NSI of all flavors and nonzero $\theta_{13}$, $\epsilon_{1,2}$ would be
much more complicated combinations of $\varepsilon_{\alpha\beta}$.	We refer to \cite{Gonzalez-Garcia:2013usa} for the full expressions
of $\epsilon_{1,2}$.
	Note that unlike the standard solar neutrino oscillation which is independent of $\theta_{23}$ and $\delta_{\rm CP}$, the NSI $P_{ee}$ depends on $\theta_{23}$ which appears in Eq.~\eqref{eq:n-4}, and also on $\delta_{\rm CP}$ which would appear in the full expressions with nonzero $\theta_{13}$~\cite{Gonzalez-Garcia:2013usa}.

\begin{figure}
	\centering
	
	\includegraphics[width=0.6\textwidth]{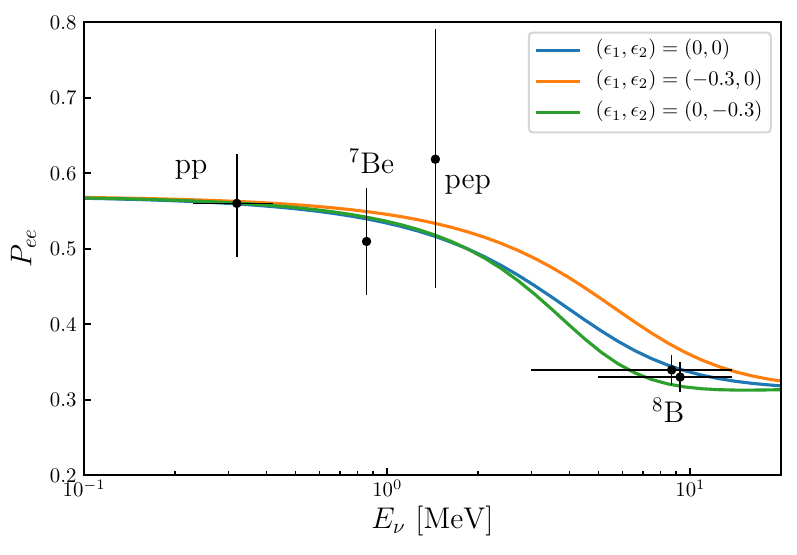}
	
	\caption{The survival probability of solar electron neutrinos $P_{ee}$ in
		the presence of NSI, computed according to Eq.~\eqref{eq:p-ee-NSI}.
		The experimental measurements (black points/bars) are the same as those in Fig.~\ref{fig:survival}.
Note that the curve and the data points are only for illustration, neglecting some subleading order effects which are discussed in the text.
		\label{fig:survival-NSI} }
\end{figure}

Figure~\ref{fig:survival-NSI} shows how NSI might change the survival
probability $P_{ee}$. Here all the curves are produced using Eq.~\eqref{eq:p-ee-NSI}
assuming the central solar density $\rho=10^{2}{\rm g}/{\rm cm}^{3}$,
$\Delta m_{21}^{2}=7.5\times10^{-5}{\rm eV}^{2}$, and $\theta_{12}=34^{\circ}$.
As shown in Fig.~\ref{fig:survival-NSI}, NSI with sizable $\epsilon_{1,2}$
can distort the standard MSW-LMA solution significantly at intermediate
energies of a few MeV (the {\it up-turn}). It implies that measurements
at the {\it up-turn} would be crucial to probing new
physics effects on solar neutrinos. Very recently, the Super-Kamiokande
collaboration performed an analysis on the $^{8}$B  solar neutrino data
collected with 277 kton-yr exposure and reported
that nonzero values of $\epsilon_{1}$ and $\epsilon_{2}$ are favored
at $1.8\sigma$ (for NSI with the $u$ quark) or $1.6\sigma$ (with
the $d$ quark)~\cite{2203.11772}.

Another interesting consequence of introducing NSI is that they can lead to the so-called  LMA-Dark (LMA-D) solution~\cite{Esteban:2018ppq,Miranda:2004nb,Bakhti:2014pva, Farzan:2015doa, Chaves:2021pey}. The LMA-D solution arises from a well-known degeneracy: performing the transformation $\theta_{12}\to \pi/2-\theta_{12}$,  $\Delta m_{31}^2\to -\Delta m_{31}^2$, and $\delta_{\rm CP}\to \pi/2-\delta_{\rm CP}$,  the Hamiltonian $H$ in vacuum changes to $-H^*$, implying that the oscillation probabilities in vacuum are invariant under the above transformation. For instance, Eq.~\eqref{eq:-6} is explicitly invariant under the transformation. In matter, the degeneracy is broken by the standard MSW effect, but 
it can be approximately
restored by NSI parameter shifting when the energy dependence of $P_{ee}$
is weak (e.g.~$10\ \text{MeV}\lesssim E_{\nu}\lesssim20\ \text{MeV}$). 
Hence the observed $^8{\rm B}$ neutrino flux could be explained either by the standard MSW-LMA solution or the LMA-D solution with large NSI and the above transformation. Currently, the LMA-D solution is disfavored by elastic neutrino scattering data at $2\sigma$ C.L.~\cite{Chaves:2021pey}.

\vspace{8pt}
\textbf{$\blacksquare$ Effect on detection}

Another effect of NSI on solar neutrinos is that they may modify the
cross section of neutrino scattering with target particles at detection.
Solar neutrinos are either detected via CC processes (e.g. $\nu_{e}+{}^{37}{\rm Cl}\to e^{-}+{}^{37}{\rm Ar}$)
or elastic scattering (e.g. $\nu_{e}+e^{-}\to\nu_{e}+e^{-}$) which
involves NC and/or CC interactions. Since the target particle has to be a nucleus for the former, only CC-like NSI with quarks could be
relevant. However, due to existing strong constraints on CC-like NSI,
most studies on the scattering effect are mainly concerned with NC-like
NSI, which modifies only elastic scattering cross sections.

For elastic $\nu_{\alpha}+e^{-}$ scattering, the cross section including
NSI contributions\footnote{Note that flavor-changing NSI also leads to $\nu_{\alpha}+e^{-}\rightarrow\nu_{\beta}+e^{-}$
	with $\beta\neq\alpha$, which is included as non-interference terms
	in Eqs.~\eqref{eq:n-5}-\eqref{eq:n-6}. } reads~\cite{Link:2019pbm}:
\begin{equation}
\frac{d\sigma}{dT}=\frac{m_{e}G_{F}^{2}}{2\pi}\left[g_{1}^{2}+g_{2}^{2}\left(1-\frac{T}{E_{\nu}}\right)^{2}-x_{12}\frac{m_{e}T}{E_{\nu}^{2}}\right],\label{eq:n-8}
\end{equation}
where $T$ is the recoil energy of the electron,  and the other parameters
are defined as
\begin{align}
g_{1}^{2} & \equiv\left(g_{V}+g_{A}+2\varepsilon_{\alpha\alpha}^{e,L}\right)^{2}+\sum_{\beta\neq\alpha}\left(2\varepsilon_{\alpha\beta}^{e,L}\right)^{2},\label{eq:n-5}\\
g_{2}^{2} & \equiv\left(g_{V}-g_{A}+2\varepsilon_{\alpha\alpha}^{e,R}\right)^{2}+\sum_{\beta\neq\alpha}\left(2\varepsilon_{\alpha\beta}^{e,R}\right)^{2},\label{eq:n-7}\\
x_{12} & \equiv\left(g_{V}+g_{A}+2\varepsilon_{\alpha\alpha}^{e,L}\right)\left(g_{V}-g_{A}+2\varepsilon_{\alpha\alpha}^{e,R}\right)+\sum_{\beta\neq\alpha}\left(2\varepsilon_{\alpha\beta}^{e,L}\right)\left(2\varepsilon_{\alpha\beta}^{e,R}\right),\label{eq:n-6}
\end{align}
with $g_{V}=2\sin^{2}\theta_{W}-1/2+\delta_{\alpha e}$, and $g_{A}=-1/2+\delta_{\alpha e}$.
The SM CC contribution to $\nu_{e}+e^{-}\to\nu_{e}+e^{-}$ is included
by $\delta_{\alpha e}$ in $g_{V}$ and $g_{A}$. 

Note that due to the interference between flavor diagonal NSI and
SM interactions, precision measurements of solar neutrinos are more
sensitive to $\varepsilon_{\alpha\alpha}$ than to $\varepsilon_{\alpha\beta}$
with $\beta\neq\alpha$. As can be seen from Eqs.~\eqref{eq:n-5}-\eqref{eq:n-6},
when expanding them in terms of $\varepsilon$, $\varepsilon_{\alpha\alpha}$
and $\varepsilon_{\alpha\beta}$ ($\beta\neq\alpha$) contribute at
${\cal O}(\varepsilon)$ and ${\cal O}(\varepsilon^{2})$ level, respectively.


\begin{table}
 	\centering
 	 	\caption{\label{tab:NSI} Constraints on NSI parameters from Borexino Phase-II
 		data~\cite{Borexino:2017rsf}, with 90\% C.L. Results taken from
 		Ref.~\cite{Coloma:2022umy}. }
	\begin{tabular}{lcc}
 		\toprule 
 		& $L$ & $R$\tabularnewline
 		\midrule 
 		$\varepsilon_{ee}^{e,L/R}$ & $[-0.03,+0.06]\oplus[-1.37,-1.29]$ & $[-0.23,+0.07]$\tabularnewline
 		$\varepsilon_{\mu\mu}^{e,L/R}$ & $[+0.58,+0.81]\oplus[-0.20,+0.13]$ & $[-0.36,+0.37]$\tabularnewline
 		$\varepsilon_{\tau\tau}^{e,L/R}$ & $[+0.45,+0.86]\oplus[-0.26,+0.26]$ & $[-0.58,+0.47]$\tabularnewline
 		$\varepsilon_{e\mu}^{e,L/R}$ & $[-0.17,+0.29]$ & $[-0.21,+0.41]$\tabularnewline
 		$\varepsilon_{e\tau}^{e,L/R}$ & $[-0.26,+0.23]$ & $[-0.35,+0.31]$\tabularnewline
 		$\varepsilon_{\mu\tau}^{e,L/R}$ & $[-0.09,+0.14]\oplus[-0.62,-0.52]$ & $[-0.26,+0.23]$\tabularnewline
 		\bottomrule
 	\end{tabular}

\end{table}

Measurements of solar neutrinos via elastic $\nu_{\alpha}+e^{-}$
scattering have produced stringent constraints on leptonic NSI~\cite{Khan:2019jvr,Coloma:2022umy}.
The latest results are summarized in Tab.~\ref{tab:NSI}, taken from Ref.~\cite{Coloma:2022umy}.
NSI with quarks could be constrained by CE$\nu$NS events,
though such events have not yet been detected successfully for solar
neutrinos. Future dark matter detectors (e.g.~multi-ton scale liquid
Xenon detectors) will be capable of detecting solar neutrinos~\cite{1705.00661,Schwemberger:2022fjl, 0812.4417, Strigari:2009bq,Anderson:2011bi,AristizabalSierra:2019ykk} 
with significant statistics and hence constrain NSI with quarks.

\subsubsection{Sterile neutrinos}

Sterile neutrinos refer to gauge singlet (i.e.~not charged under
the SM gauge symmetry) fermions that have mass mixing with the SM
left-handed neutrinos, such as right-handed neutrinos in the type
I seesaw. As indicated by the name, sterile neutrinos do not participate
in NC and CC interactions of the SM due to their singlet nature. In principle the masses of sterile neutrinos may vary rather arbitrarily
from the GUT scale to values well below the sub-eV scale
(i.e.~the case of quasi-Dirac neutrinos~\cite{deGouvea:2009fp}). However, in neutrino oscillation
phenomenology, we are mainly concerned about light ($\lesssim{\cal O}(1)$
eV) sterile neutrinos, initially motivated  by several
experimental anomalies, including the LSND and MiniBooNE excesses
(see \cite{Abazajian:2012ys} for a review) that cannot be accommodated
in the standard three-neutrino paradigm. While the sterile neutrino
explanation for these anomalies often leads to some inconsistency when
confronted with searches in neutrino experiments~\cite{Kopp:2013vaa,Gariazzo:2015rra,Lindner:2015iaa}
and cosmological observations~\cite{Hamann:2011ge,Planck:2018vyg}, the possible
existence of new oscillation modes caused by sterile neutrinos remains
far from being excluded. 

In the presence of a sterile neutrino $\nu_{s}$, one needs to generalize
the $3\times3$ PMNS mixing to
\begin{equation}
\left(\nu_{s},\ \nu_{e},\ \nu_{\mu},\ \nu_{\tau}\right)^{T}=U\left(\nu_{s}',\ \nu_{1},\ \nu_{2},\ \nu_{3}\right)^{T},\ U=U'U_{{\rm PMNS}}^{(4)}\thinspace,\label{eq:n-9}
\end{equation}
where $\nu_{s}'$ denotes the mass eigenstate approximately identical to $\nu_{s}$, $U_{{\rm PMNS}}^{(4)}={\rm diag}(1,\ U_{{\rm PMNS}})$,
and $U'$ is a unitary matrix that accounts for the small active-sterile
mixing. 
Since $\nu_{s}$ does not have NC or CC interactions, the
Hamiltonian reads~\cite{Lindner:2015iaa}
\begin{equation}
H=\frac{1}{2E_{\nu}}U\left(\begin{array}{cccc}
m_{s}^{2}\\
& m_{1}^{2}\\
&  & m_{2}^{2}\\
&  &  & m_{3}^{2}
\end{array}\right)U^{\dagger}+V_{e}\left(\begin{array}{cccc}
\frac{n_{n}}{2n_{e}}\\
& 1\\
&  & 0\\
&  &  & 0
\end{array}\right),\label{eq:n-10}
\end{equation}
where $n_{n}$ is the neutron number density\footnote{Its presence is due to the fact that the NC contribution to the MSW potential
	was subtracted for active neutrinos of all flavors in Eq.~\eqref{eq:Hamilt}.
	The subtraction should
	be added back for $\nu_{s}$, which makes no such contribution. Usually only neutrons are considered here because protons
	have a much smaller effective vector coupling to $Z$ (suppressed
	by $1-4\sin^{2}\theta_{W}$). }. 
	While Eq.~\eqref{eq:n-10} allows one to numerically solve the Schrödinger
equation to obtain the oscillation probability, the adiabatic
approximation is still valid if the mass splitting $\Delta m_s^2\equiv  m_{1}^{2}-m_{s}^{2} $ is not too small. More specifically, one needs to evaluate the   $\gamma$ parameter in Eq.~\eqref{eq:adia_valid} with $\Delta m_{12}^2\to \Delta m_s^2$ to determine whether adiabaticity is violated or not. 
For $E_{\nu}=10$ MeV,  $\gamma \ll 1$ corresponds to  $\Delta m_s^2\gg 6\times 10^{-8} {\rm eV}^2$. For  $\Delta m_s^2$ well above this value, the adiabatic
approximation can be used.


\begin{figure}
	\centering
	
	\includegraphics[width=0.6\textwidth]{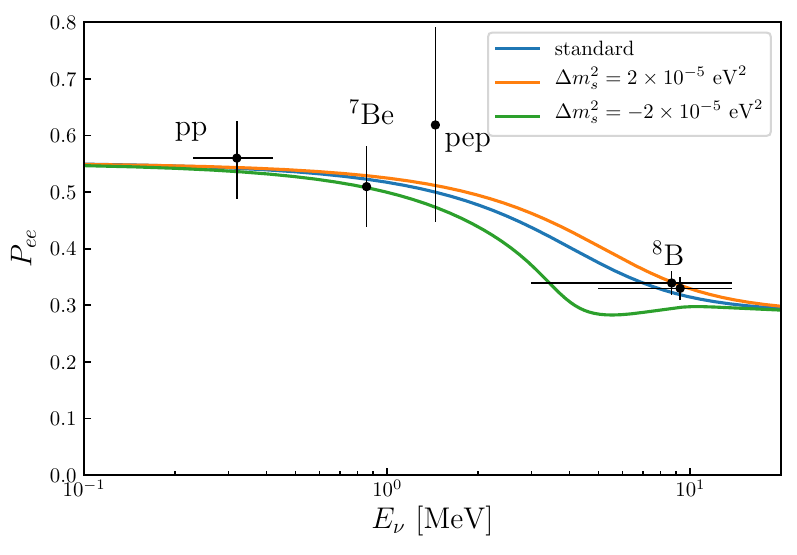}
	
	\caption{The survival probability of solar electron neutrinos $P_{ee}$ in
		the presence of sterile neutrinos, computed according to Eqs.~\eqref{eq:n-10} and \eqref{eq:n-11}.
		The orange and green curves assume  $U'_{12}=-U'_{21}=0.05$ for the sterile-active neutrino mixing.  
		The experimental measurements (black points/bars) are the same as those in Fig.~\ref{fig:survival}.
		Note that the curve and the data points are only for illustration, neglecting some subleading order effects which are discussed in the text.
		\label{fig:survival-sterile} }
\end{figure}

The impact of sterile neutrinos on solar neutrino physics has been
explored extensively in the literature~\cite{deGouvea:2009fp,Giunti:2000wt,Gonzalez-Garcia:2001hid,Bahcall:2002zh,deHolanda:2003tx,Das:2009kw,Giunti:2009xz,deHolanda:2010am,Palazzo:2011rj,Long:2013ota,Long:2013hwa,Billard:2014yka,Ankush:2018mkv,Hostert:2020oui,Goldhagen:2021kxe}.
It has been shown that  sterile
neutrinos with a mass squared difference of $(0.7-2)\times10^{-5}{\rm eV}^{2}$
and a small mixing ($|U'_{12}|^2\sim 10^{-4}-10^{-3}$) 
would modify the up-turn of
the MSW-LMA solution and might  cause a dip of
the survival probability at a few MeV~\cite{deHolanda:2003tx,deHolanda:2010am}. 
In Fig.~\ref{fig:survival-sterile}, we use Eqs.~\eqref{eq:n-10} and \eqref{eq:n-11} to reproduce such an effect of sterile neutrinos, assuming $U'_{12}=-U'_{21}=0.05$  and $\Delta m_{s}^{2}=\pm 2 \times10^{-5}\ {\rm eV}^{2}$.  
Ref.~\cite{deGouvea:2009fp}
studied the scenario of sterile neutrinos with Majorana masses well
below the sub-eV scale, rendering neutrino quasi-Dirac. Solar neutrino
data can impose strong constraints on such a scenario, and it was found
that the Majorana masses in this regime need to be below $10^{-9}$
eV.  Apart from the aforementioned cases of small mass splittings,
one can also consider sterile neutrinos at the eV scale (as possible
explanations for various short-baseline anomalies) and test them in
solar neutrino measurements~\cite{Billard:2014yka,Goldhagen:2021kxe}.
A recent study in Ref.~\cite{Goldhagen:2021kxe} shows that the current
solar neutrino data have excluded significant regions of the parameter
space responsible for some recent anomalies.

\subsubsection{Neutrino magnetic moments\label{sub:NMM}}
Despite being electrically neutral, neutrinos can interact with the
photon via loop processes. In the SM, such loop diagrams are
mediated by the $W^{\pm}$ or $Z$ boson with the photon leg attached
either to the $W^{\pm}$ boson, or to the charge fermion running in the
loop. It is well-known that these diagrams give rise to neutrino magnetic
moments \cite{Petcov:1976ff,Marciano:1977wx,Lee:1977tib,Fujikawa:1980yx,Pal:1981rm,Shrock:1982sc},
provided that neutrinos have small masses. However, neutrino magnetic
moments generated in this way are extremely small, typically around
$10^{-20}\mu_{B}$ ($\mu_{B}=0.296\ {\rm MeV}^{-1}$ is 
the Bohr magneton) for Dirac neutrinos. For Majorana neutrinos, the theoretical values
are further suppressed. In new physics models, loop interactions of
neutrinos with the photon might potentially lead to much larger magnetic
moments~\cite{Voloshin:1987qy,Barr:1990um,Barr:1990dm,Babu:1990hu,Babu:1992vq,Lindner:2017uvt,Xu:2019dxe}.
In addition to the magnetic moment, neutrinos could also possess
other electromagnetic form factors such as electric dipole moments,
charge radii, and anapoles---see~\cite{Giunti:2014ixa} for a comprehensive
review. 

Neutrino electromagnetic interactions would affect both solar neutrino
propagation and detection. Here we concentrate on the latter and leave
the former to Sec.~\ref{subsec:Spin-flavor-precession}. In fact,
constraints on neutrino magnetic moments derived from the latter are
generally much more stringent and more robust than those from the
former. 

In the presence of significant neutrino magnetic moments or other electromagnetic
form factors, the photon can mediate elastic neutrino scattering. Due to its massless feature, it could drastically enhance the cross
section in the soft-scattering limit. The cross section of elastic
$\nu+e$ scattering including the contribution of a neutrino magnetic
moment reads~\cite{Giunti:2014ixa,Vogel:1989iv}:
\begin{equation}
\frac{d\sigma}{dT}=\frac{d\sigma_{{\rm SM}}}{dT}+\frac{\pi\alpha^{2}}{m_{e}^{2}}\left(\frac{1}{T}-\frac{1}{E_{\nu}}\right)\left(\frac{\mu_{\nu}}{\mu_{B}}\right)^{2},\label{eq:n-13}
\end{equation}
where $\frac{d\sigma_{{\rm SM}}}{dT}$ denotes the SM cross section
[see Eq.~\eqref{eq:n-8} with NSI parameters set to zero], $\alpha=1/137$,
and $\mu_{\nu}$ is the neutrino magnetic moment. As is implied by
Eq.~\eqref{eq:n-13}, to gain the sensitivity to $\mu_{\nu}$, one
needs to focus on low $T$ or low $E_{\nu}$, which is the advantage
of solar neutrino data. Therefore, testing neutrino electromagnetic
interactions via elastic scattering of solar neutrinos has been investigated
in many studies~\cite{Coloma:2022umy,Schwemberger:2022fjl,Grimus:2002vb,Canas:2015yoa,Borexino:2017fbd,Huang:2018nxj,PandaX-II:2020udv,Ye:2021zso,Yue:2021vjg,Miranda:2021kre,Akhmedov:2022txm},
with some being  motivated by the recent XENON1T excess, which could
be explained by $\mu_{\nu}\in[1.4,\ 2.9]\times10^{-11}\mu_{B}$ (90\%
C.L.)~\cite{XENON:2020rca}. This value was close to the  best limit by then from Borexino
\cite{Borexino:2017fbd}: $\mu_{\nu}<2.8\times10^{-11}\mu_{B}$
at 90\% C.L.
Unfortunately, the XENON1T excess disappeared with the latest updates from LUX-ZEPLIN~\cite{LZ:2022ufs,AtzoriCorona:2022jeb} and XENONnT~\cite{XENON:2022mpc}.  Nevertheless, the investigation of  possible signals of neutrino magnetic moments in solar neutrinos have led to so far the most stringent constraints: 
$\mu_{\nu}<6.2\times10^{-12}\mu_{B}$ from  LUX-ZEPLIN~\cite{AtzoriCorona:2022jeb} and $\mu_{\nu}<6.3\times10^{-12}\mu_{B}$ from  XENONnT~\cite{XENON:2022mpc}, both at 90\% C.L.



In addition to elastic $\nu+e^{-}$ scattering, CE$\nu$NS of solar neutrinos
at dark matter detectors could be used to test neutrino electromagnetic
interactions. Due to the comparatively high momentum transfer required in order to produce observable nuclear recoils,
it is unlikely that CE$\nu$NS in future experiments will lead to stronger constraints than $\nu+e^{-}$ scattering~\cite{Harnik:2012ni,Papoulias:2018uzy,Li:2022bqr}\footnote{See Fig.~4 in Ref.~\cite{Harnik:2012ni} and Fig.~11 in Ref.~\cite{Papoulias:2018uzy},
	which implies that the CE$\nu$NS bounds would only be competitive if the
	solar neutrino floor could be measured at extremely low nuclear recoils
	($10^{-3}\sim10^{-2}$ keV). }. 

\subsubsection{Neutrino interactions with light mediators}

Neutrino interactions with light mediators such as dark gauge bosons
could be tested by elastic scattering of solar neutrinos off electrons as well~\cite{Coloma:2022umy,Schwemberger:2022fjl,Bauer:2018onh,Abdullah:2018ykz,Amaral:2020tga,Gninenko:2020xys,Li:2022jfl}.
The most extensively studied case is a neutral gauge boson similar to the $Z$ boson in the SM, often denoted by $Z'$. 
For a generic vector mediator $Z'$, one can take the cross
section in Eq.~\eqref{eq:n-8} with $g_{1}$ and $g_{2}$ modified
as follows~\cite{Link:2019pbm,Chauhan:2022iuh}:
\begin{align}
g_{1} & =g_{1}^{{\rm SM}}+\frac{g_{eL}g_{\nu}}{\sqrt{2}G_{F}\left(2m_{e}T+m_{Z'}^{2}\right)}\thinspace,\ g_{1}^{{\rm SM}}=2s_{W}^{2}-1+2\delta_{\alpha e}\thinspace,\label{eq:n-14}\\
g_{2} & =g_{2}^{{\rm SM}}+\frac{g_{eR}g_{\nu}}{\sqrt{2}G_{F}\left(2m_{e}T+m_{Z'}^{2}\right)}\thinspace,\ g_{2}^{{\rm SM}}=2s_{W}^{2}\thinspace,\label{eq:n-15}
\end{align}
where $m_{Z'}$ is the mass of $Z'$, $(g_{eL},\ g_{eR},\ g_{\nu})$
are $Z'$ couplings 
defined in the Lagrangian terms 
${\cal L}\supset Z'_{\mu}\overline{e}\gamma^{\mu}\left(g_{eL}P_{L}+g_{eR}P_{R}\right)e$ $+$ $Z'_{\mu}\overline{\nu}\gamma^{\mu}g_{\nu}P_{L}\nu$.
When $Z'$ is light,  the cross section could be significantly enhanced 
at low energies by $2m_{e}T+m_{Z'}^{2}$ in the denominators in Eqs.~\eqref{eq:n-14}
and \eqref{eq:n-15}. In fact, since among all  $\nu_{\mu}$ and $\nu_{\tau}$ sources for practical detection solar neutrinos have the lowest energy, they have been used to constrain the $L_{\mu}-L_{\tau}$
model which, after imposing all experimental constraints, can still
successfully accommodate the muon $g-2$ anomaly~\cite{Gninenko:2020xys}
for $10^{-2}\ {\rm GeV}\lesssim m_{Z'}\lesssim10^{-1}\ {\rm GeV}$~\cite{Coy:2021wfs}.
The lower bound of $m_{Z'}$ for the muon $g-2$ is mainly determined
by the Borexino data ---see \cite{Gninenko:2020xys} for a recent
update.

\subsubsection{Spin-flavor precession and solar antineutrinos\label{subsec:Spin-flavor-precession}}

In addition to the effect on elastic neutrino-electron scattering as elucidated in Sec.~\ref{sub:NMM}, neutrino magnetic moments may cause another particularly interesting effect, the spin-flavor precession~\cite{Lim:1987tk,Akhmedov:1988uk,Akhmedov:2022txm,Akhmedov2003}. When a neutrino propagates in the solar magnetic field with a nonzero  magnetic moment, the magnetic field could flip the spin of a neutrino and convert it to an antineutrino. The spin flipping effect combined with flavor oscillations results in the conversion of $\nu_{e}\rightarrow \bar{\nu}_{e}$, with the probability given by~\cite{Akhmedov:2022txm,Akhmedov2003}:
\begin{eqnarray}
P_{\nu_{e}\rightarrow\bar{\nu}_{e}}\simeq1.1\times10^{-10}\times\left[\frac{\mu_{\nu}}{10^{-12}\mu_{B}}\frac{B_{\perp}(r_{0})}{10\text{kG}}\right]^{2},
\label{eq:vebar-sfp}
\end{eqnarray}
where $\mu_{\nu}$ is the neutrino magnetic moment  and
$B_\perp(r_{0})$ represents the strength of the solar magnetic field at $r_0\approx 0.05 R_\odot$. 
Note that Eq.~\eqref{eq:vebar-sfp} is not universally valid for all neutrino energies.  For low and high energy parts
of the solar neutrino spectra one should use numerical calculations---see Ref.~\cite{Akhmedov:2022txm} for such a discussion. 

Historically the idea that the solar magnetic field could lead to the neutrino-antineutrino conversion was proposed as a solution to the solar neutrino problem~\cite{Cisneros:1970nq,Okun:1986na,Lim:1987tk,Akhmedov:1988uk}. However, this explanation has faded due to experimental confirmation of the MSW-LMA solution. 
Nevertheless, the neutrino-antineutrino conversion has motivated experimental searches for solar antineutrinos. 
It should be noted that the standard solar model can produce a highly suppressed amount of antineutrinos due to the existence of $\beta^-$ decay elements such as $^{40}$K, $^{238}$U, and $^{232}$Th. 
The expected antineutrino flux  from the standard solar model is around $200\ \text{cm}^{-2}\text{s}^{-1}$ on the Earth's surface, with energies up to 3 MeV. They are buried under the much higher flux of geo-neutrinos ($\sim 10^8\ \text{cm}^{-2}\text{s}^{-1}$)~\cite{Usman:2015yda} and the global reactor antineutrino flux (At CJPL~\cite{Wan:2016nhe}, e.g.,~this is around $10^5\text{cm}^{-2}\text{s}^{-1}$). Photofission reactions occurring in the solar interior, on the other hand, can produce a more energetic flux $\sim 10^{-3}\text{cm}^{-2}\text{s}^{-1}$ at $3$-$9$ MeV,  which is far below existing antineutrino fluxes on the Earth. Therefore, observations of solar $\bar{\nu}_e$ above the known background would be a powerful probe of new physics. 

\begin{table}
	\centering
	\caption{Summary of experimental searches for the 
		measurement of solar $\nu_e\rightarrow\overline{\nu}_e$.}
	\begin{tabular}{llll}
		\toprule
		Expt.   & Target & $E_{\overline{\nu}_e}$(MeV) & $P_{\nu_e\rightarrow\overline{\nu}_e}$(90\%C.L.)\\
		\midrule
		KamLAND~\cite{KamLAND_nu_mu} & Liquid Scintillator & 8.3-31.8 & $5.3\times10^{-5}$ \\
		Borexino~\cite{BRX_nu_mu} & Liquid Scintillator & 1.8-16.8 & $7.2\times10^{-5}$ \\
		SNO~\cite{SNO_nu_mu}     & Heavy Water  & 4.0-14.8 & $8.1\times10^{-3}$ \\
		SK-IV~\cite{sk2022_anti_solarnu}   & Pure Water        & 9.3-17.3 & $4.7\times10^{-4}$\\
		\bottomrule
		\label{tab:expt_nu_mu}
	\end{tabular}
\end{table}

Table~\ref{tab:expt_nu_mu} summarizes the results of experimental searches for $\nu_{e}\rightarrow \bar{\nu}_{e}$.
The KamLAND and Super-K experiments focus on neutrino energies above $8\sim9$ MeV to reduce the 
reactor antineutrino background. Borexino and SNO, with their much lower reactor antineutrino backgrounds, can perform such searches at lower energies, with the lower energy bounds being only limited by their detection thresholds.

The experimental searches rely crucially on the experimental ability to detect and identify $\overline{\nu}_e$ events.  In the aforementioned experiments, $\overline{\nu}_e$ is detected either by the inverse beta decay (KamLAND, Borexino, and Super-K) or the charged-current reaction on deuterium, $\overline{\nu}_e+d\rightarrow e^++n+n$ in heavy water (SNO). 
Neutron tagging at Super-K is important for the background reduction, which  can be significantly improved by adding Gd to the detector.

\subsubsection{Dark matter annihilation \label{sub:DM-anni}}

The local density of the galactic dark matter (DM) halo is known to
be around $0.4\ {\rm GeV}/{\rm cm}^{3}$ in the solar system. As the
Sun moves in the halo, it can capture DM particles that fall into
its gravitational potential well and scatter with normal matter particles
or, in the presence of DM self-interactions, with DM particles. The
theory of DM being captured by the Sun or other large celestial bodies
has been developed  since the 1980s~\cite{Spergel:1984re,Gould:1987ju,Gould:1987ir}. 

The accumulation of DM in the Sun can be used to constrain DM annihilation,
which may produce various SM particles, including quarks, leptons,
etc. While most of them cannot escape the Sun except for neutrinos,
they may decay to neutrinos and other final states. Hence neutrinos
from DM annihilation in the Sun could be used to constrain or probe
DM properties---see, e.g.,~Refs.~\cite{Blennow:2007tw,Zentner:2009is,Bell:2011sn,Rott:2012qb,Bernal:2012qh,Busoni:2013kaa,Chen:2014oaa,Guo:2015hsy,Gorham:2015rfa,Lawson:2015cla,Gaidau:2018yws,Gupta:2022lws}.
Most of these studies focused on Weakly Interacting Massive Particles
(WIMP), which is so far the most extensively studied DM candidate.
The mass of WIMP typically varies from a few GeV to hundreds of GeV,
implying that neutrinos from WIMP annihilation should have similar
energies (i.e., around the same orders of magnitude). It requires
that neutrino detectors have the capability to detect high-energy
neutrinos within the above energy window\footnote{Within this energy window, the dominant background is atmospheric
	neutrinos, which could be reduced by improving the directional resolution.
	However, 
as previously discussed, solar atmospheric neutrinos remain as an irreducible background even if the direction can be well measured.
 }. An interesting exception is Quark Nugget Dark Matter~\cite{Gorham:2015rfa,Lawson:2015cla}, which leads to neutrino signals in the 20--50 MeV range.

\begin{figure}
	\centering
	
	\includegraphics[width=0.48\textwidth]{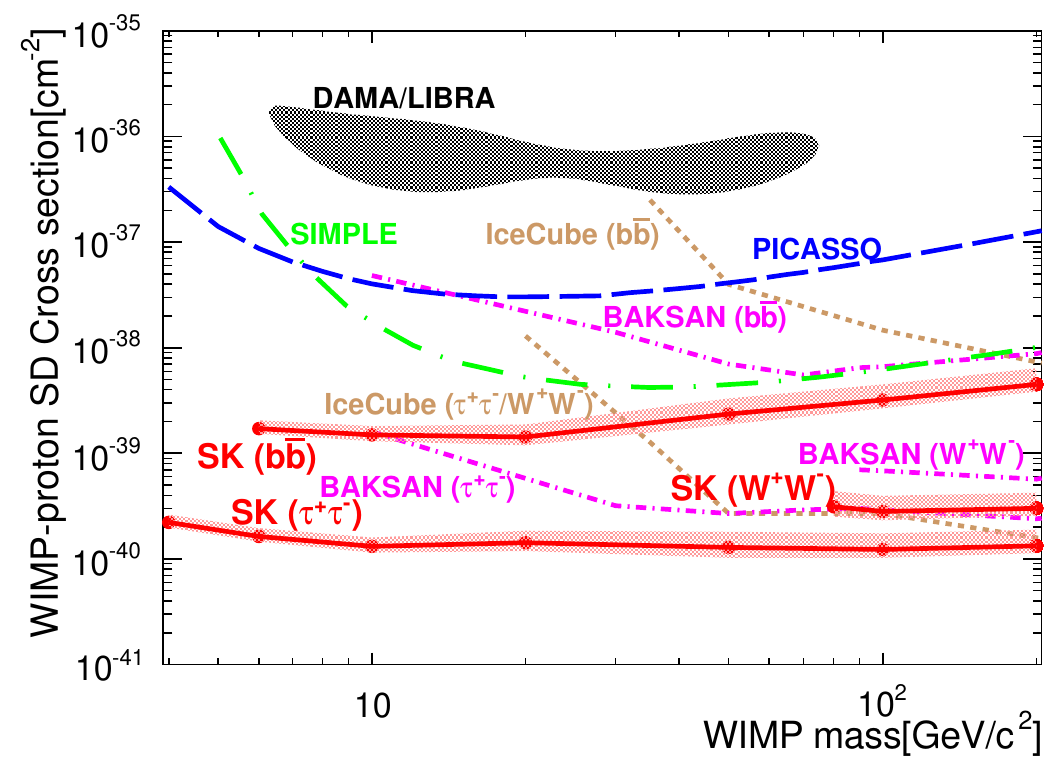}\includegraphics[width=0.465\textwidth]{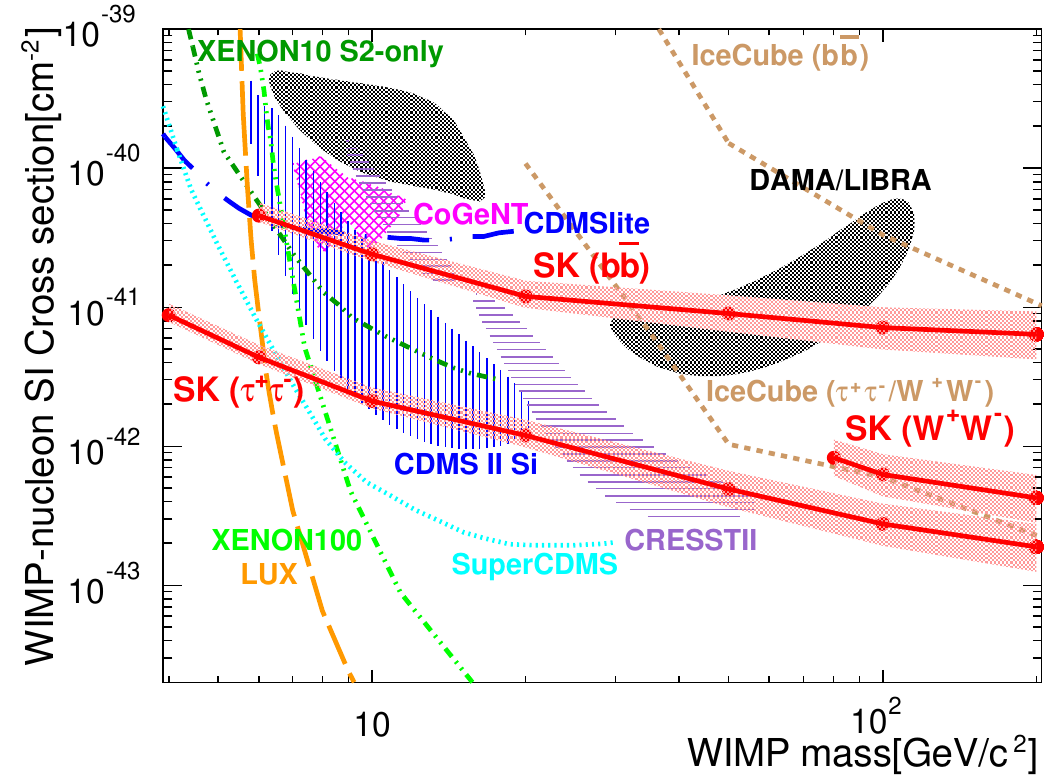}
	
	\caption{Super-Kamiokande (SK) limits of DM annihilation in the Sun on SD (left)
		and SI (right) WIMP-proton cross sections, taken from Ref.~\cite{Super-Kamiokande:2015xms}.
		\label{fig:DM}}
	
\end{figure}

Several experiments have conducted searches for neutrinos from DM
annihilation in the Sun, including Super-Kamiokande~\cite{Super-Kamiokande:2015xms},
IceCube~\cite{IceCube:2016yoy,IceCube:2016dgk}, and ANTARES~\cite{ANTARES:2016xuh}.
All experiments found no significant excess, putting stringent constraints
on WIMP-proton cross sections. In particular, the  most stringent
limits on spin-dependent (SD) WIMP-proton cross section are obtained
from these experiments. Figure~\ref{fig:DM} shows the 90\% C.L. upper
limits on SD and spin-independent (SI) WIMP-proton cross sections
reported by Super-Kamiokande~\cite{Super-Kamiokande:2015xms}. Depending
on the mass and the annihilation channel of WIMP, the limits vary
from $10^{-38}\sim10^{-40}\ {\rm cm}^{-2}$ and $10^{-40}\sim10^{-43}\ {\rm cm}^{-2}$
for the SD and SI cross sections, respectively. The SI bounds are generally
weaker than those from direct detection due to the coherent enhancement
of large nuclei in the SI case. 

\subsubsection{Neutrino decay }

Massive neutrinos are not absolutely stable. Even with pure SM interactions, due to loop-level processes, massive neutrinos can decay as $\nu_{i}\to\nu_{j}\gamma$ or $\nu_{i}\to\nu_{j}\nu_{k}\nu_{l}$,\footnote{This case is similar to lepton flavor violating decays of charged leptons
	such as $\mu\rightarrow e+\gamma$ and $\mu\rightarrow3e$, which
	are present (though highly suppressed) when neutrino masses and the
	PMNS mixing are introduced to the SM. }
though the lifetime is much longer than the universe's age~\cite{Zatsepin:1978iy,Smirnov:1981rz}.
Nevertheless, like the situation of neutrino magnetic moments, new
physics of neutrinos might potentially enhance the decay rates to an
experimentally accessible level. 

Since, so far, all neutrinos being successfully detected are relativistic,
the lifetime of a neutrino during flight, $\tau_{{\rm flight}}$,
is dilated by the Lorentz factor, which is equal to $E_{\nu}/m_{\nu}$,
i.e.~$\tau_{{\rm flight}}=\tau_{{\rm rest}}E_{\nu}/m_{\nu}$ where
$\tau_{{\rm rest}}$ is the lifetime in the rest frame. As neutrinos
decay during propagation, the flux is depleted by the factor $\exp(-L/\tau_{{\rm flight}})$
where $L$ is the distance of propagation. Increasing $L$
can drastically enhance experimental sensitivity to neutrino decay.
Hence the strongest constraints on neutrino decay are derived
from observations of supernova neutrinos (SN1987A) and solar neutrinos. 

Using solar neutrinos to constrain neutrino lifetimes was first studied
in Refs.~\cite{Joshipura:2002fb,Beacom:2002cb}, followed by several studies further exploring various aspects~\cite{Hostert:2020oui,Huang:2018nxj,Bandyopadhyay:2002qg,Cecchini:2004ym,Berryman:2014qha,Picoreti:2015ika,deGouvea:2022cmo}.
The lower bound on $\tau_{{\rm rest}}/m_{\nu}$ varies within $10^{-4}\sim10^{-3}\ {\rm sec}/{\rm eV}$~\cite{Beacom:2002cb,Berryman:2014qha},
depending on how the standard MSW-LMA solution and its uncertainties
are taken into account and also on which mass eigenstate decays. The
bound is about ten orders of magnitude weaker than that from supernova
neutrinos of SN1987A. However, the solar neutrino bounds have the merit that they can be applied to a specific mass eigenstate, while the constraint from SN1987A would be invalid if any of the three mass eigenstates is stable.

\subsubsection{Others}

In addition to those mentioned above, there have been a variety of other
new physics scenarios that solar neutrinos could probe. Below
we briefly mention some interesting examples.
\begin{enumerate}
	\item New long-range forces could be present with very weak couplings. Such
	forces could induce additional flavor-dependent effective potentials
	and hence be probed by neutrino oscillation, as has been studied in
	Refs.~\cite{Joshipura:2003jh,Grifols:2003gy,Bandyopadhyay:2006uh,GonzalezGarcia:2006vp,Nelson:2007yq,GonzalezGarcia:2008wk,Samanta:2010zh,Heeck:2010pg,Davoudiasl:2011sz,Chatterjee:2015gta,Bustamante:2018mzu,Khatun:2018lzs,Wise:2018rnb,Smirnov:2019cae,Babu:2019iml}.
	Solar neutrinos have the advantage that the new potential caused by the Sun is much larger than that
	caused by the Earth for the same strength of a long-range
	force. For a generic vector mediator with a mass $m_{A}$
	and a universal coupling $g$, solar neutrinos can probe $g\sim10^{-25}$  when $m_{A}$
	is around the inverse of the solar radius~\cite{Smirnov:2019cae}. This exceeds other known experimental bounds significantly.
	\item Dark matter-neutrino interactions could affect neutrino oscillation
	when neutrinos propagate on a DM background \cite{Berlin:2016woy,Krnjaic:2017zlz,Capozzi:2017auw,Brdar:2017kbt,Liao:2018byh,Huang:2018cwo,Lopes:2020hem}.
	For instance, Ref.~\cite{Berlin:2016woy} showed that for the fuzzy
	DM scenario, current solar neutrino data are more sensitive to the
	neutrino-DM coupling than CMB limits by more than two orders of magnitude.
	Ref.~\cite{Capozzi:2017auw} proposed a framework which connects
	DM and sterile neutrinos via a dark gauged $U(1)$ and studied the
	solar MSW effect caused by DM, dubbed Solar Dark MSW. The authors
	showed that Solar Dark MSW is characterized by comparatively large
	modifications of $^{8}$B, $^{15}$O, and $^{13}$N neutrinos, with
	the other fluxes less affected. 
	\item In addition to interactions with dark matter,  neutrinos could also
	interact with dark energy,  as exemplified by the idea of  Mass-Varying
	Neutrinos (MaVaN)~\cite{Fardon:2003eh} which has drawn considerable
	interest in cosmology.   If neutrinos are coupled to a dark scalar
	field whose background value accounts for the dark energy, then neutrino
	masses are connected to dark energy. It explains the intriguing
	coincidence of today's dark energy density $\rho_{\Lambda}\sim(2\times10^{-3}\ {\rm eV})^{4}$
	with the neutrino mass scale, $\rho_{\Lambda}^{1/4}\sim m_{\nu}$.
	The neutrino oscillation probe of MaVaN was proposed and investigated
	in Ref.~\cite{Kaplan:2004dq,Barger:2005mn,Cirelli:2005sg}. 
	\item  DM accumulated in the Sun could affect the energy transport in the Sun and modify the solar core temperature~\cite{Lopes:2001ig,Bottino:2002pd,Taoso:2010tg,Cumberbatch:2010hh,Lopes:2012af,Lopes:2012zz,Lopes:2014aoa,Casanellas:2015uga,Lopes:2018wgp}. 
	For non-annihilating DM accumulated in the Sun, the central
	temperature could be reduced by a few percent.
	Consequently, neutrino production rates in the very central region
	are reduced and and enhanced in the outer part. This feature allows
	one to use precision measurements of the solar neutrino spectrum to
	constrain GeV DM effectively---see, e.g.,~Ref.~\cite{Taoso:2010tg,Lopes:2012zz} for further details. 
	
\end{enumerate}

\newpage
\section{Detection and experimental challenges}

Neutrinos are detected when they scatter on particles such as electrons
or nuclei in a detector and generate observable signals. Neutrino
 scattering processes are mediated by charged-current (CC) or neutral-current (NC) interactions.
NC interactions are flavor universal, whereas CC interactions are, for solar neutrinos,  only
relevant to the detection of $\nu_{e}$ since
$\nu_{\mu}$ and $\nu_{\tau}$ do not have sufficient energy to produce
their corresponding heavy charged lepton partners $\mu$ and $\tau$. 

For solar neutrino scattering on electrons, it has to be elastic ($\nu+e^{-}\to\nu+e^{-}$) and only the final-state electron is observable. The elastic $\nu+e^{-}$ scattering
process is particularly important for measuring low-energy
solar neutrino fluxes due to its zero threshold. 
Its disadvantage is that  the
neutrino energy usually cannot be constructed on an event-by-event basis, unless both the electron energy and direction are well measured. 

For scattering on nuclei, there are several possibilities including
elastic scattering ($\nu+N\to\nu+N$), CC scattering ($\nu_{e}+N\to N'+e^{-}$),
and NC inelastic scattering (e.g.,~$\nu+{\rm ^{2}H}\to\text{p}+\text{n}+\nu$),
etc.  The last two types of reactions have been successfully applied to solar neutrino observations (Homestake, GALLEX/GNO, SNO).
In contrast, elastic scattering on nuclei has not yet been observed for solar neutrinos. In the foreseeable future, with the improvement of ultra-low
nuclear recoil detection in, e.g.,~DM detectors, elastic scattering
on nuclei, which is a coherent process due to the low energy of solar neutrinos, will soon become an effective way to detect solar neutrinos.

Below we focus our discussions on two predominant processes in solar
neutrino detection,  elastic scattering on electrons and CC scattering
on nucleus.

\subsection{Elastic neutrino-electron scattering }

Elastic neutrino-electron scattering applies to all three neutrino flavors: 
\begin{equation}
\nu_{\alpha}+e^{-}\to \nu_{\alpha}+e^{-}\thinspace, (\alpha=e,\ \mu,\ \tau)\,.\label{eq:3}
\end{equation}
Note that the total cross section of $\nu_e+e^-$ scattering is about 6 times greater than that of $\nu_{\mu, \tau}+e^-$ when the recoil electron is relativistic. This difference is because the former receives contributions from both CC and NC interactions, while the latter is mediated only by NC interactions. More specifically, for $E_{\nu}\gg m_{e}$, we have $\sigma(\nu_{e}+e^{-}) /10^{-46} {\rm cm}^2 \approx93s/{\rm MeV}^{2}$
and $\sigma(\nu_{\mu,\tau}+e^{-})/10^{-46} {\rm cm}^2\approx15s/{\rm MeV}^{2}$ with $s=2E_{\nu}m_{e}$~\cite{Bahcall:1986pf, Giunti:2007ry}.
Towards low energy, e.g.~from 10 MeV to 1 MeV, $\sigma(\nu_{\mu,\tau}+e^{-})/\sigma(\nu_{e}+e^{-})$ increases, and this needs to be considered for the relevant experimental study for the solar neutrino up-turn effect.

In elastic  neutrino-electron scattering, the electron recoil kinetic energy $T_{e}$ is related to the initial neutrino
energy $E_{\nu}$ by
\begin{equation}
T_{e}=\frac{2m_{e}E_{\nu}^{2}\cos^{2}\theta}{(m_{e}+E_{\nu})^{2}-E_{\nu}^{2}\cos^{2}\theta}\thinspace,\label{eq:3-1}
\end{equation}
where $\theta$ denotes the angle between the outgoing electron and
the incoming neutrino (solar) directions, varying from $0^{\circ}$ to $90^{\circ}$.
Given a fixed $E_{\nu}$, $T_{e}$ reaches its maximum, $T_{\max}$,
at $\theta=0^{\circ}$ and vanishes when $\theta=90^{\circ}$. The
maximum is given by
\begin{equation}
T_{\max}=\frac{2E_{\nu}^{2}}{2E_{\nu}+m_{e}}\thinspace.\label{eq:Tmax}
\end{equation}
Theoretically, with the measured electron scattering angle $\theta$ and its recoil $T_{e}$, we can obtain $E_{\nu}$ from 
\begin{equation}
E_{\nu}=\frac{m_{e}}{\sqrt{1+2m_{e}/T_{e}}\text{cos}\theta-1}\thinspace.\label{eq:3-2}
\end{equation}
In practice, the angle $\theta$ cannot be measured accurately at low energies.  For instance, water Cherenkov detectors have angular resolution $\sim 40^{\circ}$ ($20^{\circ}$) for $T_e=5$ (15) MeV (see Sec.~\ref{sub:E-and-angle}), while conventional liquid scintillator detectors are hardly able to measure the direction.
For this reason, the recoil energy spectrum with respect to $T_e$ is primarily used in the data analyses. The $T_e$ spectrum is related to the neutrino energy spectrum by
\begin{equation}
\frac{dN}{dT_{e}}=N_{e}t_{{\rm exposure}}\int_{0}^{E_{\nu}^{\max}}\phi(E_{\nu})\frac{d\sigma}{dT_{e}}\Theta(T_{\text{max}}-T_{e})dE_{\nu}\thinspace,\label{eq:nu-e-rate}
\end{equation}
where $\frac{dN}{dT_{e}}$ denotes the event rate; $N_{e}$ is the
total number of electrons in the detector ($N_{e}=3.3\times10^{32}$
for 1kt water); $t_{{\rm exposure}}$ is the exposure time; 
$\phi(E_{\nu})$ is the neutrino flux; 
$\frac{d\sigma}{dT_{e}}$
is the differential cross section for which we refer to Eq.~(\ref{eq:n-8});
and $\Theta(T_{\text{max}}-T_{e})$ is the Heaviside step function.

\begin{figure}
	\centering \includegraphics[width=0.49\textwidth]{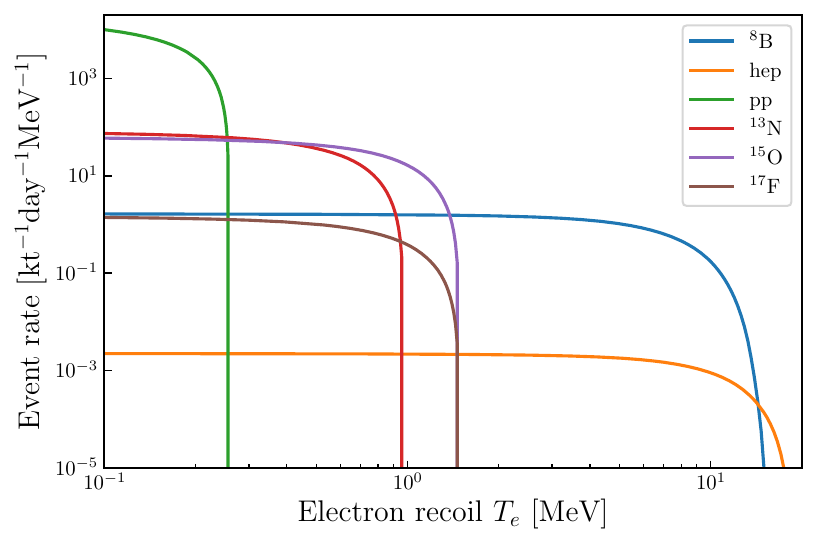}
	\includegraphics[width=0.49\textwidth]{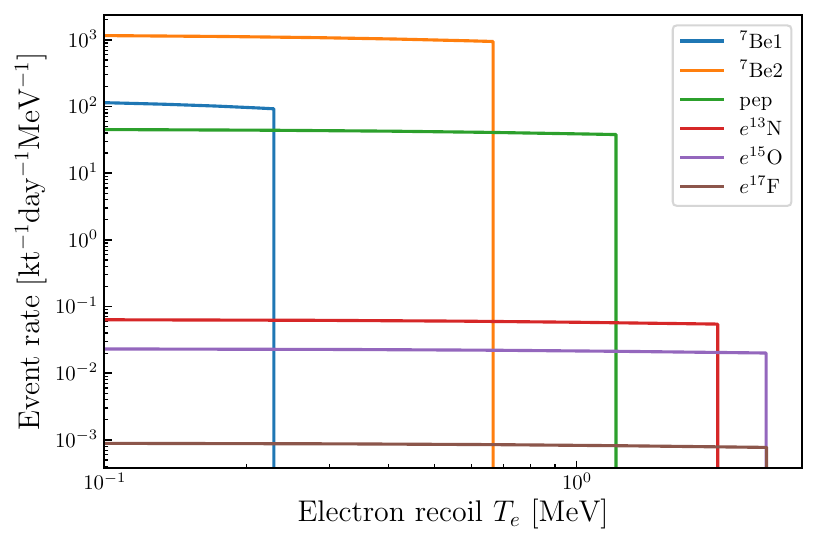}\caption{Electron recoil distributions of elastic $\nu_{e}+e^{-}$ scattering
		for continuous (left) and monochromatic (right) solar neutrino spectra. }
	\label{fig:recoilenergy} 
\end{figure}

Figure~\ref{fig:recoilenergy} shows the electron recoil spectra
of solar neutrinos obtained using the fluxes presented in Fig.~\ref{fig:solarnuflux}
and Eq.~\eqref{eq:nu-e-rate}, assuming no flavor conversion. One can
see that although the continuous and monochromatic neutrino spectra
in Fig.~\ref{fig:solarnuflux} are rather distinct from each other,
their electron recoil spectra are quite similar, all being flat at
low $T_{e}$ and quickly falling to zero when $T_{e}$ increases to
$T_{\max}$. The main difference is at the turning point where the
monochromatic case has a sharp cut-off, but resolving this difference would require very high precision measurements of the spectrum. 
In general, it is difficult to use electron recoils to distinguish between continuous and monochromatic
neutrino spectra, and unfolding the solar neutrino
spectra based on elastic scattering data is challenging. 


\subsection{CC scattering on nucleus }

\begin{table}[h]
\centering 

\caption{Thresholds of $\nu_{e}$-capture reactions. Most reactions are ground-to-ground transitions except for $\nu_{e}+{}^{40}{\rm Ar}\rightarrow{}^{40}{\rm K}^*+e^{-}$ in which the final nucleus is an excited nuclear state (further details explained in the text).
Data obtained from Ref.~\cite{iaea}.\label{tab:nu_cap_elements}}
\begin{tabular}{clc}
\toprule 
Reaction & $E_{\nu}$ threshold & Experiments\tabularnewline
\midrule 
$\nu_{e}+{}^{163}{\rm Dy}\rightarrow{}^{163}{\rm Ho}+e^{-}$ & 2.8 keV & -\tabularnewline
$\nu_{e}+{}^{205}{\rm Tl}\rightarrow{}^{205}{\rm Pb}+e^{-}$ & 50.6 keV & LOREX~\cite{Pavicevic:2018jdl,Kostensalo:2019cua}\tabularnewline
$\nu_{e}+{}^{123}{\rm Sb}\rightarrow{}^{123}{\rm Te}+e^{-}$ & 51.9 keV & -\tabularnewline
$\nu_{e}+{}^{193}{\rm Ir}\rightarrow{}^{193}{\rm Pt}+e^{-}$ & 80.2 keV & -\tabularnewline
 & $\cdots$ & \tabularnewline
$\nu_{e}+{}^{55}{\rm Mn}\rightarrow{}^{55}{\rm Fe}+e^{-}$ & 0.231 MeV & -\tabularnewline
$\nu_{e}+{}^{71}{\rm Ga}\rightarrow{}^{71}{\rm Ge}+e^{-}$ & 0.232 MeV & GALLEX~\cite{Gallex1992}, SAGE~\cite{sage2002}\tabularnewline
$\nu_{e}+{}^{73}{\rm Ge}\rightarrow{}^{73}{\rm As}+e^{-}$ & 0.345 MeV & -\tabularnewline
 & $\cdots$ & \tabularnewline
$\nu_{e}+{}^{37}{\rm Cl}\rightarrow{}^{37}{\rm Ar}+e^{-}$ & 0.814 MeV & Homestake~\cite{Davis:1968cp}\tabularnewline
$\nu_{e}+{}^{57}{\rm Fe}\rightarrow{}^{57}{\rm Co}+e^{-}$ & 0.836 MeV & -\tabularnewline
$\nu_{e}+{}^{7}{\rm Li}\rightarrow{}^{7}{\rm Be}+e^{-}$ & 0.862 MeV & Refs.~\cite{Kopylov:2008zz,Kopylov:2009aj,Fujita:2021gsu,Shao:2022yjc}\tabularnewline
$\nu_{e}+{}^{75}{\rm As}\rightarrow{}^{75}{\rm Se}+e^{-}$ & 0.865 MeV & -\tabularnewline
 & $\cdots$ & \tabularnewline
$\nu_{e}+{}^{40}{\rm Ar}\rightarrow{}^{40}{\rm K}^*+e^{-}$ &   \hspace{-0.5em}\raisebox{1em}{
\begin{minipage}[t]{13em}%
5.888 MeV (Fermi)\\
$3.8\sim4.6$ MeV (GT)%
\end{minipage}
}
& DUNE~\cite{DUNE2016} \tabularnewline
\bottomrule
\end{tabular}

\end{table}

At solar neutrino energies, CC scattering of neutrinos with a nucleus
usually cannot break the nucleus, simply converting one of the neutrons
in the nucleus to a proton, $\nu_{e}+\text{n}\rightarrow\text{p}+e^{-}$,
which we refer to as neutrino-capture beta decay ($\nu$BD). It differs from inverse beta decay (IBD), $\overline{\nu_{e}}+\text{p}\rightarrow\text{n}+e^{+}$,
which has been extensively used to detect antineutrinos. Because free
neutrons are unstable, in a practical $\nu$BD process, the neutron
has to be bound in a nucleus $_{Z}^{A}N$ which after absorbing $\nu_{e}$
becomes $_{Z+1}^{A}N'$:
\begin{equation}
\nu_{e}+{}_{Z}^{A}N\rightarrow{}_{Z+1}^{A}N'+e^{-}\thinspace.\label{eq:nuecapture}
\end{equation}
Such a process has been applied to solar neutrino detection since
the very early stage of experimental studies\footnote{In the 1930s, Crane and Halpern used such reactions to look
for neutrinos by measuring the energy of the emitted $\beta$-ray
and the recoil atom~\cite{Crane1938,Crane1939,CraneHalpern1939,Wang1941}.}. For example, the Homestake experiment employed $\nu_{e}+{}^{37}{\rm Cl}\rightarrow{}^{37}{\rm Ar}+e^{-}$,
which  is a typical $\nu$BD reaction. 

The CC cross section calculation can be found in Ref.~\cite{Bahcall1989, RevModPhys.50.881, Bahcall1966, Barinov:2017ymq}. 
Unlike elastic $\nu+e^{-}$ scattering, the neutrino energy $E_{\nu}$
in $\nu$BD can be well determined from the electron kinetic energy
$T_{e}$ and the masses of initial and final particles\footnote{This feature is rather important for the Earth matter effect measurement.}:
\begin{equation}
E_{\nu}=T_{e}+m_{N'}-m_{N}+m_{e}\thinspace,\label{eq:nueenergy}
\end{equation}
where $m_{N}$ and $m_{N'}$ are the masses of the initial and final
nuclei, respectively. 
The kinetic energy of the final nucleus is negligible
since  it is of the order $\sim E_{\nu}^{2}/m_{N'}$, much smaller
than $E_{\nu}$.  Note that when the final nucleus is in an excited state, then $m_{N'}$ denotes the mass of the excited nucleus mass, which can be obtained by adding the excitation energy to the ground-state nucleus mass. In the presence of multiple allowed transitions to different excited states, one needs to take into account their branching ratios, which depend on the corresponding nuclear matrix elements.
The experimental determination of the nuclear matrix elements is usually done with (p, n) or ($^3$He, t) reactions. More details can be found in, e.g.~Ref.~\cite{Yoshi, Yoshi2}.


The usage of $\nu$BD in solar neutrino detection concerns two limitations.
First, $\nu$BD is only applicable to the detection of $\nu_{e}$,
irrelevant to neutrinos of other flavors. Second, it has a threshold
given by 
\[
E_{\nu}^{{\rm thre}}=m_{N'}-m_{N}+m_{e}=m_{N'}^{({\rm atom})}-m_{N}^{({\rm atom})},
\]
where $m_{N}^{({\rm atom})}$ and $m_{N'}^{({\rm atom})}$ denote
the masses of $N$ and $N'$ atoms. For some final state isotopes, the nuclear excitation energy levels can be rather complicated and 
some low-energy excitation states may have very low transition rates. In such cases, the complicated energy levels might smear the threshold. 

In Tab.~\ref{tab:nu_cap_elements}, we present a list of low-threshold
$\nu$BD processes obtained
by looking for electron capture (i.e.,~the inverse of $\nu$BD) interactions
with low $Q$ values. 
Due to technical difficulties, many low-threshold $\nu$BD processes have not been used in solar neutrino detection. The successful
examples  $^{71}{\rm Ga}$ and $^{37}{\rm Cl}$ have a common feature:
they can be used in liquid form, and the final-state nuclei can be
extracted and counted using radiochemical methods.  In the Homestake
experiment, $^{37}{\rm Cl}$ was used in the form of tetrachloroethylene
($\text{C}_{2}\text{Cl}_{4}$), which is liquid at room temperature.
Metal $^{71}{\rm Ga}$ melts at $29.8\thinspace^{\circ}{\rm C}$ and
was used in the SAGE experiment. In the GALLEX experiment, $^{71}{\rm Ga}$
is contained in the detector as an aqueous solution of gallium chloride. 

In addition, we also include in Tab.~\ref{tab:nu_cap_elements} the process  $\nu_{e}+{}^{40}{\rm Ar}\rightarrow{}^{40}{\rm K}^*+e^{-}$ which is import to DUNE. Its threshold depends on the excited states of $^{40}\text{K}^{*}$~\cite{Ormand:1994js,Trinder:1997xr}. The Fermi transition of $^{40}\text{Ar} (0^+)$ to the second $0^+$ excited state of $^{40}\text{K}$ has the largest nuclear matrix element (hence the largest cross section for sufficiently high $E_{\nu}$). The threshold is given by  $E_{\nu}^{\rm thre}=E_{\nu0}^{\rm thre}+E_i$ where $E_{\nu0}^{\rm thre}= 1.504$ MeV is the would-be threshold if the ground-state transition were allowed and  $E_i=4.384$ MeV is the excitation energy. Apart from the Fermi transition, several Gamow-Teller (GT) transitions to $^{40}\text{K}(1^+)$ with the excitation energy $E_i= 2.290, 2.730$, and $3.110$ MeV have lower thresholds but  smaller nuclear matrix elements. 

J. Bahcall proposed that lithium could be used to detect solar neutrinos
in 1964~\cite{Bahcall1964}. This possibility has recently been investigated
in Refs.~\cite{Kopylov:2008zz,Kopylov:2009aj,Fujita:2021gsu,Shao:2022yjc}.
The ground-state-to-ground-state transition $\nu_{e}+{}^{7}{\rm Li}\rightarrow{}^{7}{\rm Be}+e^{-}$
has a threshold of $E_{\nu}^{{\rm thre}}=0.862$ MeV. In addition,
the final-state nucleus can be in its first excited state: $\nu_{e}+{}^{7}{\rm Li}\rightarrow{}^{7}{\rm Be}^{*}+e^{-}$,
which has a threshold of $E_{\nu}^{{\rm thre}}=1.291$ MeV.  Both
GT
 and Fermi transitions contribute to the ground-state
reaction, while for the excited case, only the GT transition is possible~\cite{Fujita:2021gsu}.
 The cross section of this reaction is about 60 times the cross section
of elastic $\nu_{e}+e^{-}$ scattering when applied to $^{8}{\rm B}$
neutrino detection~\cite{Shao:2022yjc}. A lithium detector might
be possible by exploiting the high solubility of LiCl, $74.5{\rm g}$
per 100g of water at $10\ ^{\circ}{\rm C}$. An initial test reported
in Ref.~\cite{Shao:2022yjc} indicates that a saturated LiCl solution
shows excellent optical transparency. The attenuation length of the
solution under 430nm LED light is measured to be $11\pm1$ m. Hence
the use of LiCl solution in a 10-m diameter detector seems promising.

Some other isotopes such as  $\rm{{}^{11}B}$~\cite{boron} and $\rm{{}^{115}In}$~\cite{indium} 
have been well discussed and explored by experimentalists. The naturally high radioactivity limits the usage of $\rm{{}^{115}In}$. 
Xenon as a dark matter detection medium has been considered as a neutrino target~\cite{Haselschwardt}.
People have also thought about finding delayed coincidence to reduce the experimental difficulty for $^{115}$In~\cite{Raghavan:1976yc}, $^{100}$Mo~\cite{Ejiri:1999rk}, $^{176}$Yb~\cite{Raghavan:1997ad}, $^{116}$Cd~\cite{Zuber}, and
$\rm{{}^{71}Ga}$~\cite{Wang}.

%

\subsection{Coherent elastic neutrino-nucleus scattering (CE$\nu$NS) \label{subsec:CEvNS}}

At $E_{\nu}\lesssim50$ MeV, elastic scattering of a neutrino with
a nucleus via NC interactions is generally considered to be coherent,
which implies that the cross section can be substantially enhanced
by the large number of nucleons in the nucleus. This process is known
as coherent elastic neutrino-nucleus scattering (CE$\nu$NS), with the following differential cross section~\cite{Freedman:1973yd,Lindner:2016wff}:
\begin{equation}
\frac{d\sigma}{dT}=\frac{G_{F}^{2}\left[N-(1-4s_{W}^{2})Z\right]^{2}F^{2}(T)M}{4\pi}\left(1-\frac{T}{T_{\max}}\right),\label{eq:coh}
\end{equation}
where $N$ and $Z$ are the neutron and proton numbers of the nucleus;
$T$ and $M$ denote the recoil energy and the mass of the nucleus;
$T_{\max}$ is the maximal value of $T$ determined by
\begin{equation}
T_{\max}=\frac{2E_{\nu}^{2}}{M+2E_{\nu}}\thinspace.\label{eq:coh-1}
\end{equation}
The form factor $F$, which quantifies the coherency of the scattering,
can be computed using Helm's approximate formula~\cite{Helm:1956zz}:
\begin{equation}
F=3\frac{j_{1}(qr_{n})}{qr_{n}}e^{-(qs)^{2}/2}\thinspace,\label{eq:coh-2}
\end{equation}
where $q\approx\sqrt{2TM}$, $s\approx0.9$ fm,  $r_{n}\approx1.14A^{1/3}$
fm with $A=N+Z$, and $j_{1}$ is the spherical Bessel function of
the first kind. At $q=0$, the form factor should be $F=1$. For solar
neutrino scattering on Xe nuclei, the deviation $1-F$ typically varies
around $5\%$. 

CE$\nu$NS has received increasing interest since the first successful
observation by the COHERENT experiment in 2017~\cite{COHERENT:2017ipa}.
Despite the large cross section of CE$\nu$NS, the main challenge
of detecting neutrinos via this process is the low nucleus recoil
energy ($\sim$keV for 10 MeV neutrinos) which is basically invisible
in liquid scintillator or water Cherenkov detectors. Dark matter detectors
are generally sensitive to such low nucleus recoils. With the rapid
development of dark matter detectors (larger scales and lower backgrounds),
CE$\nu$NS becomes a promising process to detect solar neutrinos
in the near future---see Sec.~\ref{sub:DM} for further discussions
on the experimental progress.

\subsection{Energy and direction measurements\label{sub:E-and-angle}}

As elucidated above, after neutrino scattering with particles in a
detector, the neutrino energy is partially (for elastic scattering)
or fully (for CC scattering on nucleus) transferred to charged particles
such as electrons and nuclei. The kinetic energy of a final-state
nucleus is negligible since it is typically below keV. Here we concentrate
on the energy and directional measurements of electrons. 

Electrons produced from solar neutrino scattering can only travel
a very short distance in the detector medium before it stops. For MeV
electrons in water, the propagation distance is around $0.5\ {\rm cm}\times(T_{e}/{\rm MeV)}$~\cite{LANNUNZIATA2007119}.
 Within this short distance, electrons lose energy due to ionization,
bremsstrahlung, and Cherenkov radiation. 

The effects of ionization and Cherenkov radiation are employed in
modern neutrino detectors filled
with water (Super-K, SNO) or liquid scintillator (Borexino) and equipped with PMTs. Water-based
Cherenkov detectors  can provide important directional information
using Cherenkov light, whereas liquid scintillator detectors excel
at energy resolution using uniform scintillation light caused by ionization. 

\vspace{5pt}

\textbf{$\blacksquare$ Water Cherenkov detectors}

\begin{figure}
\centering

\includegraphics[width=0.8\textwidth]{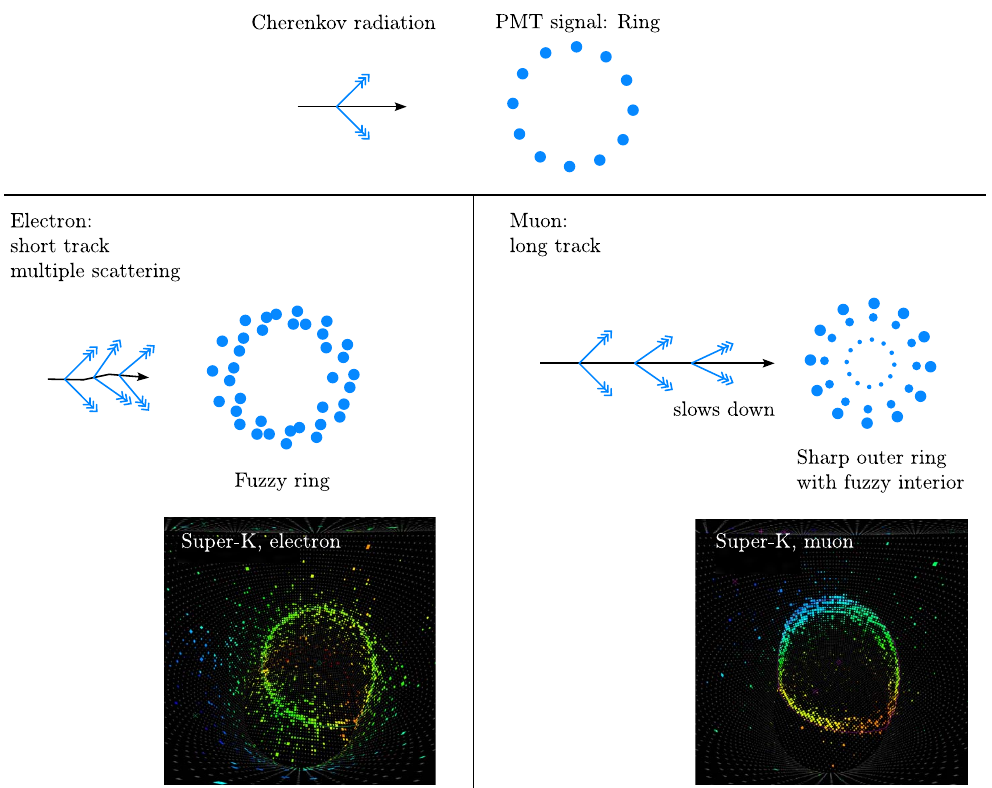}

\caption{Cherenkov rings caused by relativistic electrons and muons. A relativistic
electron leaves a short track deflected by multiple soft scattering,
leading to a fuzzy ring. By comparison, a muon track is straight and
much longer, with a sharp outer ring and a fuzzy interior due to the
slow-down of the muon. \label{fig:Cherenkov}}

\end{figure}

In water Cherenkov detectors, the Cherenkov light emitted by relativistic
charged particles such as electrons propagates as a cone in the water
and reaches the surrounding PMTs as a ring---see Fig.~\ref{fig:Cherenkov}.
The Cherenkov cone has an opening angle given by
\begin{equation}
\theta_{c}=\arccos\frac{1}{\beta n}\approx41^{\circ}\thinspace,\label{eq:3-3}
\end{equation}
 where $n\approx1.33$ is the refractive index for water, and
$\beta\approx1$ is the speed of the particle. Theoretically, the
emission of Cherenkov light from a moving electron only requires $\beta>1/n$,
corresponding to $E_{e}>m_{e}/\sqrt{1-1/n^{2}}$ or $T_{e}>0.26$
MeV.  In practice, the threshold of detecting electrons via Cherenkov
light is higher (e.g.,~$3.49$ MeV at Super-K~\cite{Super-Kamiokande:2016yck}) due to a sharp increase in event rate caused by radioactive backgrounds and PMT dark noises, as described in 
Sec.~\ref{sub:cos-bkg}.

The light emission follows from the Frank--Tamm formula:
\begin{equation}
\frac{d^{2}E}{dLd\omega}=\alpha\omega\sin^{2}\theta_{c}\thinspace,\label{eq:3-4}
\end{equation}
where $\alpha\approx1/137$ is the fine-structure constant, $\omega$
is the frequency of the Cherenkov light, and $L$ is the propagation
distance.  For electrons, the total energy of Cherenkov radiation
takes only a very minor fraction of the kinetic energy since the major
energy loss is ionization. The electron energy can be inferred from
the number of Cherenkov photons.  The electron energy determined
from Cherenkov light is less accurate than that from the 
measurement of ionization energy, which so far is only possible
in liquid scintillator. 

The directional measurement plays an essential role in event reconstruction
and background reduction. Due to its small mass, the electron undergoes
multiple soft (i.e. the momentum transfer is much smaller than the
electron energy) scattering processes during Cherenkov radiation.
As illustrated in Fig.~\ref{fig:Cherenkov}, each of the multiple
scattering processes deflects the direction of the electron and hence
the direction of the Cherenkov cone. Consequently, the signal arriving
at surrounding PMTs is a fuzzy ring. This is to be compared with a
muon track which can hardly be deflected due to $m_{\mu}\gg m_{e}$
and features a sharp outer ring with fuzzy interior rings caused by
the slow-down of the muon. 



In the Super-K detector, the energy
resolution varies from $10\%$ (for $E_{e}\approx40$ MeV) to $20\%$
(for $E_{e}\approx4$ MeV) and the angular resolution varies from
$20^{\circ}$ (for $E_{e}\approx18$ MeV) to $40^{\circ}$ (for $E_{e}\approx4$
MeV)~\cite{Super-Kamiokande:2010tar}.   The energy resolution improves when $E_{e}$ increases because of
the statistics of photoelectrons detected by PMTs increases for larger
$E_{e}$.  The angular resolution improves when $E_{e}$ increases because
of the aforementioned multiple scattering, which occurs more often
for low-energy electrons and reduces the capability of Cherenkov detectors
to measure the electron direction.

\vspace{5pt}

\textbf{$\blacksquare$ Liquid scintillator detectors }

Because  most of the electron kinetic energy deposited in the detector
is transferred to ionization instead of Cherenkov radiation, the electron
kinetic energy can be more straightforwardly (and hence more precisely)
measured from the ionization energy. Liquid scintillator (LS) offers an
effective way to measure the energy deposited in ionization by converting
it to optical photons. 

The working mechanism of LS is the subject of fluorescence.
The most widely used LS is organic solvent doped
with fluorophores (often briefly referred to as fluor). Both the solvent and fluorophore molecules possess
aromatic rings. The ionization leads to excitation of the aromatic
solvent molecules, which then transfer their excited energy to fluorophore
molecules. Due to an effect known as the Stokes shift, the fluorophore, 
when undergoing deexcitation, emits photons with considerably larger
wavelengths than the photon spectrum of the solvent. 

The primary advantage of LS is its high light yield,
$\sim10^{4}$ photons/MeV for deposit energy, which is about 50 times
higher than that in water Cherenkov detectors~\cite{Borexino:2013zhu}.
Only a small fraction of these photons are detected due to the attenuation length of LS and   
the coverage and detection
efficiency of PMTs. 
For example,  the Borexino detector receives $\sim$500 photoelectrons per MeV.

The energy resolution of a LS detector mainly relies on the
number of  photoelectrons, $N_{pe}$, of which the statistical fluctuation
is $\sqrt{N_{pe}}$. For Borexino, $N_{pe}\sim500E_{e}/{\rm MeV}$
implies that the energy resolution is approximately $1/\sqrt{N_{pe}}\sim5\%\sqrt{E_{e}/{\rm MeV}}$~\cite{Borexino:2019wln}. 

Measuring directions in LS is challenging 
since the scintillation light emission from ionization is isotropic. An electron path length is about $0.5\ {\rm cm}\times(T_{e}/{\rm MeV)}$~\cite{LANNUNZIATA2007119},
which is sufficiently short to be treated as nearly point-like in comparison to
the resolution of position reconstruction ($\sim10$ cm in Borexino~\cite{Borexino:2013zhu}). 
The  Cherenkov light emission is directional but usually overwhelmed by the scintillation light. Observation of Cherenkov light in LS might be possible if the detector is capable of making use of the time separation between them (the Cherenkov light is emitted almost instantaneously while the scintillation light is emitted at the nanosecond level) or if the scintillation light is reduced (e.g.,~in water-based LS). The former has recently been demonstrated feasible by Borexino~\cite{BOREXINO:2021efb}.

\subsection{Backgrounds}

The solar neutrino spectrum spans from ${\cal O}(0.1)$ MeV (pp neutrinos)
to $18.77$ MeV (the endpoint of hep neutrinos). Within this range, there are two categories of backgrounds: cosmogenic and detector-related. The latter comes from external (e.g., surrounding rock) and internal (contamination in the fiducial volume) radioactivity.

\subsubsection{Cosmogenic backgrounds\label{sub:cos-bkg}}

\begin{figure}
	\centering
	
	\includegraphics[width=0.96\textwidth]{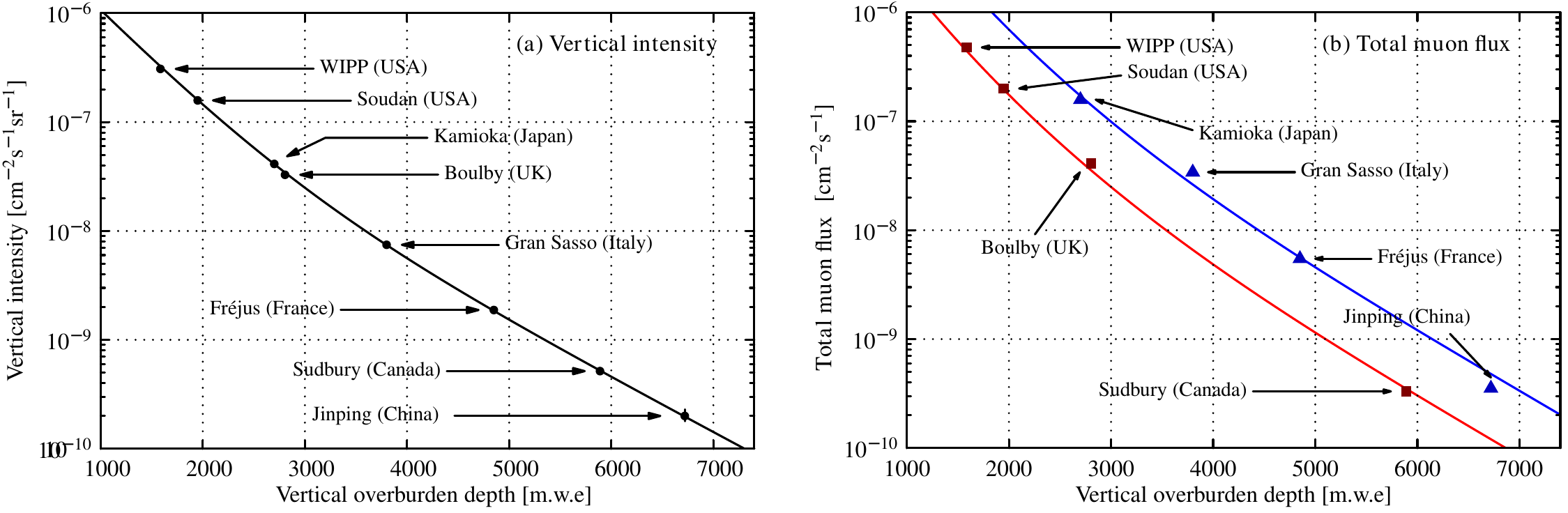} \caption{Measured muon fluxes in underground laboratories compared with theoretical calculations. The left/right panel is for vertical/total muon fluxes. 
		 The black, red, and blue lines represents fits using the empirical formula. The difference between red and blue lines in the right panel is caused by the shape of the overburden. A flat overburden (e.g.,~Sudbury) with the same depth can shield better from large-zenith-angle cosmic muons.
		 Figure taken and modified from Ref.~\cite{JNE:2020bwn}.
	 }
	\label{fig:muon_flux}
\end{figure}

Cosmic-ray muons on the ground can trigger the data acquisition system of a neutrino detector, causing a sharp increase in event rate and a production of radioactive background, which are not desirable for a rare-event experiment. Therefore, neutrino experiments usually go underground to utilize the thick rock overburden to shield these muons as much as possible. 
The flux of cosmic-ray muons can be parametrized by Gaisser’s formula~\cite{Gaisser} and further modifications~\cite{Guan:2015vja,JNE:2020bwn}. 
Figure~\ref{fig:muon_flux} shows how the muon flux decreases with the depth in units of  {\it meter of water equivalent} (m.w.e).
 
\begin{figure}
	\centering \includegraphics[width=0.6\textwidth]{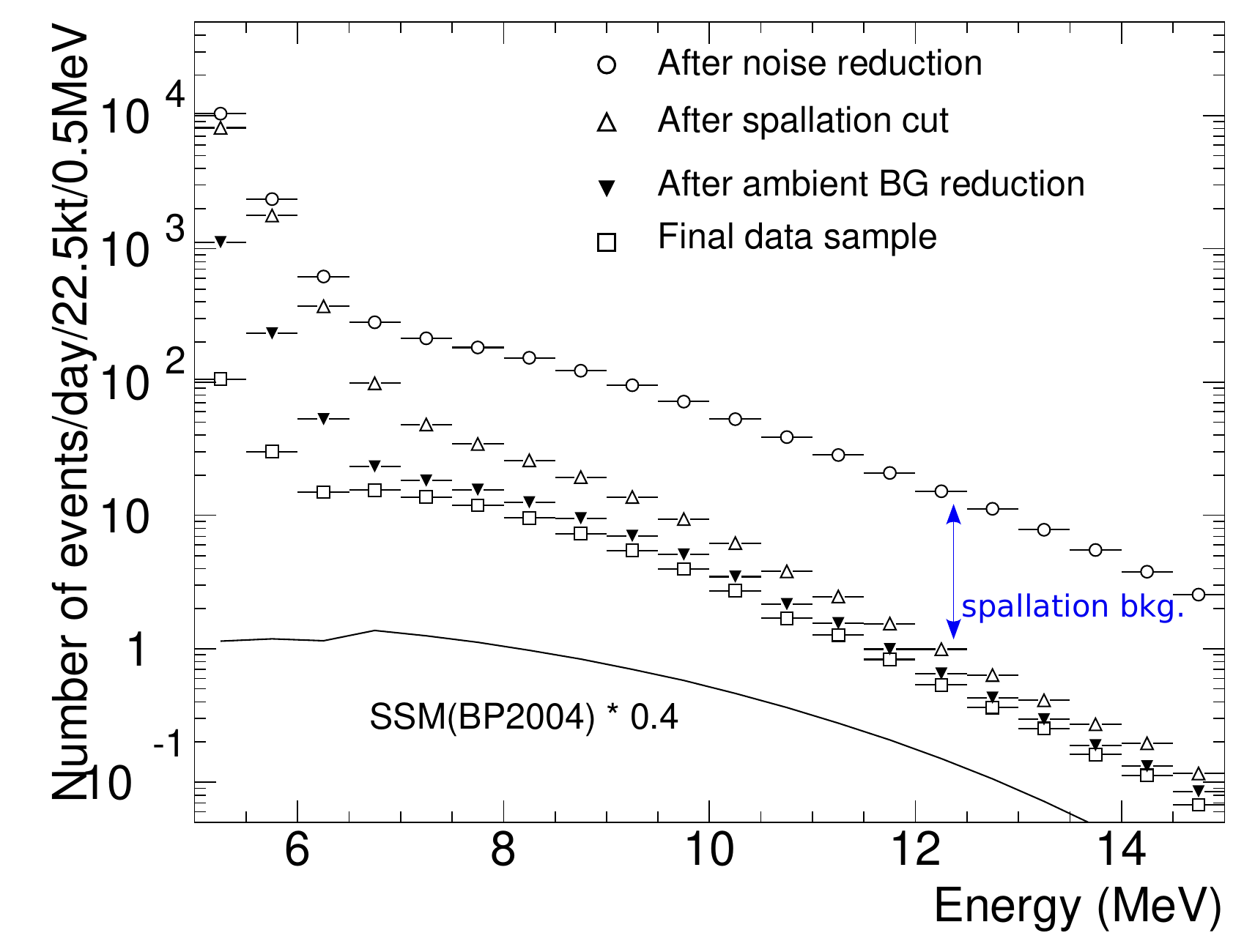} \caption{Background reduction at multiple steps in Super-Kamiokande~\cite{Super-Kamiokande:2005wtt}.
		Above $\sim$6 MeV, the spallation background which comes from decays
		of radioactive isotopes ($^{12}{\rm B}$, $^{12}{\rm N}$, $^{9}{\rm Li}$
		, $^{8}{\rm Li}$, etc.) generated by cosmic muons scattering off
		$^{16}{\rm O}$ becomes dominant. }
	\label{fig:skenergy}
\end{figure}

\begin{figure}
	\centering 
\includegraphics[width=0.6\textwidth]{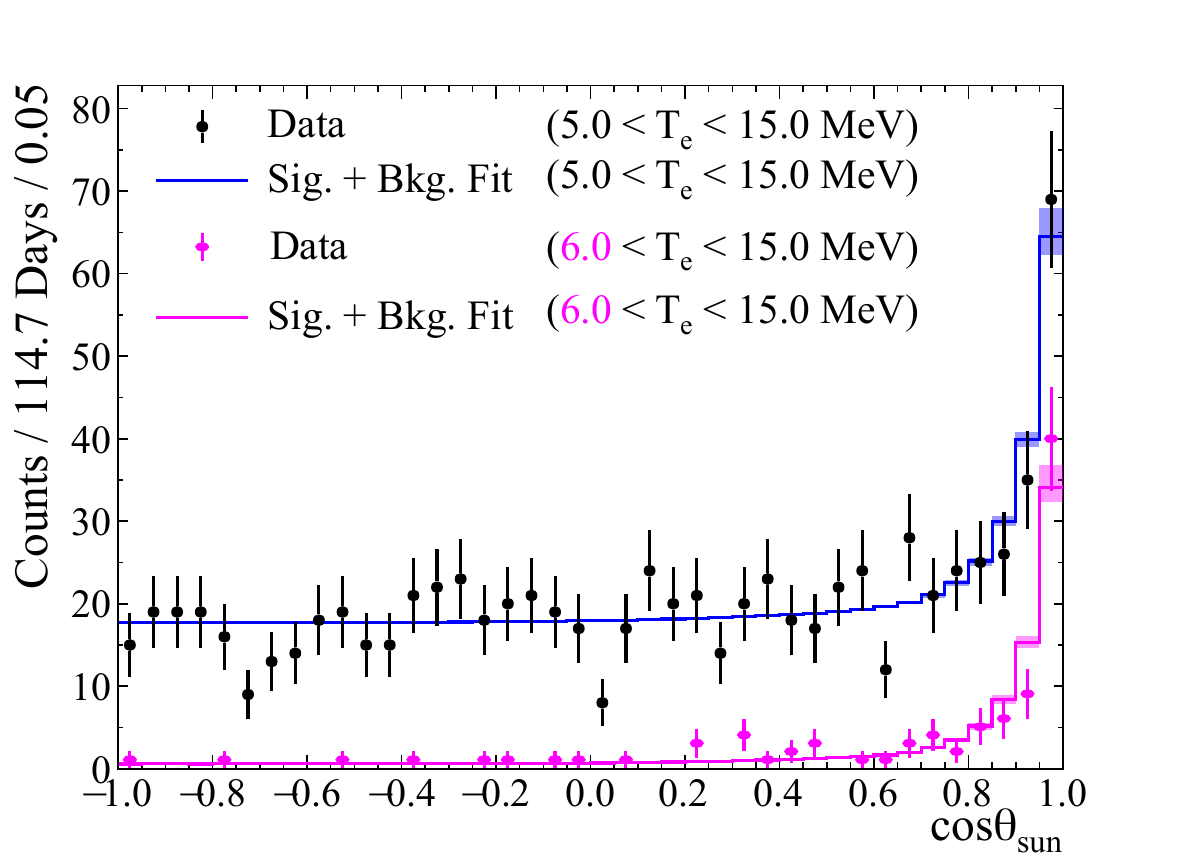}	
	\caption{Angular distributions of solar neutrino-electron scattering events
		in 
		SNO+. 
		The original figures are taken from
		Ref.~\cite{SNO:2018fch}, we combine both figures to give a better illustration. The elastic scattering rate peaks
		at $\cos\theta_{{\rm sun}}=1$, and the non-vanishing flat part with
		$\cos\theta_{{\rm sun}}<0$ can be viewed as either the spallation background or the detector-related background since, theoretically, the solar neutrino signal always has $\cos\theta_{{\rm sun}}>0$. See more discussion in the text.
	}
	\label{fig:sk-sno-angle} 
\end{figure}

When an energetic cosmic muon impinges on a nucleus,  it may break up the nucleus and cause a spallation process, thereby generating many short-lived radioactive isotopes. These isotopes then undergo $\alpha$-,
$\beta$- or $\gamma$- decays to produce radiations, mimicking solar
neutrino signals~\cite{Back2006,ABe2010,Zhang2016}.

In water-based detector, since the heaviest element abundantly contained in the fiducial volume is $^{16}$O,  radioactive isotopes generated via spallation
are lighter than $^{16}{\rm O}$. For example, $\mu^{\pm}+{}^{16}{\rm O}\to\mu^{\pm}+{}_{5}^{12}{\rm B}+3{\rm p}+{\rm n}$
 gives rise to $_{5}^{12}{\rm B}$ which decays in the $\beta^{-}$
mode within $\sim10^{-2}$ sec. In Super-K, the significant background
from cosmic muon spallation arises from $^{12}{\rm B}$, $^{12}{\rm N}$,
$^{9}{\rm Li}$, $^{8}{\rm Li}$, $^{15}{\rm C}$, etc. These isotopes
decay and emit $\beta^{\pm}$ rays of ${\cal O}(10)$ MeV~\cite{Super-Kamiokande:2005wtt}.
In addition, the muon capture process $\mu^{-}+{}^{16}{\rm O}\to\text{\ensuremath{\nu_{\mu}}}+{}^{16}{\rm N}$
generates $^{16}{\rm N}$ which has a relatively long lifetime, $7.13$
sec. In Super-K, this background is reduced by imposing a cut based
on spatial and time correlations of the stopping muon with $^{16}{\rm N}$
decays~\cite{Super-Kamiokande:2005wtt}.   Figure~\ref{fig:skenergy}
shows the event rates at multiple steps of background reduction~\cite{Super-Kamiokande:2005wtt}.
As shown in the figure,  the background above 6 MeV majorly comes from spallation.  

The spallation background has no angular preference because the generated non-relativistic radioactive isotopes decay isotropically.
 Hence the spallation background and solar neutrino signals can be
differentiated in angular distributions of  event rates.  Figure~\ref{fig:sk-sno-angle}
shows angular distributions of solar neutrino-electron scattering
events in 
SNO+~\cite{SNO:2018fch}.
Here $\theta_{\text{sun}}$ is defined as the angle between the reconstructed
recoil electron direction and the expected neutrino direction, which
is known from the Sun's position at the event time. The elastic
scattering rate peaks at $\cos\theta_{{\rm sun}}=1$ and becomes flat
at smaller $\cos\theta_{{\rm sun}}$ (corresponding to large $\theta_{{\rm sun}}$).
Theoretically, elastic neutrino-electron scattering does not allow
a negative $\cos\theta_{{\rm sun}}$. Therefore, the non-vanishing
flat rate with $\cos\theta_{{\rm sun}}<0$ indicates the background
level. 
For SNO+, thanks to its substantial overburden (2000m below a flat ground), the cosmic spallation background is reduced to a signal-to-noise level of four for the energy region above 6 MeV. 

\begin{figure}[h]
\centering \includegraphics[width=0.6\textwidth]{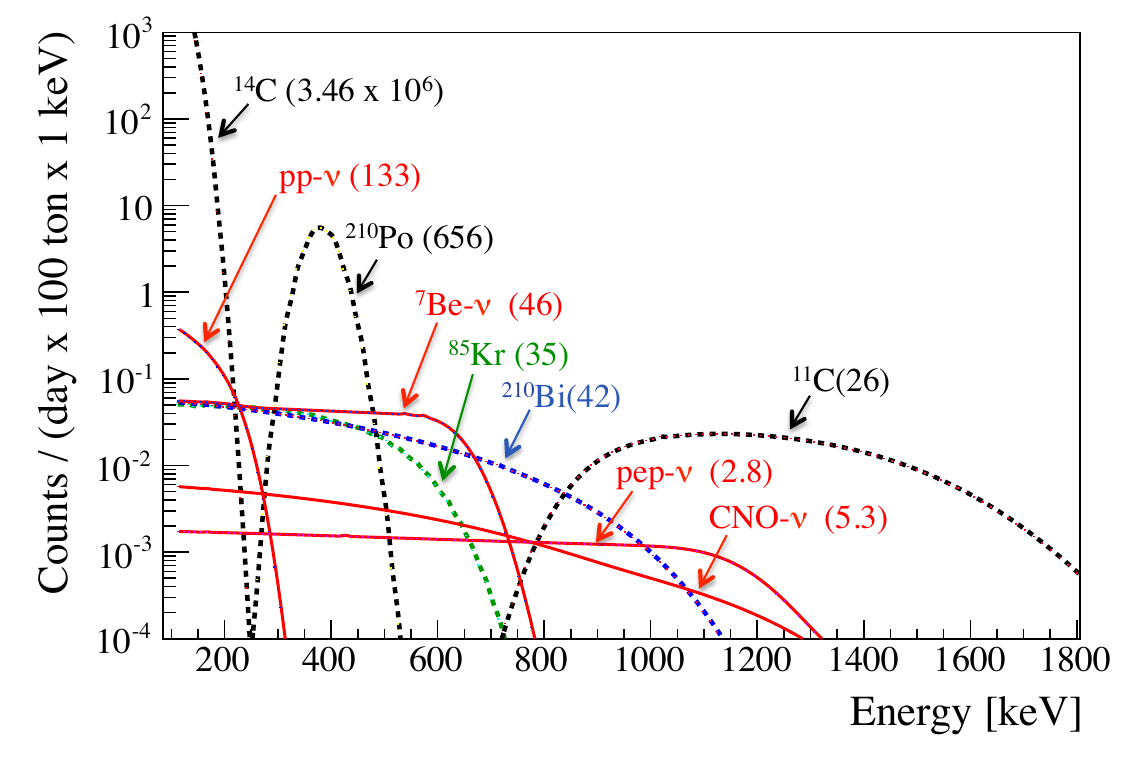}
\caption{Borexino backgrounds (dashed lines) compared with solar neutrino signals
(solid lines) in the low energy regime~\cite{Borexino:2013zhu}.
}
\label{fig:bore-bkg}
\end{figure}

For liquid scintillator experiments, the dominant cosmogenic background
comes from carbon spallation: $\mu+{}^{12}{\rm C}\rightarrow\mu+{}^{11}{\rm C}+n$.
The neutron in the above spallation process is captured by hydrogen
in liquid scintillator, releasing a 2.225 MeV $\gamma$-ray. The unstable
isotope $^{11}{\rm C}$ decays with a half-life time of 20.34 minutes.
Its dominant\footnote{The electon capture process is also possible ($0.19-0.23\%$): $^{11}{\rm C}+e^{-}\rightarrow{}^{11}{\rm B}+\nu_{e}$
with a $Q$ value of $1.98$ MeV.} decay channel is positron emission: $^{11}\text{C}\rightarrow{}^{11}{\rm B}+e^{+}+\nu_{e}$
with a $Q$ value of $0.96$ MeV. As $e^{+}$ will eventually annihilate
with $e^{-}$ in the detector, the total energy released from $^{11}\text{C}$
decay in the detector ranges from $2m_{e}=1.02$ MeV to $2m_{e}+Q=1.98$
MeV. As shown in Fig.~\ref{fig:bore-bkg}, the $^{11}{\rm C}$ background
in Borexino dominates the event rates in this range (the energy resolution effect is taken into account), posing a substantial challenge
to the measurement of pep and CNO neutrinos~\cite{Borexino:2013zhu}. 

Due to its relatively long lifetime, $^{11}{\rm C}$ is difficult
to tag. Borexino has developed a Three-Fold Coincidence (TFC) method
for $^{11}{\rm C}$ tagging through the spatial and time coincidence
among (i) the positron from $^{11}\text{C}$ decay, (ii) the parent
muon, and (iii) the neutron capture. Using this method has successfully reduced the $^{11}\text{C}$
background by $\sim10\%$ (see Fig.~40 in Ref.~\cite{Borexino:2013zhu}).

Deeper underground laboratories such as SNO or CJPL can almost eliminate
the cosmogenic backgrounds. A precise $^{8}$B measurement and search for
hep neutrinos can go down to the level at which the low-energy atmospheric neutrino background starts to be significant. 

\subsubsection{Detector-related backgrounds}

Detector-related backgrounds can be categorized as internal or external. 

\begin{table}
	\centering \caption{Summary of natural radioactive background levels for water and liquid
		scintillator detectors. Note that radioactive background levels are based on studies for SNO and Borexino. Hence they should not be considered a priori as generic expectations for future detectors.
	 }
	\label{tab:radiative} 
	\begin{tabular}{lrr}
		\toprule 
		& Water (g/gH$_{2}$O) & Liquid Scintillator (g/gLS)\tabularnewline
		\midrule 
		$^{238}$U Chain & $6.6\times10^{-15}$\cite{Okeefe2008} & $1.6\times10^{-17}$\cite{Gatti1996,Borexino2008}\tabularnewline
		$^{232}$Th Chain & $8.8\times10^{-16}$\cite{Okeefe2008} & $6.8\times10^{-18}$\cite{Gatti1996,Borexino2008}\tabularnewline
		$^{40}$K & $6.1\times10^{-16}$\cite{Balata1996} & $1.3\times10^{-18}$\cite{Gatti1996,Borexino2009}\tabularnewline
		\bottomrule
	\end{tabular}
\end{table}

External backgrounds come from the radioactivity of surrounding materials,
mainly from the glass of PMTs, the vessel, the support structure, rock, and cement.
Typical radioactive backgrounds include $\alpha$, $\beta$, $\gamma$ rays and   neutrons.
Among them, $\gamma$ rays and neutrons need a careful detector design.  
A buffer layer (e.g., water, mineral oil, or LS with quenched material) can significantly
reduce the external background.

\begin{figure}[t]
	\centering \includegraphics[width=0.95\textwidth]{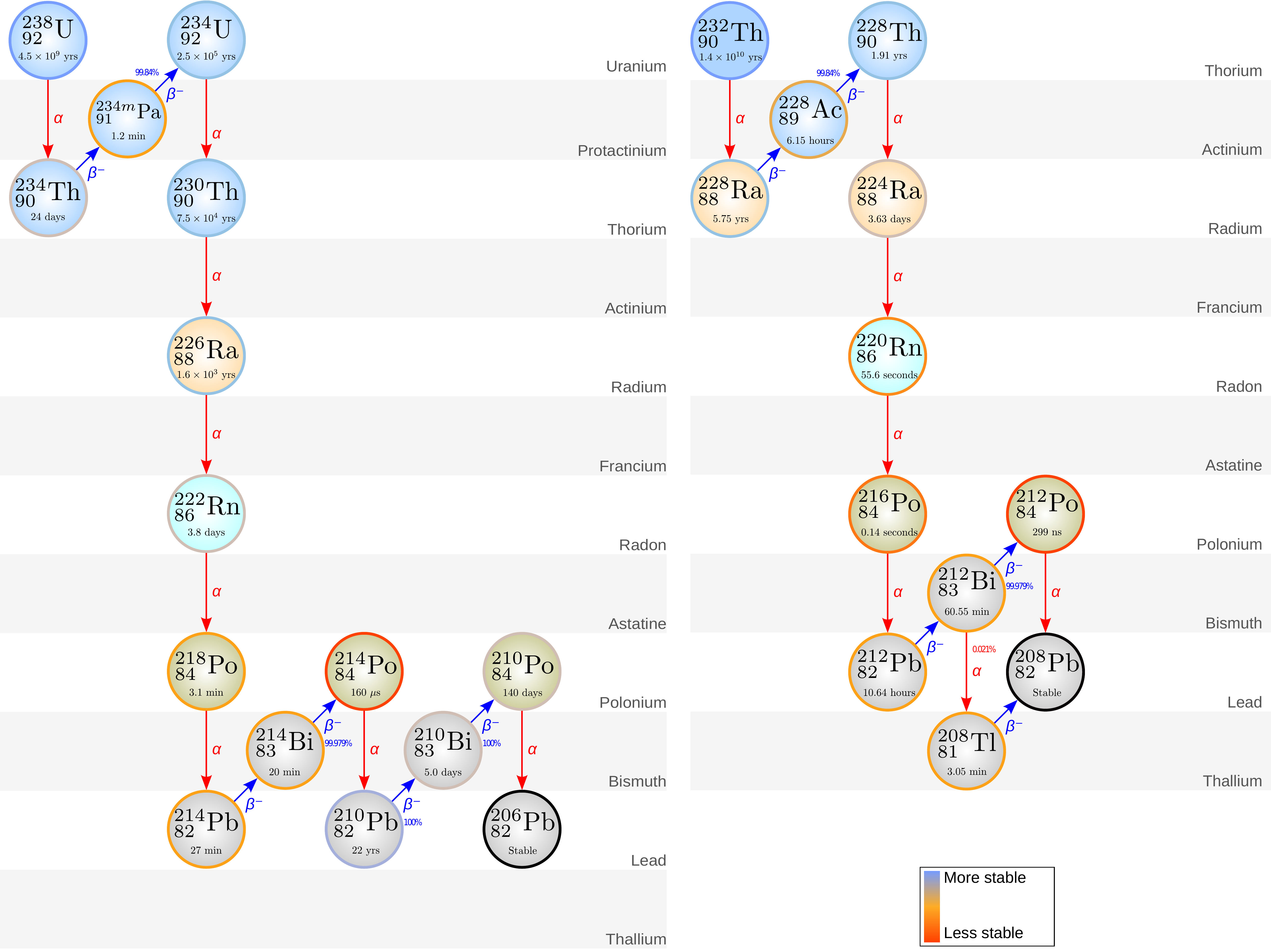}
	\caption{Decay chains of $^{238}$U and $^{234}$Th, with he half-life time indicated below each element.
	}
	\label{fig:chains}
\end{figure}

Internal backgrounds come from radioactive isotopes in the fiducial
volume. Radioactive ions (e.g.,~$^{40}{\rm K}$), noble gas ($^{85}$Kr,
$^{39}$Ar), and products of $^{238}$U and $^{234}$Th decay chains (see Fig.~\ref{fig:chains})
can be dissolved in water or liquid scintillator. 

Table~\ref{tab:radiative} summarizes natural radioactive background
levels for water and liquid scintillator detectors, according to studies for SNO and Borexino~\cite{Okeefe2008,Gatti1996,Borexino2008,Balata1996,Borexino2009}. 
Backgrounds from $^{238}$U and $^{232}$Th chains and $^{40}$K in
water are significantly higher (roughly by two orders of magnitude)
than those in liquid scintillator. This is because the binding ability
of water molecules with inorganic ions is stronger than that of organic
liquid scintillator. Hence the solubility of inorganic ions
in water is higher than that in liquid scintillator.

It is important to note that both the $^{238}\text{U}$ and $^{232}\text{Th}$
chains contain the inert gas element radon ($^{222}{\rm Rn}$, $^{220}{\rm Rn}$), which is difficult to be removed chemically. The radon isotopes can
continuously enter the fiducial volume emanation from external materials. 

Tagging some of decays in the $^{238}\text{U}$ and $^{232}\text{Th}$
chains is possible by exploiting a fast decay sequence existing in
each of the chains:
\begin{align}
^{214}\text{Bi}\thinspace\xrightarrow{28.4\thinspace{\rm min}}\thinspace^{214}\text{Po}+e^{-}+\overline{\nu_{e}},\ \  & ^{214}\text{Po}\thinspace\xrightarrow{236\thinspace{\rm \mu s}}\thinspace{}^{210}\text{Pb}+\alpha\ \ (\text{in the \ensuremath{^{238}\text{U}} chain})\thinspace,\label{eq:3-5}\\
^{212}\text{Bi}\thinspace\xrightarrow{87.4\thinspace{\rm min}}\thinspace^{212}\text{Po}+e^{-}+\overline{\nu_{e}},\ \  & ^{212}\text{Po}\thinspace\xrightarrow{413\thinspace{\rm ns}}\thinspace{}^{208}\text{Pb}+\alpha\ \ (\text{in the \ensuremath{^{232}\text{Th}} chain})\thinspace.\label{eq:3-6}
\end{align}
Here $^{214}\text{Bi}$ and $^{212}\text{Bi}$ are $\beta$ emitters,
and their decays are followed by $\alpha$ decays with a mean life
of 236 ${\rm \mu s}$ or 413 ${\rm ns}$. This feature allows one
to evaluate the amount of radon in the detector and to infer the contamination
by isotopes in the $^{238}\text{U}$ and $^{232}\text{Th}$ chains.
However, successful tagging $^{214}\text{Bi}$-$^{214}\text{Po}$
and $^{212}\text{Bi}$-$^{212}\text{Po}$ does not imply that the
background caused by other isotopes in the $^{238}\text{U}$ and $^{232}\text{Th}$
chains can be effectively suppressed. For example, the $^{210}\text{Bi}$-$^{210}\text{Po}$
decay sequence has mean-life times of 7.23 and 200 days. Hence the
background of $^{210}\text{Bi}$ and $^{210}\text{Po}$ (shown in
Fig.~\ref{fig:bore-bkg}) can not be reduced by time coincidence.


\newpage

\section{Prospects of solar neutrino experiments}

Future neutrino experiments will feature higher statistics, lower
backgrounds, better energy and angular resolution, and new detection
channels. Table~\ref{tab:nudetectors} summarizes past, present
and future (including under-construction and proposed) neutrino detectors for solar neutrino measurements. 

\begin{table}[h]
	\centering \caption{Summary of past, present, and future solar neutrino detectors. }
	\label{tab:nudetectors} 
	
	\begin{tabular}{lccrclc}
		\toprule 
		Detectors & Depth  & Cosmic $\mu^{\pm}$ flux & Type  & Fiducial mass  & Live period  & Location\tabularnewline
		& {[}m{]} & {[}$\text{cm}^{-2}\text{s}^{-1}${]} &  & {[}ton{]} &  & \tabularnewline
		\midrule 
		Homestake~\cite{Cleveland:1998nv} & 1478  & 4.4$\times10^{-9}$~\cite{Theia:2019non} & ${\rm C}_{2}{\rm Cl}_{4}$ & 615 & 1968-1994 & South Dakota, USA \tabularnewline
		GALLEX/GNO~\cite{GNO:2005bds} & 1400  & $3.3\times10^{-8}$~\cite{Borexino:2019wln} & $\text{Ga}\text{Cl}_{3}$ & 
		$30.3_{\rm (Ga)}$ 
		& 1991-2003 & Gran Sasso, Italy\tabularnewline
		SAGE~\cite{SAGE:2009eeu} & 2100  & 3$\times10^{-9}$~\cite{SAGE:2009eeu} & Ga metal & 50 & 1989-2007 & Baksan, Russia\tabularnewline
		SNO~\cite{SNO:2020gqd} & 2092  & $3.3\times10^{-10}$~\cite{SNO:2009oor} & ${\rm D}_{2}{\rm O}$ & 1k & 1999-2006 & Sudbury, Canada\tabularnewline
		SK I-IV~\cite{Super-Kamiokande:2016yck}
		& 1000  & $\sim10^{-7}$~\cite{Abe:2011ts} & Water & 22.5k & 1996-2018 & Kamioka, Japan \tabularnewline
		KamLAND~\cite{KamLAND:2011fld,KamLAND:2014gul} & 1000 & $\sim10^{-7}$~\cite{Abe:2011ts} & LS & 1k & 2002-2011 & Kamioka, Japan\tabularnewline
		Borexino~\cite{Borexino:2013zhu}  & 1400  & $3.3\times10^{-8}$~\cite{Borexino:2019wln} & LS & 278 & 2007-2021 & Gran Sasso, Italy \tabularnewline
		SNO+~\cite{SNO:2021xpa} & 2092 & $3.3\times10^{-10}$~\cite{SNO:2009oor} & LS & 800 & 2018- & Sudbury, Canada\tabularnewline
		SK-GD~\cite{MartiMagro:2021ygs,Goldsack:2022fkr} & 1000  &  $\sim10^{-7}$~\cite{Abe:2011ts} & Water & 22.5k & 2020- & Kamioka, Japan \tabularnewline
		JUNO~\cite{JUNO:2021vlw} & 700  & 4$\times10^{-7}$~\cite{JUNO:2021vlw} & LS & 20k & Future & Jianmeng, China\tabularnewline
		Hyper-K~\cite{Hyper-Kamiokande:2018ofw,Abe:2011ts} & 650  & $\sim10^{-6}$~\cite{Abe:2011ts} & Water & 187k & Future & Kamioka, Japan\tabularnewline
		DUNE~\cite{DUNE2016,DUNE:2020ypp} & 1500 & 4.4$\times10^{-9}$~\cite{Theia:2019non} & LAr & 40k & Future & South Dakota, USA\tabularnewline
		THEIA~\cite{Theia:2019non} & 1500 & 4.4$\times10^{-9}$~\cite{Theia:2019non} & WbLS & 25k/100k & Future & South Dakota, USA\tabularnewline
		JNE~\cite{Jinping:2016iiq} & 2400 & $2.6\times10^{-10}$~\cite{JNE:2020bwn} & SLS & 2k & Future & Jinping, China\tabularnewline
		\bottomrule
	\end{tabular}
\end{table}

As can be seen from the table, water Cherenkov and liquid scintillator (LS) detectors are still taking their roles in current and future solar neutrino observations.
The low-background LS experiment Borexino has succeeded dramatically
as a solar neutrino observatory. The measurement of $^{7}\text{Be}$
neutrinos, which was its primary scientific goal, has achieved excellent
precision ($\sim5\%$)~\cite{Bellini:2011rx}. Moreover, it measured the pp neutrino flux with a precision of $\sim10\%$~\cite{BOREXINO:2014pcl}. It has
been able to resolve (at 7$\sigma$ C.L.)
the component of CNO neutrinos from complex backgrounds and other
solar signals~\cite{Borexino:2022pvu}.

It is worth mentioning that these remarkable achievements were accomplished
by Borexino with only 278 ton LS in the fiducial volume, while its
successors, including SNO+ and JNE, will have significantly larger fiducial
masses (800 tons and 2 kilotons, respectively) and lower backgrounds (by two
orders of magnitude in terms of cosmic muon flux)---see Tab.~\ref{tab:nudetectors}
and Fig.~\ref{fig:muon_flux}. KamLAND, as a long-baseline reactor
neutrino experiment, has played an important role in both
the measurements of solar neutrino mixing parameters ($\theta_{12}$
and $\Delta m_{12}$) and solar neutrino fluxes~\cite{KamLAND:2011fld,KamLAND:2014gul}, and 
JUNO can be viewed as its successor, with the fiducial mass 20 times as large as that of KamLAND. 

On the other hand, the capability of water-based detectors to detect
solar neutrinos will  be substantially improved with the Super-Kamiokande
detector running in the Gd-doped phase (since 2020), and the future
upgrade---Hyper-Kamiokande. 
THEIA, a Water-based Liquid Scintillator
(WbSL) experiment, has been proposed, with a 25 kt fiducial mass
at the initial phase and a possible upgrade to 100 kt~\cite{Theia:2019non}.
WbSL features the directionality of water Cherenkov detectors, 
LS-like energy measurement (low energy thresholds, high energy resolution), 
and high chemical solubility for loading
isotopes like $^{7}\text{Li}$ to measure low-energy neutrino
spectroscopy. 

In addition, the technology of liquid noble gas (liquid Ar, liquid
Xe) detectors advances rapidly, and the scale of such detectors will
expand considerably in the upcoming years. It might provide 
new insight into solar neutrino physics. For example, the low-threshold elastic scattering in such detectors can be sensitive to new physics related to neutrino magnetic moments and light mediators, as reviewed in Sec.~\ref{sub:new-ph}.

Below, we elucidate the improvement and novelty of the next-generation
neutrino detectors and discuss the expected gains in physics from these experiments.

%

\subsection{Water Cherenkov detectors}

Historically, 
Kamiokande~\cite{Kamiokande-II:1989hkh,Kamiokande-II:1992hns,Kamiokande-II:1987idp,Kamiokande:1996qmi} and Irvine-Michigan-Brookhaven (IMB)~\cite{Casper:1990ac,IMB:1988suc}
experiments based on the water Cherenkov technique have made groundbreaking contributions to solar and atmospheric neutrino observations and the discovery of supernova 1987A neutrinos. Today water-based Cherenkov detectors, due to their low costs, mature technologies, and the feature of being easy to expand, are still a good option for future neutrino experiments.

\subsubsection{SK-Gd}

As the successor of Kamiokande,  Super-Kamiokande (SK) is a gigantic water Cherenkov detector located 1 km underground in the Kamioka mine in Hida City, Gifu Prefecture,
Japan. It consists of a cylindrical tank measuring 39.3 m in diameter
and 41.4 m in height, filled with 50 kilo-ton water, and equipped with 11129
inner\footnote{Initially, the number of inner PMTs was 11146, reduced to 5182 after
	an accident of chain-reaction PMT implosions in 2001, and finally replenished
	to 11129 in 2006. } and 1885 veto PMTs~\cite{Super-Kamiokande:2016yck}.  Since 1996,
it has undergone four data-taking phases, SK-I to SK-IV, and accomplished
a series of profound measurements of solar, atmospheric, and accelerator
neutrino oscillations, which are crucial to the now-established framework
of three neutrino mixing. 

Since 2020, SK has entered the SK-Gd phase in which Gadolinium (Gd) in
the form of $\text{Gd}_{2}(\text{SO}_{4})_{3}\cdot8\text{H}_{2}\text{O}$
has been added to the pure water for high efficiency
of neutron tagging, as initially proposed by Beacom and Vagins~\cite{Beacom:2003nk}.
The neutron capture by Gd has a much larger cross section than that by Hydrogen (which would be the case if it is pure water) and also produces
a clear signal presenting as an 8 MeV gamma cascade. With only a 0.02\%
concentration of Gd sulfate octahydrate in water, the neutron-Gd capture rate
can reach approximately the same as the neutron-Hydrogen capture rate~\cite{MartiMagro:2021ygs}. For a $0.2\%$ concentration,
$90\%$ of neutrino captures will be on Gd.  The high efficiency
of neutron tagging will drastically enhance SK's capability to detect
the diffuse supernova background (DSNB)~\cite{Super-Kamiokande:2019xnm}\footnote{Also referred to as Supernova Relic Neutrinos (SRN) in the literature\,---\,see, e.g., SK measurements~\cite{Super-Kamiokande:2002hei,Super-Kamiokande:2011lwo} and   early theoretical calculations~\cite{Fukugita:2002qw,Ando:2002zj,Strigari:2003ig}. 
The name of DSNB occurred later~\cite{Beacom:2005it,Strigari:2005hu,Beacom:2010kk} but has become more frequently used in recent years.
}. 

For solar neutrinos, this operation also makes an impact. With the loading of
Gd, SK will be able to separate well solar $^{8}{\rm B}$ neutrino
events from cosmogenic background events, which are usually accompanied by the production of neutrons in $^{16}\text{O}$ spallation---see
Sec.~\ref{sub:cos-bkg}. Hence the measurement of solar $^{8}{\rm B}$
neutrinos will be improved, due to the reduced background. 
Consequently, the measurements of the upturn due to the MSW effect in the Sun and the day-night asymmetry due to
the Earth's matter effect are also expected to be improved.
The SK-Gd detector, with its reduced background, might be able to see a hint of hep neutrinos
since the expected event rate of hep neutrinos in the 15.5-19.5 MeV range is about 0.064 event/year/kt~\cite{Super-Kamiokande:2016yck}, high enough for the 22.5 kt detector to detect a few events within several years (In this case, the background reduction is crucial).
In addition, searches for solar antineutrinos
(see Sec.~\ref{sub:new-ph}) will benefit substantially from the high
efficiency of neutrino tagging in an IBD event.

\subsubsection{Hyper-Kamiokande}

\begin{figure}
	\centering
	\includegraphics[width=0.6\textwidth]{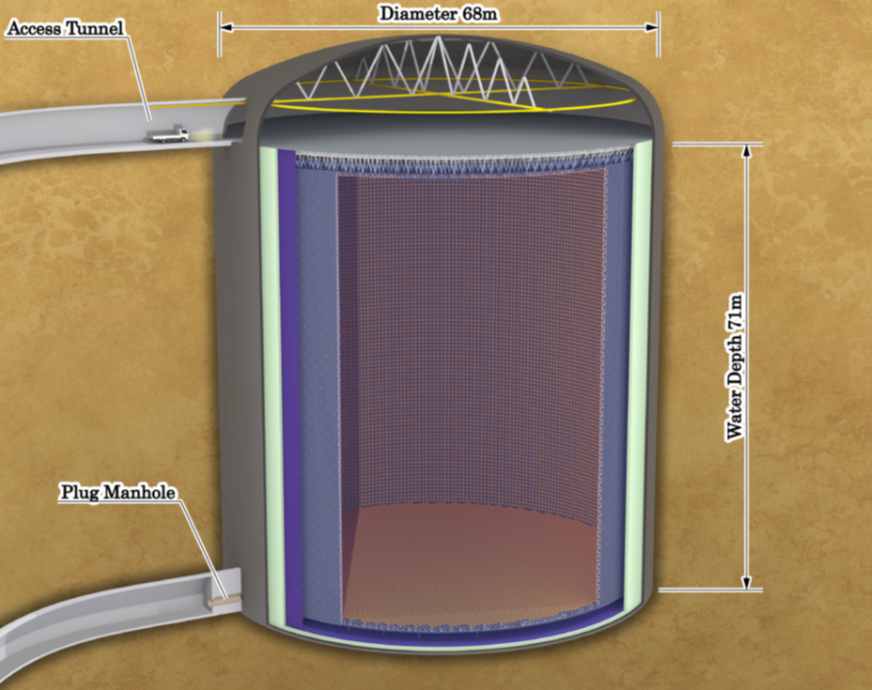}\caption{The schematic view of the Hyper-Kamiokande water Cherenkov detector~\cite{Hyper-Kamiokande:2018ofw}.}
	\label{fig:hyperk}
\end{figure}

The Hyper-Kamiokande (HK), located 8 km south of the SK site, is a next-generation water Cherenkov detector. It will expand the scale enormously to promote its goals for various physics topics~\cite{Hyper-Kamiokande:2018ofw}.
The primary goal is to study CP violation in neutrino oscillation. Meanwhile it also serves as one of the most important experiments for high-statistics solar neutrino observations and proton decay searches. 
Figure~\ref{fig:hyperk} shows a schematic view of the HK detector. 
The cylindrical tank of HK has a double-cylinder structure, with the outer (inner) cylinder measuring 74 m (70.8 m) in diameter and 60 m (54.8 m) in height. It will be  filled with 258 kt water (fiducial mass 187 kt) and equipped with 40,000 inner and 6,700 outer PMTs~\cite{Hyper-Kamiokande:2018ofw}.
The construction has been undertaken since
2020 and is expected to collect data starting from 2027~\cite{Tashiro:2022bgd}. 

HK uses PMTs with higher single-photon
detection efficiency, 24\% (to be compared with 12\% in SK), and better
single-photon timing resolution, 1 ns (to be compared with 2-3 ns
in SK)---see Table VII in Ref.~\cite{Hyper-Kamiokande:2018ofw}.
Therefore, the energy threshold of detecting low-energy electrons
might be further improved, possibly below 3.5 MeV\footnote{At pointed out in Ref.~\cite{Hyper-Kamiokande:2018ofw}, page 185,
	the background caused by $^{214}\text{Bi}$ beta decay which has an
	endpoint energy of 3.27 MeV may severely limit the energy threshold.}. With this threshold, the up-turn predicted by the MSW effect could
be probed at 3$\sigma$ (5$\sigma$) C.L. within 2 (11) years of exposure
to solar $^{8}{\rm B}$ neutrinos~\cite{Hyper-Kamiokande:2018ofw}. 
However, due to the thinner rock overburden, the cosmic-ray spallation background will be higher, affecting the solar neutrino study. A specific analysis of this background is necessary and has already been performed at SK.
The sensitivity of the day-night asymmetry measurement is within the range of 4-8$\sigma$ depending on the systematics for 10 years of exposure.
With a 10-year exposure, HK would be able to measure hep solar neutrinos at 3.2$\sigma$ C.L. if the spallation background could be neglected. 


\subsection{Liquid scintillator detectors}

Liquid scintillator, due to its high light yield, is commonly used
for MeV neutrino detection. In Borexino, the LS used for detection
is a solution of 2,5-diphenyloxazole (PPO) in pseudocumene (PC, 1,2,4-trimethylbenzene)~\cite{Borexino:2013zhu}.
Now linear-akyl-benzene (LAB) is more commonly adopted as the solvent
(e.g.~the JUNO LS is a solution of 2.5 g/L PPO in LAB, with additional
bis-MSB 
 as a wavelength shifter) due to a number of optical
advantages as well as the comparability with acrylic vessels~\cite{SNO:2021xpa,JUNO:2021vlw}.


\subsubsection{JUNO}

\begin{figure}
	\centering 
	
	\includegraphics[viewport=0cm 0cm 1191bp 842bp,width=0.8\textwidth]{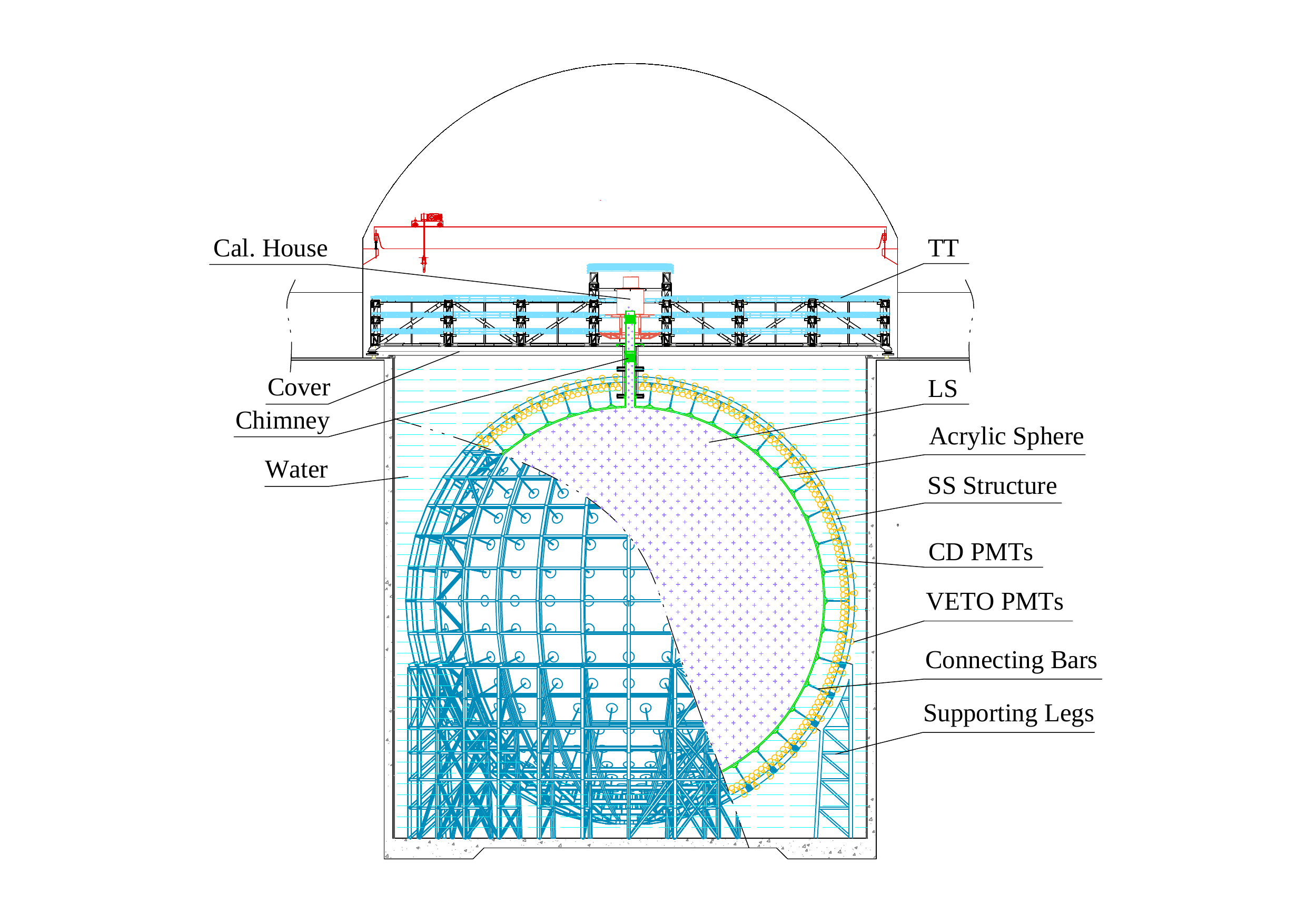}
	\caption{A schematic view of the JUNO detector~\cite{JUNO:2021vlw}.}
	\label{fig:juno} 
\end{figure}

The Jiangmen Underground Neutrino Observatory (JUNO) is currently
the largest LS detector ever built, containing 20 kt LS in a spherical
acrylic vessel with an inner diameter of 35.4 m. The acrylic vessel
is submerged in a water pool and roofed by Top Tracker (TT) for muon
veto---see Fig.~\ref{fig:juno} for a schematic view of the detector~\cite{JUNO:2021vlw}.

The detector is equipped with 15,000 Microchannel-Plate (MCP) PMTs
and 5,000 dynode-type PMTs. The PMTs have high photon detection efficiencies: 28.9\% for
MCP PMTs and 28.1\% for dynode PMTs. Due to the high performance
of the PMTs deployed, together with 80\% PMT coverage, JUNO
features an unprecedented energy resolution. The expected yield of photoelectrons can reach 1345 per MeV, which is significantly higher than
Borexino ($\sim500$ photoelectrons/MeV). Consequently, the relative
energy resolution, mainly determined by the statistics of
photoelectrons, will reach an unprecedented level, $3\%/\sqrt{E/{\rm MeV}}$~\cite{JUNO:2021vlw}.

Like KamLAND, JUNO is primarily a long-baseline reactor neutrino experiment
that is sensitive to the long oscillation mode (subject to $\theta_{12}$
and $\Delta m_{21}^{2}$) of reactor neutrinos. The oscillation parameters,
$\sin^{2}\theta_{12}$ and $\Delta m_{21}^{2}$, which are relevant
to  the flavor conversion of solar neutrinos,
will be measured to $0.5\sim0.7\%$ relative precision~\cite{JUNO:2015zny}.
	The high precision measurements of $\theta_{12}$ and $\Delta m_{21}^{2}$ at JUNO are very important to future solar neutrino physics because they can be used to calibrate solar neutrino oscillations.

JUNO has the advantage of conducting a high-statistics measurement of solar neutrino fluxes. However, the cosmogenic background is high due to the relatively shallow overburden (700 m). This shortcoming, which was also a concern in KamLAND's measurement of solar neutrinos, can be partially compensated by stringent muon spallation cuts, enormously high statistics, and significantly improved energy resolution in JUNO. 

The primary channel for solar neutrino detection in JUNO is elastic
$\nu+e^{-}$ scattering. According to the study in  Ref.~\cite{JUNO:2020hqc},
a 10-year data taking will generate 60,000 recoil electrons and 30,000 background events. 
Therefore, in addition to the aforementioned precision measurements of $\theta_{12}$ and $\Delta m^2_{21}$ using reactor neutrinos, JUNO is also capable to measure them using solar neutrinos. The precision is expected to be around $7\%$ and $16\% \sim 21\%$ for  $\sin^2\theta_{12}$ and $\Delta m^2_{21}$ respectively~\cite{JUNO:2020hqc}.

In addition to elastic scattering, a considerable
amount of $^{13}\text{C}$ nuclei in LS ($9\times10^{30}$ per 20kt) can also allow JUNO to detect solar neutrinos via the CC and NC reactions~\cite{Suzuki:2012aa,Suzuki:2019cra}:
\begin{align}
{\rm CC}:\  & \nu_{e}+{}^{13}\text{C }\to e^{-}+{}^{13}\text{N}\thinspace,\ \ \ \ \ \ \ \ \ E_{\nu}^{{\rm thre}}=2.2\text{MeV}\thinspace,\label{eq:JUNOCC}\\
{\rm NC}:\  & \nu+{}^{13}\text{C }\to\nu+{}^{13}\text{C}^{*}(3^{-}/2)\thinspace,\ \ E_{\nu}^{{\rm thre}}=3.7\text{MeV}\thinspace.\label{eq:JUNONC}
\end{align}
A preliminary estimation assuming $100\%$ detection efficiency and
a 200 kt$\cdot$year exposure indicates that the CC (NC) channel would
observe 3768 (3165) and 14 (13.5) events for $^{8}\text{B}$ and hep
neutrinos, respectively~\cite{JUNO:2021vlw}. However,
background-driven cuts imposed on them may substantially reduce the
actual numbers of signal events, which requires a more dedicated study. 


\subsubsection{SNO+}

Upgraded from SNO, the SNO+ experiment replaces the heavy water in SNO with LAB-PPO LS
(PPO concentration 2g/L). It recycles SNO's acrylic vessel, PMTs,
support structure, light water system, and electronics and trigger
system~\cite{SNO:2021xpa,SNOPLUS2016}. 
Although the primary scientific goal is to search for neutrino-less double beta decay, SNO+ can still continue its  solar neutrino study. One of
the most significant advantages that SNO+ inherits from SNO is the ultra-low cosmogenic
background, which is almost negligible above 6~MeV for solar neutrino events---see previous
discussions in Sec.~\ref{sub:cos-bkg} and Fig.~\ref{fig:sk-sno-angle}. 

%


Serving as a low-background solar neutrino detector, SNO+  can provide precision measurements of solar neutrinos. 
	According to the sensitivity study in Ref.~\cite{SNOPLUS2016}, with one year of data in the LS phase, the $^7$Be and $^8$B fluxes can be measured to 4\% and 8\%, respectively.
Due to its low cosmogenic background and large target volume compared to Borexino, SNO+ may significantly contribute to observing CNO neutrinos. But it will rely on how well the radioactive background can be obtained and if the direction information can be extracted as what Borexino has achieved. 
Assuming a conservative uncertainty for separating the $^{210}$Bi background and CNO, the predicted uncertainty of the CNO flux measurement is 15\%~\cite{SNOPLUS2016}.
It should be emphasized that even in the phase of neutrino-less double beta decay searches with Te loaded,  SNO+ can still detect $^{8}$B neutrinos down to the endpoint of Te double beta decay, 2.53 MeV. Precision measurements of $^{8}$B neutrinos above this endpoint allows SNO+ to probe the up-turn predicted by the MSW mechanism. 

\subsection{Hybrid detectors }

Can the 
virtues 
of water Cherenkov detectors (good directionality) and pure LS detectors (good energy resolution, low-energy thresholds) be combined in MeV-scale neutrino detectors? Recent efforts in developing water-based LS (WbLS) and slow LS will make the combination possible in future hybrid detectors. 

\subsubsection{THEIA}

THEIA (after the Titan Goddess of light)~\cite{Theia:2019non} is a proposed large-scale (25-100 kt) multi-purpose detector
with a broad range of physics goals, including measurements of long-baseline
neutrino oscillations (e.g.,~serving as one of DUNE's far detectors),
solar neutrinos, supernova neutrinos, neutrinoless double beta
decay, etc. The experiment intends to adopt WbLS, which has been actively investigated as a future alternative to LS~\cite{Yeh:2011zz}.   The basic idea of WBLS
is to mix LS with water so that ionization (the dominant energy deposited by a recoil electron) can be effectively
converted to optical signals.
Meanwhile, the observability of Cherenkov light is retained with new PMTs or ultrafast LAPPDs (Large Area Picosecond Photo-Detectors). The amount of LS added to water typically varies
from 1\% to 10\%, depending on the designed light yield,
the PMTs' capability of Cherenkov/scintillation separation,  as well
as the cost and environmental concerns. After considering these issues, Ref.~\cite{Theia:2019non}
suggested that the light yield can be as high as 10\% of LS. 
 Hence WbLS can partially inherit the excellent energy resolution from LS. 

With the capability to perform both direction and energy measurements, THEIA would be a powerful solar
neutrino observatory featuring high-statistics, low-threshold
(MeV-scale) observations. For instance, it could improve the CNO neutrino measurement to the precision of $4\%\sim10\%$, assuming 100
kt WbLS with 5\% LS is used---see Fig.~9 and Tab.~4 in Ref.~\cite{Theia:2019non}.
Such a high precision measurement would allow THEIA to  resolve
the solar metallicity problem conclusively. 
The collaboration is also  studying the physics potential of adding $^7$Li to the detector, and the impact of isotope loading.


\subsubsection{JNE}

\begin{figure}
	\centering 
	
	\includegraphics[width=0.95\textwidth]{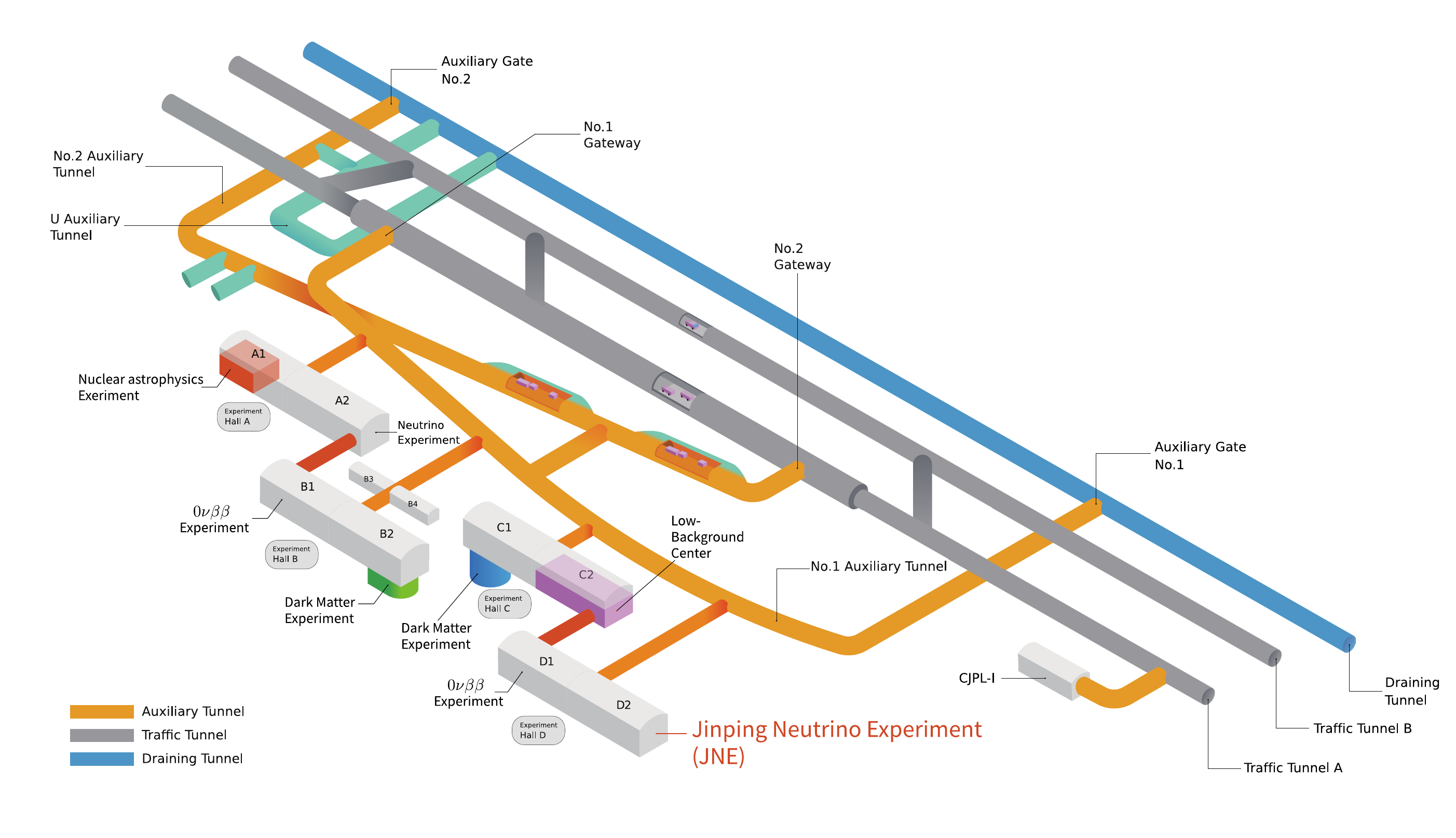} 
	\caption{The layout of CJPL, including CJPL-I and CJPL-II. It may be subject to changes with new hall excavations. Figure provided by CJPL. 
	}
	\label{fig:CJPL} 
\end{figure}

\begin{figure}
	\centering 
	
	\includegraphics[width=0.6\textwidth]{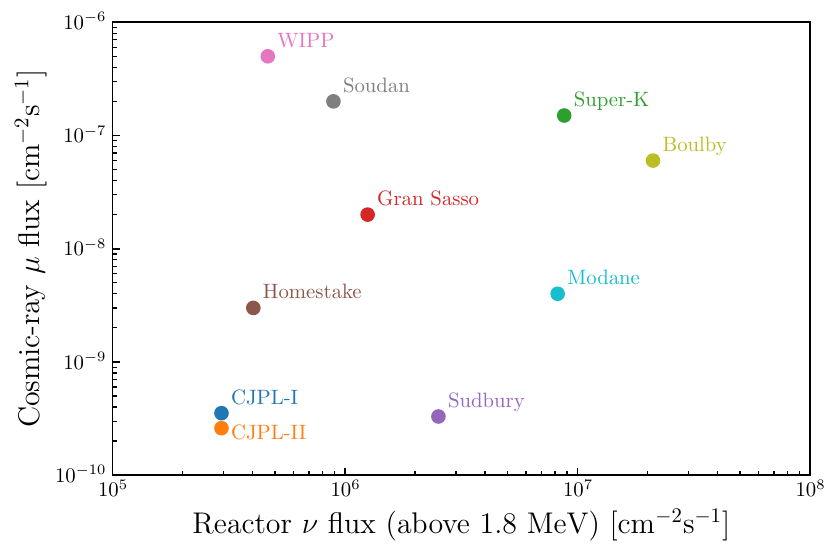} \caption{A comparison of cosmic-ray muon fluxes and reactor neutrino fluxes
		among underground laboratories. The muon flux data is taken from Ref.~\cite{Jinping:2016iiq} and  updated according to Ref.~\cite{JNE:2020bwn}. The reactor neutrino data is taken from Ref.~\cite{IAEA-reactor}.
  }
	\label{fig:jinping_flux}
\end{figure}

The Jinping Neutrino Experiment (JNE)~\cite{Jinping:2016iiq,JNE:2020bwn,JNE:2021cyb}
is a neutrino observatory for low-energy neutrino physics, astrophysics, 
and geophysics, to be built in the China JinPing underground Laboratory
(CJPL)~\cite{Cheng:2017usi},  located around 2400 m below
Jinping Mountain, Sichuan Province, China. 
 CJPL was constructed in two phases (CJPL-I and CJPL-II) and completed all the tunnel excavation in 2016, as shown in Fig.~\ref{fig:CJPL}.  
Among all underground laboratories, CJPL has the lowest vertical cosmic muon flux and also the lowest reactor neutrino background. The total cosmic-ray muon flux at CJPL-I is measured to be $(3.53\pm0.22_{{\rm stat.}}\pm0.07_{{\rm sys.}})\times 10^{-10}\text{cm}^{-2}\text{s}^{-1}$~\cite{JNE:2020bwn}. The expected flux for the four major lab halls at CJPL-II varies with the individual hall location. Using the CJPL-I measurement extrapolated to CJPL-II  gives  about $2.6\times 10^{-10}\text{cm}^{-2}\text{s}^{-1}$. 
Figure~\ref{fig:jinping_flux} compares its cosmic-ray
muon flux and reactor neutrino flux with other underground laboratories, 
using data from Refs.~\cite{Jinping:2016iiq, JNE:2020bwn,IAEA-reactor}.
The low background makes CJPL an ideal site for low-energy neutrino observations. 

JNE plans to employ a new type of LS, known as slow LS, which features
both high light yield and the capability of Cherenkov/scintillation
separation~\cite{Guo:2017nnr,Biller:2020uoi}. By reducing the
concentration of the primary fluor, the pulse shape of the scintillation
light is stretched to several tens of nanoseconds, allowing the prompt
Cherenkov light component to be distinguishable. This new technique will enable JNE to measure both the electron direction and its energy, which are crucial in solar neutrino experiments.

Suppose JNE can fulfill its proposal of multi-kilo-ton fiducial target mass. In that case, it will provide high-precision measurements of solar neutrino fluxes, particularly the low-energy components like pp and $^{7}\text{Be}$ neutrinos. Assuming 500 photoelectrons per MeV can be attained (depending on the light yield and the PMT efficiency), these two neutrino fluxes can be measured within 0.5\% and 0.4\% statistical uncertainties, respectively---see Tab.~3.5 in Ref.~\cite{Jinping:2016iiq}. 
	In addition, JNE will also be able to observe CNO neutrinos with more than $5\ \sigma$ statistical significance. Assuming the oscillation parameters are well determined with high precision, the high and low metallicity hypotheses can be resolved at more than $5\ \sigma$ C.L.~\cite{Jinping:2016iiq}.
 These achievements require demonstrating the Cherenkov scintillator technique in a large-volume neutrino detector. It needs the PMT waveform output and analysis in both online and offline stages. Despite the lowest cosmogenic background, low energy radioactive background control will be another challenging issue.  
The group is also investigating the possibility of loading lithium~\cite{Shao:2022yjc} or gallium~\cite{Wang} to enrich its solar neutrino research program.

\subsection{DUNE}

\begin{figure}
	\centering 	
	\includegraphics[width=0.8\textwidth]{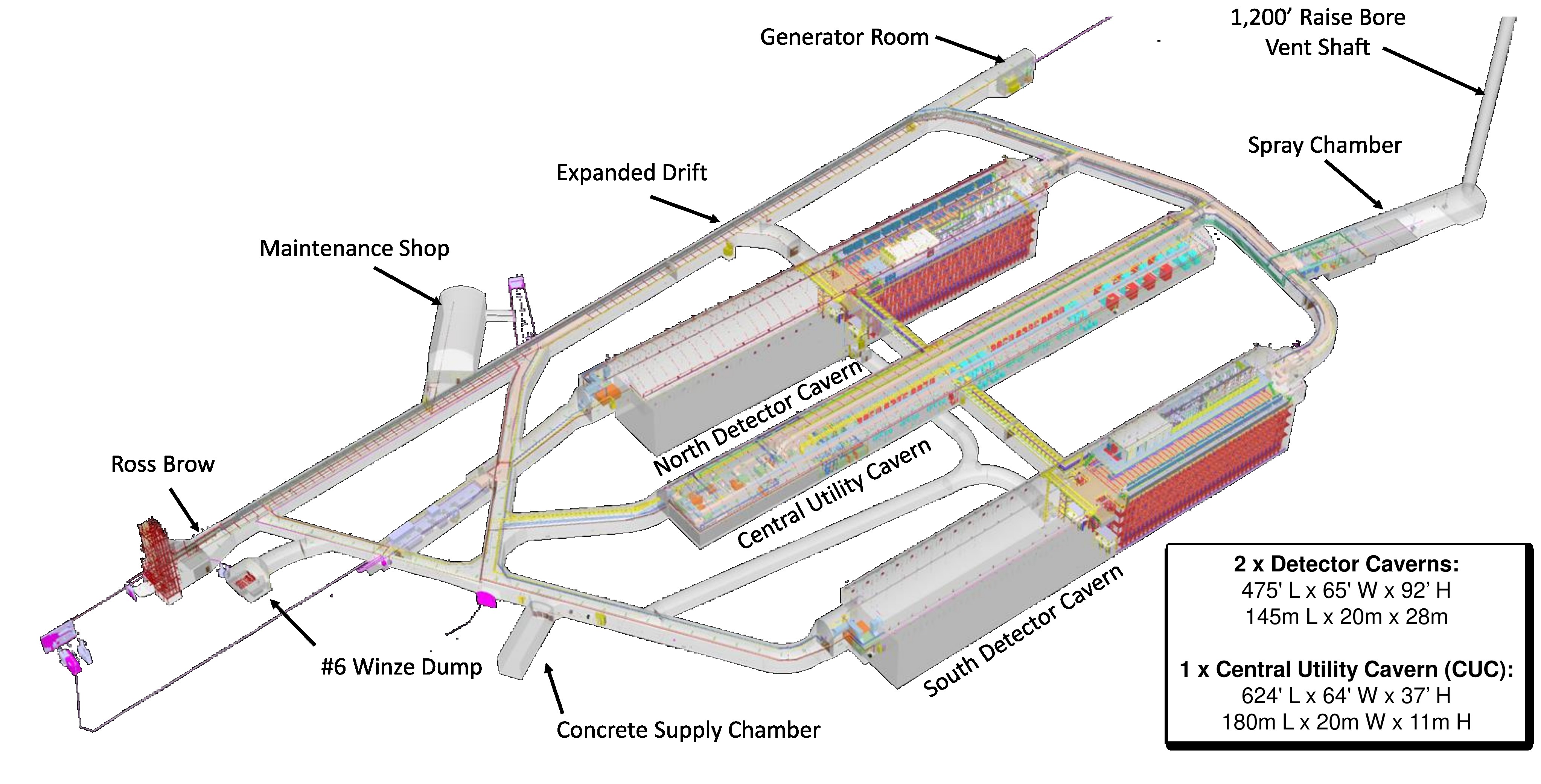} \caption{
 Far-site underground caverns for DUNE at SURF, with an overburden of 1478 m (~4300 m.w.e). The vertical shaft is on the right.  
 Phase I will have two detector modules.
 Figure taken from Ref.~\cite{DUNE:2020ypp}}
	\label{fig:dune_farsite}
\end{figure}

The Deep Underground Neutrino Experiment (DUNE)~\cite{DUNE2016,DUNE:2020ypp} is an accelerator-based long-baseline experiment at Sanford Underground Research Facility (SURF) in South Dakota, with the primary goal of discovering matter-antimatter asymmetries in neutrino flavor mixing. 
It consists of near and far detectors and is currently under the construction phase.
According to the design, DUNE will eventually be equipped with four Liquid
Argon (LAr) TPC\footnote{TPC stands for Time-Projection Chamber, while LAr TPC was first
	proposed by C. Rubbia~\cite{Rubbia1977} and
	realized in the ICARUS experiment~\cite{Cennini1994,ICARUS2003}.  Due to its tracking ability 
	and high  target material density, LAr TPC was first proposed for $^{8}\text{B}$ solar neutrino detection in the 1980s~\cite{Chen:1982wi}.  } far detectors. 
In practical considerations, DUNE will be built in two phases.
In Phase I, the far site will accommodate two 17-kt detector modules, as shown in Fig.~\ref{fig:dune_farsite}. Each will have at least a 10-kt fiducial liquid argon target mass. The first physics results will appear around 2030. Phase II will fulfill the whole DUNE design goal within the next decade after the completion of Phase I~\cite{DUNE:2022aul}.  

\begin{figure}
	\centering 	
	\includegraphics[width=0.6\textwidth]{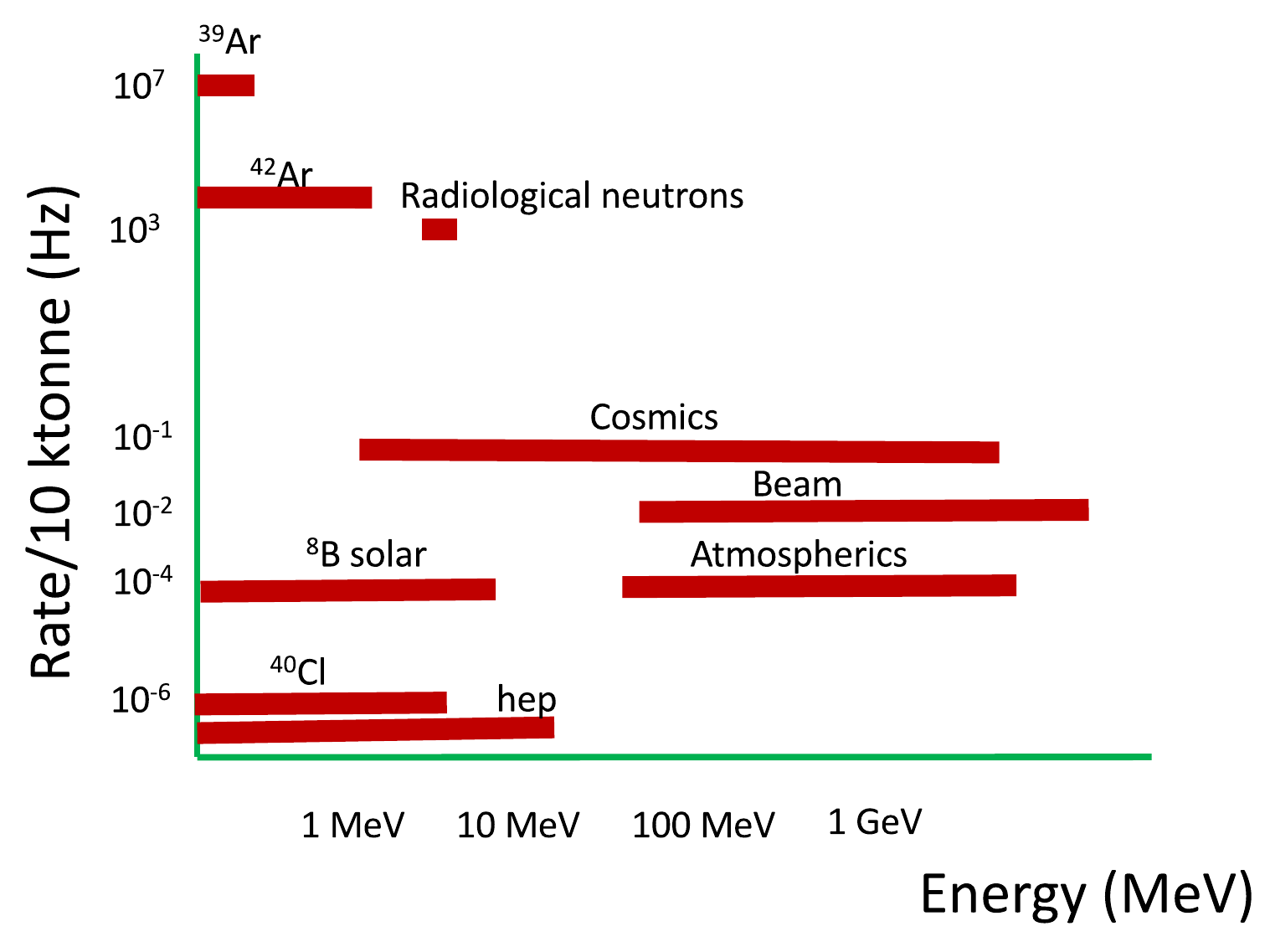} \caption{
 Expected solar neutrino event rates ($^{8}\text{B}$ solar and hep) in DUNE far detectors and compared to other neutrinos, cosmic-ray and radioactive background sources. 
 Figure taken from Ref.~\cite{DUNE:2020txw}.}
	\label{fig:dune_rates}
\end{figure}

As a large-scale neutrino experiment with the next generation of advanced detection technology, DUNE has apparent advantages for solar neutrinos above several MeV. 
In liquid argon, 
solar neutrinos can be detected
either via the elastic $\nu+e^{-}$ scattering or the CC process $\nu_{e}+{}^{40}\text{Ar}\rightarrow e^{-}+{}^{40}\text{K}^{*}$,
which has a threshold of around $4\sim 5$ MeV~\cite{DUNE2016} (see also Tab.~\ref{tab:nu_cap_elements})
The 
large 
fiducial mass gives DUNE an opportunity for future solar neutrino physics study, in particular, for the measurements of $^{8}\text{B}$ and hep solar neutrino fluxes~\cite{Capozzi:2018dat}. 
Figure~\ref{fig:dune_rates} shows the estimated event rates in one module adopting the single-phase far detector, indicating that DUNE would be able to detect $^8$B neutrinos at an event rate of about several counts/day/kT.  If DUNE can well reconstruct the MeV-scale electron track, the measurement of $^8$B and hep fluxes can reach 2.5\% and 11\%,  according to the study in Ref.~\cite{Capozzi:2018dat}. Using the solar neutrino data, the neutrino mixing parameters $\sin^2 \theta_{12}$ and $\Delta m_{21}^2$ can be measured to $3\%$ and $6\%$, respectively, better than the combination of all solar experiments to date. 

Since the CC process of ${}^{40}\text{Ar}$ has a larger cross section than $\nu+e^{-}$ scattering and can directly relate the electron energy to the  neutrino energy,  
DUNE will have the best chance to improve $^8$B flux precision and discover hep neutrinos --- 
the last undiscovered solar neutrino component in the pp chain.
However, these prospects rely on the performance of LAr TPC working in a large target volume and MeV energy scale. The challenging issues include the suppression of the radioactive background and the improvement of calibration at low energies~\cite{Parsa:2022mnj}.  

\subsection{Other experimental approaches}

 Despite many efforts to build large solar neutrino experiments, there are other experimental approaches to solar neutrino observations using small detectors. The smallness is a consequence of either large neutrino cross sections (in dark matter detectors) or higher neutrino fluxes (in space-based detectors).

%
%
%
%

\subsubsection{Dark matter detectors\label{sub:DM}}

\begin{figure}
	\centering 
	
	\includegraphics[width=0.7\textwidth]{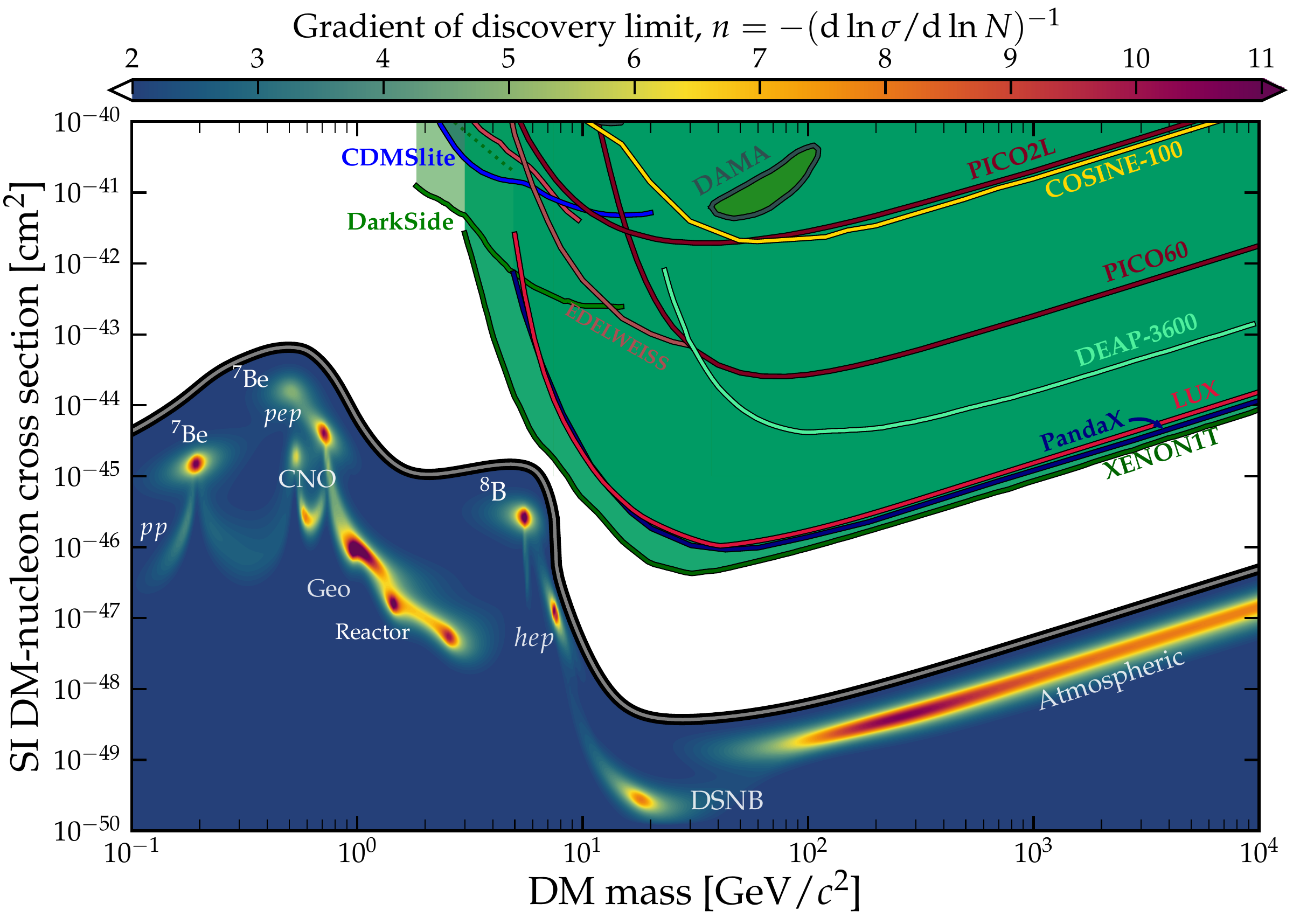} \caption{Solar neutrinos in the neutrino floor for DM direct detection. Upcoming
		liquid Xe experiments will soon be able to detect $^{8}\text{B}$
		solar neutrinos. Figure from Refs.~\cite{OHare:2021utq,OHare:2022jnx}. \label{fig:DM-detection}}
	
	\includegraphics[width=0.6\textwidth]{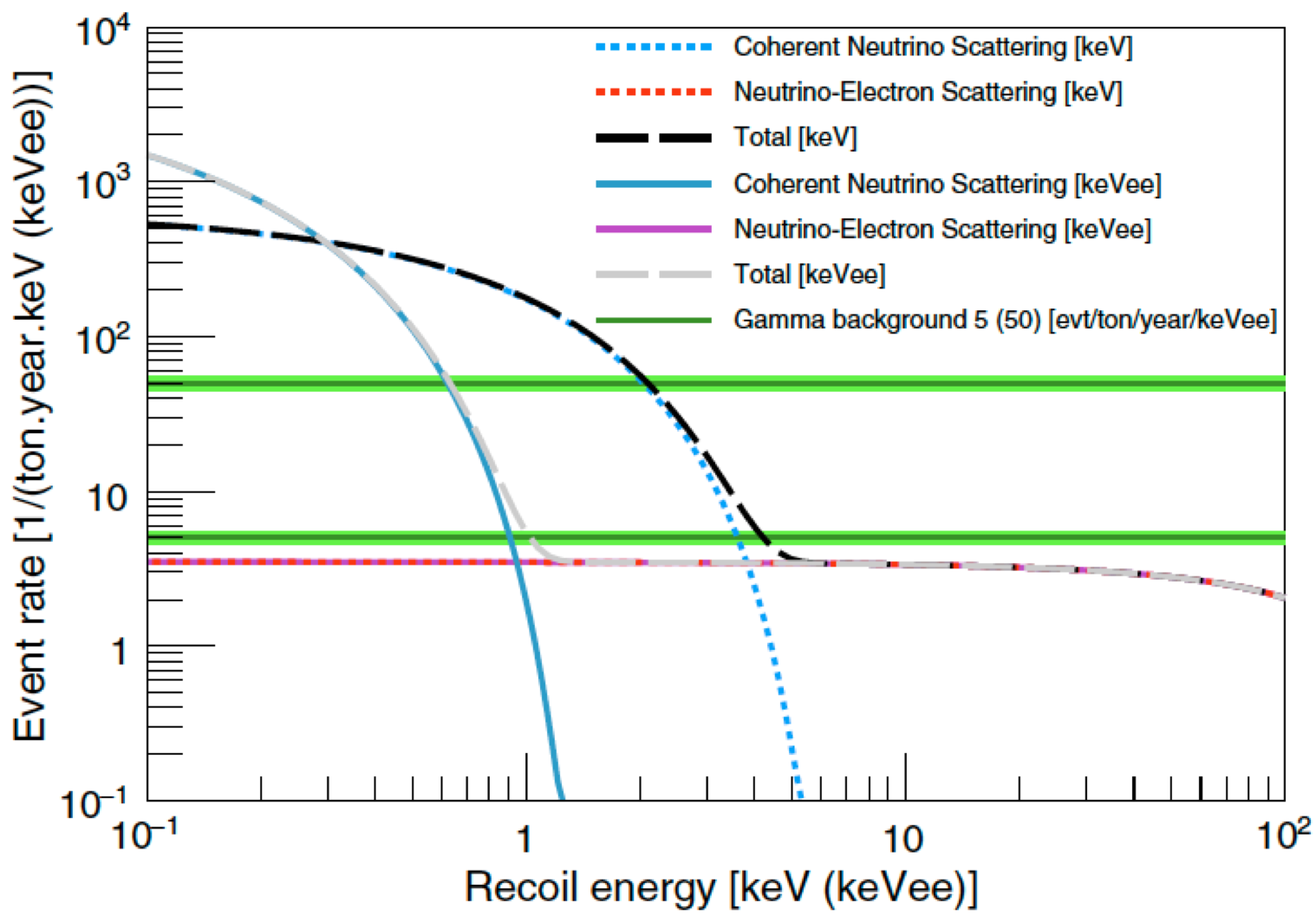} \caption{Solar neutrino event rates in a low-threshold Ge DM detector. The
		true recoil energy is presented in units of keV and the corresponding
		ionization energy, which is smaller than the true recoil energy due
		to quenching, is in units of keVee. Figure taken from Ref.~\cite{Billard:2014yka}.
		\label{fig:DM-detection-2} }
\end{figure}
Dark matter (DM) detectors have the potential to measure solar neutrinos, which comprise the low-energy part of the ultimate background for DM direct detection, known as {\it the neutrino floor} in the literature. 
As the scales and detection thresholds of DM detectors 
have been improved drastically in recent years, they will soon be able to touch the solar neutrino
floor---see Ref.~\cite{OHare:2022jnx} for a recent review. 
Figure~\ref{fig:DM-detection} shows the status as of 2021.  

It is known that the neutrino floor does not strictly stop DM detectors from detecting DM signals underneath the floor.
First, the recoil spectra of DM and neutrinos do not have exactly the same shape. Second, their cross-sections with nuclei may have different dependencies on the neutron and proton composition. These differences could be used to distinguish between DM and neutrino signals if the statistics are high~\cite{Ruppin:2014bra}. 
In addition, for the solar neutrino floor, one could employ the directionality of events to separate DM and neutrino signals~\cite{Mayet:2016zxu}. Finally, annual or diurnal modulation might be present in DM signals~\cite{Davis:2014ama,Sassi:2021umf}. 
All these possibilities of searching for DM signals below the neutrino floor imply an important interplay between DM and solar neutrino detection in the future.


Detecting solar neutrinos in DM detectors relies on elastic neutrino 
scattering with electrons or nuclei. The latter is always
coherent for solar neutrinos and the process is referred to as CE$\nu$NS---see discussions in Sec.~\ref{subsec:CEvNS}.
 In the latest results of XENON-1T~\cite{XENON:2020gfr} and PandaX-4T~\cite{PandaX:2022aac} searching for solar neutrinos via CE$\nu$NS, no significant excess of events has been observed.

CE$\nu$NS generically has a much larger cross section than elastic
$\nu+e^{-}$ scattering, but the recoil energy of a nucleus is much lower than that of an electron. Regarding this aspect, ultra-low
threshold detectors such as those based on Germanium semiconductors might have certain advantages. 
In semiconductor detectors, nuclear recoils lead to ionization and generate electronic signals. The ionization (measured in units of
keVee) is quantitatively related to the recoil energy (in units of
keV) via quenching models, which have not been well calibrated yet
in the sub-keV regime. Figure~\ref{fig:DM-detection-2} shows the
event rates of solar neutrinos expected in a Ge detector using the Lindhard ionization quenching factor model~\cite{Billard:2014yka}. 

Elastic $\nu+e^{-}$ scattering has a relatively small cross section. Hence detecting solar neutrinos via this process would require a large fiducial mass. 
Regarding this aspect, liquid Xe (e.g.,~LUX, PandaX, XENONnT) and Ar (DarkSide-20k~\cite{DarkSide-20k:2017zyg}) detectors  have a significant advantage over semiconductor detectors because the former can be easily scaled up to very large fiducial volumes (multi-ton scales or higher). 


Successful detection of solar neutrinos in DM detectors is foreseeable for the upcoming multi-ton scale experiments. It will not only be crucial to the background study
of DM detectors but also provides an important cross-check on conventional
measurements of solar neutrino fluxes. Potential discrepancies would
imply new physics such as quark NSI, neutrino magnetic moments, etc. 

\subsubsection{Space-based detectors}

The Neutrino Solar Observatory ($\nu$SOL) has been proposed~\cite{Solomey:2022gja}. The idea is to send a neutrino detector with
Gadolinium-Aluminum-Gallium Garnet scintillating crystal into space and operate in orbit close to the Sun. According to the inverse radius square law, the detector can detect the high-intensity solar neutrinos with a not-too-large target volume  practicable for space programme.  At $r=9R_{\astrosun}$, which has been proven feasible by NASA's Parker Solar Probe launched in 2018~\cite{parker},  the solar neutrino intensity is increased by a factor of $600$. Further enhancement up to $10^4$ (corresponding to $r=10^{-2} {\rm AU}\approx 2.2 R_{\astrosun}$) might be possible~\cite{2016SPD}.


Due to the enhanced flux, such a detector can be orders of magnitude smaller than those terrestrial neutrino detectors and yet reach the same high statistics. However, the background would be the severest concern since high-energy cosmic rays cannot be passively shielded in an effective way, though the iron shield can stop low-energy particles (e.g.~electrons below 15 MeV). Active veto rejection techniques are required for background reduction. The technology developed in such an experiment might also be useful for space-based DM direct detection if DM detectors are delivered to locations far away from the Sun to evade the limitation of the neutrino floor. 

The physics gain of space-based solar neutrino detectors is yet to be investigated. One interesting possibility would be to probe off-axis solar neutrinos, which would allow us to verify the distribution of the neutrino production rate, as previously shown in Fig.~\ref{fig:nuradius}. If neutrinos from the outer layers of the Sun (e.g.~$r> 0.3 R_{\astrosun}$) are observed, it might be an important hint of DM accumulating in the Sun and annihilating to neutrinos---see discussions in Sec.~\ref{sub:DM-anni}.

\section{Summary}
``{\it For centuries the heavens have been a natural laboratory to test the classical laws of motion, and more recently to test Einstein's theory of gravity. Today, astrophysics has become a vast playing ground for applications of the laws of microscopic physics, especially the properties of elementary particles and their interactions.}''\footnote{Quoted from Georg G. Raffelt's book {\it Stars as Laboratories for Fundamental Physics}~\cite{Raffelt1996}.}
Indeed, the great achievement of solar neutrino observations, which eventually led to one of the most profound and surprising discoveries in particle physics---neutrino masses, is a perfect example. 

With the upcoming next-generation neutrino detectors featuring  higher statistics, lower backgrounds, and novel detection technologies, can we make another surprising discovery in future solar neutrino observations? The finding of neutrino masses implies the existence of new physics beyond the SM,  which includes a variety of possibilities. Among them, some could potentially modify the standard MSW-LMA solution to the solar neutrino problem, as we have reviewed in Sec.~\ref{sec:th}.  In addition,  a few experimental anomalies and inconsistencies among existing measurements might be hints of the underlying new physics. Moreover, dark matter, with its abundant astrophysical and cosmological evidence of the existence and yet rather elusive particle physics nature, may have been affecting solar neutrino observations in an unnoticed way, calling for further experimental and theoretical investigations.

We can see that some predictions are still waiting for experimental confirmation, even within the most conservative framework. For example, the present experiments are still struggling to justify the day-night asymmetry and the up-turn expected from the standard neutrino oscillation theory. In addition, several components of the solar neutrino spectrum, including both old ones from the pp chain (hep, ${}^{7}{\rm Be}$-II) and new ones from the CNO cycle (e.g.~$^{17}{\rm F}$ decay, $e{}^{13}{\rm N}$ electron capture, etc.), still await experimental probe. 
Observations of CNO neutrinos have just started, with the first success reported recently by Borexino. Future precision measurements of CNO neutrinos will help resolve the long-standing solar metallicity problem and also be of great importance to studying stellar nucleosynthesis in large-mass stars.  

In summary, future solar neutrino observations will produce exciting new results and might make another surprising discovery. As history has demonstrated, studying solar neutrino physics may offer unique insight into not only our nearest star but also the most fundamental physics laws.

\section*{Acknowledgments}
We gratefully acknowledge useful discussions with Evgeny Akhmedov, John F. Beacom, Steven Biller, Alexei Yu.~Smirnov, Ya-Kun Wang and Hanyu Wei.  
This work is funded by the National Natural Science Foundation of China, under grant No. 12127808 and No. 12141501.

\bibliography{mybibfile}

\end{document}